\documentclass[aps,pra,amsmath,two column,amssymb,showpacs]{revtex4}
\usepackage{txfonts}
\usepackage{mathbbold}
\usepackage{epsfig,bm,dcolumn}
\usepackage{graphicx}
\usepackage{color}
\usepackage{overpic}

\begin{document}

\title{BCS-BEC crossover in three-dimensional Fermi gases with spherical spin-orbit coupling}

\author{Lianyi He}
\email{lianyi@itp.uni-frankfurt.de}

\author{Xu-Guang Huang}
\email{xhuang@itp.uni-frankfurt.de}

\affiliation{ Frankfurt Institute for Advanced Studies and Institute
for Theoretical Physics, J. W. Goethe University, 60438 Frankfurt am
Main, Germany}

\date{\today}

\begin{abstract}
We present a systematic theoretical study of the BCS-BEC crossover
problem in three-dimensional atomic Fermi gases at zero temperature
with a spherical spin-orbit coupling which can be generated by a
synthetic non-Abelian gauge field coupled to neutral fermions. Our
investigations are based on the path integral formalism which is a
powerful theoretical scheme for the study of the properties of the
bound state, the superfluid ground state, and the collective
excitations in the BCS-BEC crossover. At large spin-orbit coupling,
the system enters the BEC state of a novel type of bound state
(referred to as rashbon) which possesses a non-trivial effective
mass. Analytical results and interesting universal behaviors for
various physical quantities at large spin-orbit coupling are
obtained. Our theoretical predictions can be tested in future
experiments of cold Fermi gases with three-dimensional spherical
spin-orbit coupling.
\end{abstract}


\maketitle

\section{Introduction}
It has been widely accepted for a long time that, by tuning the
attractive strength in a Fermi gas,  one can realize a smooth
crossover from the Bardeen--Cooper--Schrieffer (BCS) superfluidity
at weak attraction to Bose--Einstein condensation (BEC) of difermion
molecules at strong
attraction~\cite{Eagles,Leggett,BCSBEC1,BCSBEC2,BCSBEC3,BCSBEC4,BCSBEC5,BCSBEC6,BCSBEC7}.
For a dilute Fermi gas in three dimensions where the effective range
$r_0$ of the short-range interaction is much smaller than the
inter-particle distance, the system can be characterized by a
dimensionless parameter $1/(k_{\rm F}a_s)$, where $a_s$ is the
$s$-wave scattering length of the short-range interaction and $k_{\rm
F}$ is the Fermi momentum in the absence of interaction. The BCS-BEC
crossover occurs when the parameter $1/(k_{\rm F}a_s)$ is tuned from
negative to positive values, and the BCS and BEC limits correspond
to the cases $1/(k_{\rm F}a_s)\rightarrow-\infty$ and $1/(k_{\rm
F}a_s)\rightarrow+\infty$, respectively. The BCS-BEC crossover has
also become an interesting issue for the studies of dense nuclear
and quark matter which may exists in the core of compact stars
\cite{BCSBECNM,BCSBECHe,BCSBECQM}.

This BCS-BEC crossover phenomenon has been successfully demonstrated
in ultracold fermionic atoms, where the $s$-wave scattering length and
hence the parameter $1/(k_{\rm F}a_s)$ were tuned by means of the
Feshbach resonance~\cite{BCSBECexp1,BCSBECexp2,BCSBECexp3}. At the
resonant point or the so-called unitary point where
$a_s\rightarrow\infty$, the only length scale of the system is the
inter-particle distance ($\sim k_{\rm F}^{-1}$). Therefore, the
properties of the system at the unitary point $1/(k_{\rm F}a_s)=0$
become universal, i.e., independent of the details of the
interactions. All physical quantities, scaled by their counterparts
for the non-interacting Fermi gases, become universal constants.
Determining these universal constants has been one of the most
intriguing topics in the research of the cold Fermi gases
\cite{Unitary,Unitary2}.

While the BCS-BEC crossover triggered by tuning the attraction
strength between fermions from weak to strong ($1/(k_{\rm F}a_s)$
from $-\infty$ to $+\infty$) has been comprehensively studied both
theoretically and experimentally, it is always interesting to look
for other mechanisms to realize the BCS-BEC crossover. Recent
experimental breakthrough in generating synthetic non-Abelian gauge
field \cite{SOC} has opened up the opportunity to study the
spin-orbit coupling (SOC) effect in cold atomic
gases~\cite{SOC01,SOC02,SOC03,SOC04,3DSOC}. For fermionic atoms, it
may provide an alternative way to realize the BCS-BEC
crossover~\cite{3DBCSBEC}. Apart from engineering cold-atom
analogs to known Hamitonians such as Rashba SOC, synthetic
non-Abelian gauge field can generate SOC that has no known analog
in condensed matter systems.

The spin-orbit coupling for neutral fermions can be generated by a
synthetic SU(2) gauge field. In general, the synthetic vector
potential ${\bf A}$ for spin-1/2 fermions takes the form ${\bf
A}=-(\lambda_x\sigma_x{\bf e}_x+\lambda_y\sigma_y{\bf
e}_y+\lambda_z\sigma_z{\bf e}_z)$~\cite{SOC03,SOC04,3DSOC,3Dbound},
where $\sigma_i$ ($i=x,y,z$) are the Pauli matrices. From the
minimum coupling scheme, the resulting Hamiltonian for spin-1/2
fermions moving in a gauge potential ${\bf A}$ reads ${\cal H}= {\bf
p}^2/(2m)+\mbox{\boldmath{$\sigma$}}\cdot\mbox{\boldmath{$\xi$}}_\lambda$
where
$\mbox{\boldmath{$\xi$}}_\lambda=(\lambda_xp_x,\lambda_yp_y,\lambda_zp_z)$.
The term
$\mbox{\boldmath{$\sigma$}}\cdot\mbox{\boldmath{$\xi$}}_\lambda$ can
be regarded as a generalized Rashba SOC. The gauge field strengths
$\lambda_i$ ($i=x,y,z$) characterize the spin-orbital coupling
constants. The problem of the difermion bound state in the three-dimensional (3D) case in
the presence of SOC has been studied in Ref. \cite{3Dbound}. Three
special cases were considered: (1) $\lambda_x=\lambda_y=0$ and
$\lambda_z=\lambda$ (called extreme prolate (EP)); (2)
$\lambda_x=\lambda_y=\lambda$ and $\lambda_z=0$ (called extreme
oblate (EO)); (3) $\lambda_x=\lambda_y=\lambda_z=\lambda$ (called
spherical (S)). The EO-type SOC is physically equivalent to the
Rashba SOC which is interesting for condensed matter physics. For
EO- and S-type SOCs, it was shown that the difermion bound state
exists even for $a_s<0$ where the bound state does not exist in the
absence of SOC. With increased SOC, the binding energy is generally
enhanced\cite{3Dbound}. The bound state also possesses a non-trivial
effective mass which is generally larger than twice of the fermion
mass $m$ \cite{3D1,rashbon,Iskin}. Such a novel bound state caused
by the SOC is now referred to as rashbons in the literatures
\cite{rashbon}. For the two-dimensional (2D) case, the bound state exists for
arbitrarily small attraction. It was shown in Ref. \cite{2Dhe} that the
EO-type SOC or the Rashba SOC enhances the binding energy and the
bound state also has a non-trivial effective mass. This is analogous
to the catalysis of the dynamical mass generation by an external
non-Abelian gauge field in quantum field theory~\cite{NJL}.

Because of the presence of novel bound state with SOC, it has been
proposed that a dilute Fermi gas with EO- and S-type SOCs can
undergo a smooth crossover from the BCS superfluid state to the
Bose-Einstein condensation of rashbons (RBEC) even for negative
values of $1/(k_{\rm F}a_s)$ if the SOC constant $\lambda$ is tuned
from small to large values \cite{3DBCSBEC}. Due to the presence of
SOC constant $\lambda$, the 3D BCS-BEC crossover problem depends on
two dimensionless parameters: $1/(k_{\rm F}a_s)$ and $\lambda/k_{\rm
F}$ (we set $m=1$ in this paper). The BCS-BEC crossover problem and
anisotropic superfluidity in 3D Fermi gases with EO-type SOC has
been extensively studied \cite{3D1,3D2}. It was shown that the
system enters the RBEC regime at $\lambda/k_{\rm F}\sim1$ for
EO-type SOC for negative values of $1/(k_{\rm F}a_s)$. The BCS-BEC
crossover in 2D Fermi gases with EO-type SOC was also studied
\cite{2Dhe,chen}. Similar conclusions were found for the 2D case.

In this paper, we present a systematic theoretical study of the
BCS-BEC crossover in 3D Fermi gases at zero temperature with S-type
SOC. Especially, we will study the properties of the collective
modes along the BCS-BEC crossover and the effective interaction
among the rashbons in the RBEC regime. As far as we know, in the
presence of SOC, these two interesting issues have not yet been
studied (See the \emph{Note added}). For S-type SOC, the superfluid
ground state is isotropic, which brings much convenience to the
computations, and enables us to obtain various analytical results
and universal behaviors at large SOC.

This paper is organized as follows. In Sec. II, we set up the
functional path integral formalism for the BCS-BEC crossover problem
with a spherical SOC. Then we first determine the binding energy and
the effective mass of the rashbon at vanishing density and
temperature (the vacuum in the presence of SOC) in Sec. III. The
ground state properties, such as the solution of the gap and number
equations, fermion momentum distribution, the condensate fraction,
and the superfluid density are discussed in Sec. IV. We derive
the Gross-Pitaevskii free energy for the weakly interacting rashbon
condensate at large SOC and determine the rashbon-rashbon scattering
length in Sec. V. The properties of the collective excitations,
such as the gapless Goldstone mode and the massive Anderson-Higgs
mode, are investigated in Sec. VI. We summarize in Sec. VII.

\section{Model and Effective Potential}
For neutral atoms, the spin-orbit coupling can be generated by a
synthetic non-Abelian gauge potential ${\bf A}$. For instance, the
well-known Rashba spin-orbit coupling in solid-state systems can be
generated via a 2D synthetic vector potential \cite{SOC03,SOC04}
\begin{equation}
{\bf A}=-\lambda(\sigma_x{\bf e}_x+\sigma_y{\bf e}_y).
\end{equation}
For spin-1/2 fermions moving in three spatial dimensions, this
results in an anisotropic (but circular in $x$-$y$ plane) ground
state.

In this paper, we are interested in a 3D extension of the Rashba
spin-orbit coupling. A 3D synthetic vector potential ${\bf A}$ can
be produced by laser-induced coupling to link four internal atomic
states with a tetrahedral geometry \cite{3DSOC}. The synthetic 3D
vector potential takes the form \cite{3DSOC}
\begin{equation}
{\bf A}=-\lambda_\perp(\sigma_x{\bf e}_x+\sigma_y{\bf
e}_y)-\lambda_{\parallel}\sigma_z{\bf e}_z,
\end{equation}
which includes all three components of the Pauli matrices. The
single-particle Hamiltonian describing spin-1/2 fermions moving in
three spatial dimensions in the synthetic gauge field is given by
\begin{equation}
{\cal H}_{\text{GF}} = \frac{\left(\hat{\bf p}-{\bf
A}\right)^2}{2m}\label{EQ001},
\end{equation}
where $\hat{\bf p}=-i\hbar\nabla$ is the momentum operator. In
the following we use the natural units $\hbar=k_{\text B}=m=1$. We
are interested in the fully spherical case,
$\lambda_\perp=\lambda_\parallel\equiv\lambda$. The single-particle
Hamiltonian can be reduced to
\begin{equation}
{\cal H}_{\text{GF}} =\frac{\hat{\bf
p}^2}{2}+\lambda\mbox{\boldmath{$\sigma$}}\cdot\hat{\bf p},
\end{equation}
where an irrelevant constant $\lambda^2/2$ has been omitted. The
resulting spherical SOC term
$\lambda\mbox{\boldmath{$\sigma$}}\cdot\hat{\bf p}$ can be called a
Weyl spin-orbit coupling \cite{3DSOC} in analogy to the Weyl
fermions \cite{weyl}. Here the sign of the gauge field strength
$\lambda$ is not important, since the physical quantities depend
only on the parameter $\lambda^2$ as we will show in the following.
Therefore, we set $\lambda>0$ without loss of generality.

The symmetry properties of the Hamiltonian ${\cal H}_{\text{GF}}$
can be summarized as follows: (i) It has a global rotational
symmetry generated by the total angular momentum ${\bf j}={\bf
l}+{\bf s}$ with ${\bf l}$ being the orbital angular momentum and
${\bf s}=\mbox{\boldmath{$\sigma$}}/2$ being the spin angular
momentum; (2) Since the operator
$\mbox{\boldmath{$\sigma$}}\cdot\hat{\bf p}$ is parity odd, spatial
inversion symmetry does not hold; (3) Time reversal symmetry holds;
(4) The Galilean invariance in the absence of SOC is broken by the
SOC term. However, as it will be shown, the Galilean invariance can
emerge at low energy for sufficiently large $\lambda$.

The spin degeneracy is lifted by the SOC term. For $\lambda\neq0$,
the Hamiltonian ${\cal H}_{\text{GF}}$ has two eigen-energies
$\epsilon_{\bf k}^\pm={\bf k}^2/2\pm\lambda|{\bf k}|$, which are
rotationally symmetric in the momentum space. The corresponding
orthogonal eigen-states can be expressed as \cite{dilute}
\begin{eqnarray}
&&|{\bf k}+\rangle=\alpha_{\bf k}^+|{\bf
k}\uparrow\rangle+\alpha_{\bf k}^-e^{i\phi_{\bf k}}|{\bf
k}\downarrow\rangle,\nonumber\\
&&|{\bf k}-\rangle=\alpha_{\bf k}^-|{\bf
k}\uparrow\rangle-\alpha_{\bf k}^+e^{i\phi_{\bf k}}|{\bf
k}\downarrow\rangle,
\end{eqnarray}
where $\alpha_{\bf k}^\pm=\sqrt{(1\pm k_z/|{\bf k}|)/2}$ and
$e^{i\phi_{\bf k}}=(k_x+ik_y)/\sqrt{k_x^2+k_y^2}$. Since the SOC
term includes all Pauli matrices, there does not exist a simple, ${\bf k}$-independent,
matrix which maps the state $|{\bf k}+\rangle$ to $|{\bf k}-\rangle$
and vice versa. For the 2D Rashba SOC, this matrix is simply given
by $\sigma_z$.

Now we turn to the many-body Hamiltonian. We consider a homogeneous
Fermi gas. We define the Fermi momentum $k_{\text F}$ through the
fermion density $n=N/V=k_{\text F}^3/(3\pi^2)$, and the Fermi energy
is $\epsilon_{\text F}=k_{\text F}^2/2$. For the purpose of studying
the BCS-BEC crossover, we turn on a short-range s-wave attractive
interaction in the spin-singlet channel. In the attractive strength
can be tuned by means of the Feshbach resonance \cite{catom}. In the
dilute limit $k_{\text F}r_0\ll1$ ($r_0$ is effective range of the
interaction), the interaction Hamiltonian can be modeled by a
contact interaction. The many-body Hamiltonian of the system can be
written as
\begin{eqnarray}
H &=&\int d^3 {\bf r}\psi^{\dagger}({\bf r}) \left({\cal
H}_0+\cal{H}_{\rm{so}}\right) \psi({\bf r})\nonumber\\
&-&U\int d^3 {\bf r}^{\phantom{\dag}}\psi^\dagger_{\uparrow}({\bf
r})\psi^\dagger_{\downarrow}({\bf
r})\psi^{\phantom{\dag}}_{\downarrow}({\bf
r})\psi^{\phantom{\dag}}_{\uparrow}({\bf r}),
\end{eqnarray}
where $\psi({\bf r})= [\psi_\uparrow({\bf r}), \psi_\downarrow({\bf
r})]^{\rm T}$ represents the two-component fermion fields, ${\cal
H}_0=\hat{\bf p}^2/2-\mu$ is the free single-particle Hamiltonian
with $\mu$ being the chemical potential, ${\cal H}_{\rm{so}}
=\lambda\mbox{\boldmath{$\sigma$}}\cdot\hat{\bf p}$ is the SOC term,
and $U>0$ denotes the attractive s-wave interaction between unlike
spins. For the validity of such a contact interaction, another
dilute condition $\lambda r_0\ll 1$ should be satisfied
\cite{dilute}.

In the functional path integral formalism, the partition function of
the system is
\begin{eqnarray}
{\cal Z} = \int \mathcal{ D} \psi
\mathcal{D}\bar{\psi}\exp\left\{-{\cal S}[\psi,\bar{\psi}]\right\},
\end{eqnarray}
where
\begin{eqnarray}
{\cal S}[\psi,\bar{\psi}]=\int_0^\beta d\tau\int d^3{\bf r}
\bar{\psi}\partial_\tau \psi+\int_0^\beta d\tau H(\psi,\bar{\psi}).
\end{eqnarray}
Here $\beta=1/T$ and $H(\psi,\bar{\psi})$ is obtained by replacing
the field operators $\psi^\dagger$ and $\psi$ with the Grassmann
variables $\bar{\psi}$ and $\psi$, respectively. To decouple the
interaction term we introduce the auxiliary complex pairing field
$\Phi(x) = -U\psi_\downarrow(x)\psi_\uparrow(x)$~$[x=(\tau,{\bf
r})]$ and apply the Hubbard-Stratonovich transformation.  Using the
four-component Nambu-Gor'kov spinor
$\Psi(x)=[\psi_\uparrow,\psi_\downarrow,\bar{\psi}_\uparrow,\bar{\psi}_\downarrow]^{\rm
T}$, we express the partition function as
\begin{eqnarray}
{\cal Z}&=&\int {\cal D}\Psi{\cal D}\bar{\Psi}{\cal D}\Phi {\cal
D}\Phi^{\ast} \exp\Bigg\{-\frac{1}{U}\int dx|\Phi(x)|^2\nonumber\\
&+&\frac{1}{2}\int dx\int dx^\prime\bar{\Psi}(x){\bf
G}^{-1}(x,x^\prime)\Psi(x^\prime)\Bigg\},
\end{eqnarray}
where the inverse single-particle Green's function ${\bf
G}^{-1}(x,x^\prime)$ is given by
\begin{eqnarray}
{\bf G}^{-1}=\left(\begin{array}{cc}-\partial_{\tau}-{\cal
H}_0-\cal{H}_{\rm{so}} &i\sigma_y\Phi(x)\\  -i\sigma_y\Phi^*(x)&
-\partial_{\tau}+{\cal
H}_0-\cal{H}_{\rm{so}}^{\ast}\end{array}\right)\delta(x-x^\prime).
\end{eqnarray}
Integrating out the fermion fields, we obtain $\mathcal {Z}=\int
\mathcal{D} \Phi \mathcal{D} \Phi^{\ast} \exp \big\{- {\cal
S}_{\rm{eff}}[\Phi, \Phi^{\ast}]\big\}$, where the effective action
reads
\begin{eqnarray}
{\cal S}_{\rm{eff}}[\Phi, \Phi^{\ast}] = \frac{1}{U}\int dx
|\Phi(x)|^{2} - \frac{1}{2}\mbox{Trln} [{\bf G}^{-1}(x,x^\prime)].
\end{eqnarray}

\section{Two-Body Problem}
In this section, we study the two-body problem at vanishing density.
We will determine the binding energy and effective mass of difermion
bound state formed in the non-Abelian gauge field. The systematic
way to study the two-body problem in presence of a nonzero
spin-orbit coupling $\lambda$ is to consider the Green's function
$\Gamma(Q)$ of the fermion pairs, where $Q=(i\nu_n,{\bf q})$ with
$\nu_n=2n\pi T$ ($n$ integer) being the bosonic Matsubara frequency.
For zero density, we need to consider the case $\Phi=0$. In the
functional path integral formalism, $\Gamma^{-1}(Q)$ can be obtained
from its coordinate representation defined as
\begin{eqnarray}
\Gamma^{-1}(x,x^\prime)= \frac{1}{\beta V}\frac{\delta^2{\cal
S}_{\rm{eff}}[\Phi, \Phi^{\ast}]}{\delta \Phi^{\ast}(x)\delta
\Phi(x^\prime)}\bigg|_{\Phi=0}.
\end{eqnarray}
For $\Phi=0$, the single-particle Green's function ${\bf G}(K)$
reduces to its non-interacting form
\begin{eqnarray}
{\cal G}_0(K)=\left(\begin{array}{cc}g_+(K)&0\\
0& g_-(K)\end{array}\right),
\end{eqnarray}
where $K=(i\omega_n,{\bf k})$ with $\omega_n=(2n+1)\pi T$ being the
fermionic Matsubara frequency. The matrix elements $g_\pm(K)$ read
\begin{eqnarray}
&&g_+(K)=\frac{1}{i\omega_n-\xi_{\bf k}-\xi_{\text{so}}},\nonumber\\
&&g_-(K)=\frac{1}{i\omega_n+\xi_{\bf k}-\xi_{\text{so}}^*},
\end{eqnarray}
where $\xi_{\bf k}=\epsilon_{\bf k}-\mu$ with $\epsilon_{\bf k}={\bf
k}^2/2$, $\xi_{\text{so}}=\lambda\mbox{\boldmath{$\sigma$}}\cdot{\bf
k}$ and
$\xi_{\text{so}}^*=\lambda\mbox{\boldmath{$\sigma$}}^*\cdot{\bf k}$.
Here $\mbox{\boldmath{$\sigma$}}^*=(\sigma_x,-\sigma_y,\sigma_z)$.
The inverse in $g_\pm(K)$ can be worked out and we obtain
\begin{eqnarray}
&&g_+(K)=\frac{i\omega_n-\xi_{\bf k}+\xi_{\text{so}}}{(i\omega_n-\xi_{\bf k})^2-\lambda^2{\bf k}^2},\nonumber\\
&&g_-(K)=\frac{i\omega_n+\xi_{\bf
k}+\xi_{\text{so}}^*}{(i\omega_n+\xi_{\bf k})^2-\lambda^2{\bf k}^2}.
\end{eqnarray}
The single-particle excitation spectrum therefore has two branches,
$\xi_{{\bf k}}^\pm=\xi_{\bf k}\pm\lambda |{\bf k}|$, due to the
spin-orbit coupling.

Using the free fermion propagators $g_\pm(K)$, $\Gamma^{-1}(Q)$ can
be expressed as
\begin{eqnarray}
\Gamma^{-1}(Q)=\frac{1}{U}+\frac{1}{2}\sum_K\text{Tr}\left[g_+(K+Q)\sigma_yg_-(K)\sigma_y\right].\label{A1}
\end{eqnarray}
Completing the Matsubara frequency sum and making the analytical
continuation $i\nu_n\rightarrow \omega+i0^+$, the real part of
$\Gamma^{-1}(\omega+i0^+,{\bf q})$ takes the form
\begin{eqnarray}
&&\Gamma_{\text R}^{-1}(\omega,{\bf
q})\equiv\text{Re}\Gamma^{-1}(\omega+i0^+,{\bf q})\nonumber\\
&=&\frac{1}{U}-\frac{1}{4}\sum_{\alpha,\gamma=\pm}\sum_{{\bf
k}}\frac{1-f(\xi_{{\bf k}+{\bf q}/2}^\alpha)-f(\xi_{{\bf k}-{\bf
q}/2}^\gamma)}{\xi_{{\bf k}+{\bf q}/2}^\alpha+\xi_{{\bf k}-{\bf
q}/2}^\gamma-\omega}\left(1+\alpha\gamma {\cal
T}_{\bf{kq}}\right),\nonumber\\
\end{eqnarray}
where $f(E)=1/(e^{\beta E}+1)$ is the Fermi-Dirac distribution
function, and ${\cal T}_{\bf{kq}}$ is defined as
\begin{eqnarray}
{\cal T}_{\bf{kq}}=\frac{({\bf k}+{\bf q}/2)\cdot({\bf k}-{\bf
q}/2)}{|{\bf k}+{\bf q}/2||{\bf k}-{\bf q}/2|}.
\end{eqnarray}
We use the notations $\sum_K=T\sum_n\sum_{\bf k}$ and $\sum_{\bf
k}=\int d^3{\bf k}/(2\pi)^3$ throughout this paper. Note that
$\Gamma^{-1}(Q)$ takes the form similar to that of the relativistic
systems~\cite{BCSBECHe}, due to the fact that ${\cal H}_{\text{so}}$
behaves like a Dirac Hamiltonian.

The integral over the fermion momentum ${\bf k}$ is divergent and
the contact coupling $U$ needs to be regularized. For a short range
interaction potential with its s-wave scattering length $a_s$, it is
natural to regularize $U$ by means of the two-body problem in the
absence of SOC. We have
\begin{eqnarray}
\frac{1}{U}=-\frac{1}{4\pi a_s}+\sum_{\bf k}\frac{1}{2\epsilon_{\bf
k}}.
\end{eqnarray}
In cold atom experiments, the s-wave scattering length can be tuned
by means of the Feshbach resonance \cite{catom}.

For the pure two-body problem at vanishing density and temperature,
we discard the Fermi-Dirac distribution function. The
energy-momentum dispersion $\omega_{\bf q}$ of the pair excitation
is defined as the solution $\omega+2\mu=\omega_{\bf q}$ of the
two-body equation $\Gamma_{\text R}^{-1}(\omega,{\bf q})=0$ . After
some manipulations, the two-body equation becomes
\begin{eqnarray}
\sum_{\bf k}\left(\frac{1}{{\bf k}^2}-\frac{{\cal E}_{\bf kq}}{{\cal
E}_{\bf kq}^2-4\lambda^2{\bf k}^2-\frac{4\lambda^4{\bf k}^2{\bf
q}^2\sin^2\varphi}{{\cal E}_{\bf kq}^2-\lambda^2{\bf
q}^2}}\right)=\frac{1}{4\pi a_s}.\label{A2}
\end{eqnarray}
Here $\varphi$ is the angle between ${\bf k}$ and ${\bf q}$, and
${\cal E}_{\bf kq}=\epsilon_{{\bf k}+{\bf q}/2}+\epsilon_{{\bf
k}-{\bf q}/2}-\omega_{\bf q}={\bf k}^2+{\bf q}^2/4-\omega_{\bf q}$.

\subsection{Bound state and binding energy}

We are interested in whether there exist difermion bound state in
the presence of SOC. For this purpose, we first consider zero
center-of-mass momentum ${\bf q}$ and determine the energy regime
where the imaginary part of $\Gamma^{-1}(\omega+i0^+,{\bf q}=0)$
vanishes. We have
\begin{eqnarray}
&&\text{Im}\Gamma^{-1}(\omega+i0^+,{\bf q}=0)\nonumber\\
&=&-\frac{1}{4\pi}\sum_{\alpha=\pm}\int_0^\infty
k^2dk\delta(k^2+2\alpha\lambda k-\omega-2\mu).
\end{eqnarray}
Therefore, a bound state exists if the equation $\Gamma_{\text
R}^{-1}(\omega,{\bf q}=0)=0$ has a solution in the regime
$-\infty<\omega+2\mu<-\lambda^2$.

\begin{figure}[!htb]
\begin{center}
\includegraphics[width=8.1cm]{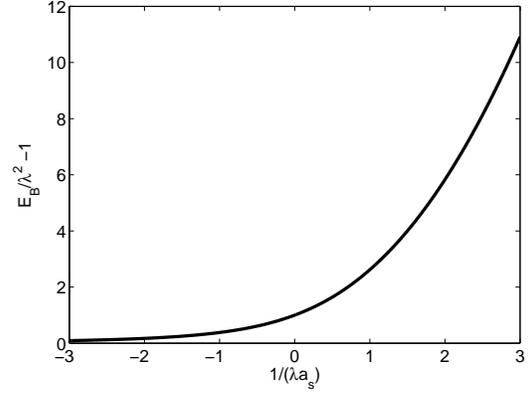}
\caption{The quantity $E_{\text B}/\lambda^2-1$ as a function of the
dimensionless parameter $\kappa=1/(\lambda a_s)$.
 \label{fig1}}
\end{center}
\end{figure}

The binding energy $E_{\text B}$ in the presence of nonzero SOC is
determined by the solution of $\omega+2\mu=-E_{\text B}$ for the
equation $\Gamma_{\text R}^{-1}(\omega,{\bf q}=0)=0$. From the
imaginary part of $\Gamma^{-1}(\omega+i0^+,{\bf q}=0)$, the binding
energy $E_{\text B}$ must be larger than a threshold
$E_{\text{th}}=\lambda^2$. The equation determining $E_{\text B}$
reads
\begin{eqnarray}
\int_0^\infty k^2dk\left[\frac{1}{k^2}-\frac{k^2+E_{\text
B}}{(k^2+E_{\text B})^2-4\lambda^2k^2}\right]=\frac{\pi}{2 a_s}.
\end{eqnarray}
Completing the integrals analytically, we obtain a simple algebraic
equation for $E_{\text B}$,
\begin{eqnarray}
\frac{E_{\text B}-2\lambda^2}{\sqrt{E_{\text
B}-\lambda^2}}=\frac{1}{a_s}. \label{EB}
\end{eqnarray}
We find that, for arbitrary scattering length $a_s$, there always
exists a solution $E_{\text B}>\lambda^2$ . Therefore, the difermion
bound state can form in the presence of SOC even for $a_s<0$ where
no bound state exists in the absence of SOC.

The solution of Eq. (\ref{EB}) can be analytically expressed as
\begin{eqnarray}
E_{\text
B}=\lambda^2+\frac{1}{4}\left(\frac{1}{a_s}+\sqrt{\frac{1}{a_s^2}+4\lambda^2}\right)^2.
\end{eqnarray}
Therefore, the quantity $E_{\text B}/\lambda^2$ depends only on the
dimensionless parameter $\kappa=1/(\lambda a_s)$. We have
\begin{eqnarray}
\frac{E_{\text B}}{\lambda^2}={\cal J}\left(\kappa\right),
\end{eqnarray}
where the function ${\cal J}(\kappa)$ is defined as
\begin{eqnarray}
{\cal
J}(\kappa)=1+\frac{1}{4}\left(\kappa+\sqrt{\kappa^2+4}\right)^2.
\end{eqnarray}
We are interested in the case $\lambda a_s\rightarrow\infty$ or
$\kappa=0$. This happens when $a_s\rightarrow\infty$ (unitary point
of the Feshbach resonance) for fixed $\lambda$ or
$\lambda\rightarrow\infty$ for fixed $a_s$. In this case, we have
${\cal J}=2$ and a very simple result
\begin{eqnarray}
E_{\text B}(\lambda a_s\rightarrow \infty)=2\lambda^2.
\end{eqnarray}
In general, the numerical result for the quantity $E_{\text
B}/\lambda^2-1={\cal J}(\kappa)-1$ is shown in Fig. \ref{fig1}.

Since the Hamiltonian has rotational symmetry generated by the total
angular momentum ${\bf J}={\bf L}+{\bf S}$, the bound state should
be a $J$ singlet. Therefore, the bound state wave function can be
expressed as \cite{3Dbound}
\begin{eqnarray}
\Psi({\bf r})=\psi_0({\bf
r})|\uparrow\downarrow-\downarrow\uparrow\rangle+\psi_1({\bf
r})|\uparrow\downarrow+\downarrow\uparrow\rangle,
\end{eqnarray}
where the spin quantization axis is chosen to be along ${\bf r}$,
the relative radius of the two fermions. $\psi_0({\bf r})$ is an $L=
0$ orbital state, while $\psi_1({\bf r})$ an $L=1$ orbital state.
The spatial wave functions can be evaluated as \cite{3Dbound}
\begin{eqnarray}
\psi_0({\bf
r})&=&\frac{e^{-br}}{r}\left(\frac{\lambda}{b}\sin{\lambda
r}+\cos{\lambda r}\right)\nonumber\\
\psi_1({\bf
r})&=&i\frac{e^{-br}}{r}\left(\left(1+\frac{1}{br}\right)\sin{\lambda
r}-\frac{\lambda}{b}\cos{\lambda r}\right),
\end{eqnarray}
where $b=\sqrt{E_{\rm B}-\lambda^2}$. In the absence of SOC, the
bound state exists only for $a_s>0$. We have $\psi_1({\bf r})=0$ and
the known result $\psi_0({\bf r})=(1/r)e^{-r/a_s}$ for spin-singlet
bound state. However, in the presence of SOC, the bound state is a
mixture of spin-singlet and spin-triplet components. This will have
a significant impact on the many-body problem, where the pair wave
function possesses both spin-singlet and spin-triplet components.

\subsection{Molecule effective mass}

\begin{figure}[!htb]
\begin{center}
\includegraphics[width=8cm]{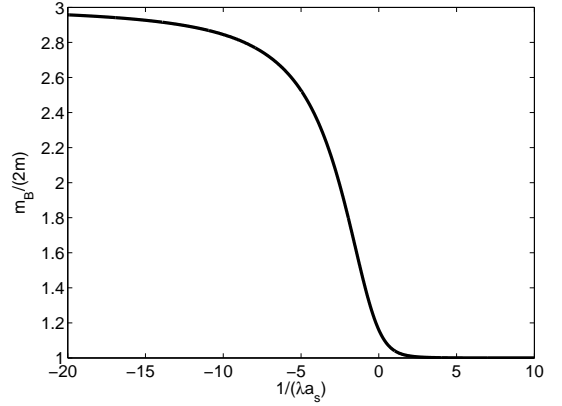}
\caption{The molecule effective mass $m_{\text B}$ (divided by $2m$)
as a functions of the dimensionless parameter $\kappa=1/(\lambda
a_s)$.
 \label{fig2}}
\end{center}
\end{figure}

For small nonzero center-of-mass momentum ${\bf q}$, the solution
for $\omega_{\bf q}$ can be written as $\omega_{\bf q}=-E_{\text
B}+{\bf q}^2/(2m_{\text B})$, where $m_{\text B}$ is referred to as
the effective mass of the bound state. Substituting this dispersion
into the equation $\Gamma_{\text R}^{-1}(\omega,{\bf q})=0$ and
expanding the equation to the order $O({\bf q}^2)$, we obtain
\begin{eqnarray}
&&\left(1-\frac{2m}{m_{\text B}}\right)\int_0^\infty
k^2dk\frac{(k^2+E_{\text B})^2+4\lambda^2k^2}{\left[(k^2+E_{\text
B})^2-4\lambda^2k^2\right]^2}\nonumber\\
&=&\frac{4}{3}\int_0^\infty k^2dk\frac{8\lambda^4k^2}{(k^2+E_{\text
B})\left[(k^2+E_{\text B})^2-4\lambda^2k^2\right]^2}.
\end{eqnarray}
Defining a new variable $x=k/\lambda$, this equation becomes
\begin{eqnarray}
&&\left(1-\frac{2m}{m_{\text B}}\right)\int_0^\infty dxx^2
\frac{(x^2+{\cal J})^2+4x^2}{[(x^2+{\cal J})^2-4x^2]^2}\nonumber\\
&=&\frac{4}{3}\int_0^\infty dxx^2\frac{8x^2}{(x^2+{\cal
J})[(x^2+{\cal J})^2-4x^2]^2}.
\end{eqnarray}
Completing the integrals analytically, we obtain
\begin{eqnarray}
\frac{2m}{m_{\text B}}=\frac{7}{3}-\frac{4}{3}\left(\frac{{\cal
J}-1}{{\cal J}}\right)^{3/2}-\frac{2}{{\cal J}}.
\end{eqnarray}
The effective mass therefore depends only on the combined parameter
$\kappa=1/(\lambda a_s)$. We have
\begin{eqnarray}
\frac{2m}{m_{\text
B}}&=&\frac{7}{3}-\frac{4}{\kappa^2+4+\kappa\sqrt{\kappa^2+4}}\nonumber\\
&-&\frac{4}{3}\left(1-\frac{2}{\kappa^2+4+\kappa\sqrt{\kappa^2+4}}\right)^{3/2}.
\end{eqnarray}
The numerical result for $m_{\rm B}/2m$ is shown in Fig. \ref{fig2}.
We find analytically that $m_{\rm B}\rightarrow 2m$ in the limit
$\kappa\rightarrow +\infty$ and $m_{\rm B}\rightarrow 6m$ in the
limit $\kappa\rightarrow -\infty$. For the case $\lambda
a_s\rightarrow\infty$ or $\kappa=0$, the effective mass reads
\begin{eqnarray}
\frac{m_{\text B}(\lambda
a_s\rightarrow\infty)}{2m}=\frac{3(4+\sqrt{2})}{14}=1.16.
\end{eqnarray}

In summary, the difermion bound state forms in the presence of SOC
for arbitrary value of s-wave scattering length $a_s$. The bound
state possesses a non-trivial binding energy $E_{\text B}$ and a
non-trivial effective mass $m_{\text B}>2m$. Such type of bound
state is referred to as rashbon in the previous literatures. Due to
the formation of bound state at the BCS side of the resonance
($a_s<0$) and the enhancement of binding energy in the presence of
SOC, we expect that there will be a crossover from the BCS
superfluid state to the Bose-Einstein condensation of rashbons if
the spin-orbit coupling $\lambda$ can be tuned from small to large
values.

\section{Superfluid Ground State: Mean Field Theory}

For the many-body problem, we first consider the properties of the
superfluid ground state ($T=0$) in the self-consistent mean-field
theory. In the superfluid ground state, the pairing field $\Phi(x)$
acquires a nonzero expectation value $\langle\Phi(x)\rangle=\Delta$,
which serves as the order parameter of the superfluidity. Without
loss of generality, we set $\Delta$ to be real. Then, we can express
the pairing field as $\Phi(x)=\Delta+\phi(x)$, where $\phi(x)$ is
the fluctuation around the mean field. The effective action ${\cal
S}_{\text{eff}}[\Phi,\Phi^*]$ can be expanded in powers of the
fluctuation,
\begin{eqnarray}
{\cal S}_{\text{eff}}[\Phi,\Phi^*]={\cal
S}_{\text{eff}}^{(0)}(\Delta)+{\cal
S}_{\text{eff}}^{(2)}[\phi,\phi^*]+\cdots,
\end{eqnarray}
where ${\cal S}_{\text{eff}}^{(0)}(\Delta)\equiv{\cal S}_{\rm
eff}[\Delta,\Delta]$ is the saddle-point or mean-field effective
action with the superfluid order parameter determined by the saddle-point condition $\partial{\cal
S}_{\text{eff}}^{(0)}/\partial\Delta=0$.

In the mean-field approximation, the grand potential $\Omega={\cal
S}_{\rm eff}[\Delta,\Delta]/(\beta V)$ can be expressed as
\begin{eqnarray}
\Omega=\frac{\Delta^2}{U}-\frac{1}{2}\frac{1}{\beta}\sum_n\sum_{\bf
k}\text{lndet}{\cal G}^{-1}(i\omega_n,{\bf k}),
\end{eqnarray}
where the inverse fermion Green's function reads
\begin{eqnarray}
{\cal G}^{-1}(i\omega_n,{\bf
k})=\left(\begin{array}{cc}i\omega_n-\xi_{\bf k}-\xi_{\text{so}}&i\sigma_y\Delta\\
-i\sigma_y\Delta& i\omega_n+\xi_{\bf
k}-\xi_{\text{so}}^*\end{array}\right).
\end{eqnarray}
Using the formula for block matrix, we first work out the
determinate and obtain
\begin{eqnarray}
\text{det}{\cal G}^{-1}(i\omega_n,{\bf
k})=\left[(i\omega_n)^2-(E_{\bf
k}^+)^2\right]\left[(i\omega_n)^2-(E_{\bf k}^-)^2\right],
\end{eqnarray}
where $E_{\bf k}^\pm=\sqrt{(\xi_{\bf k}\pm\lambda |{\bf
k}|)^2+\Delta^2}$ are quasiparticle excitation spectra. Then
completing the Matsubara frequency sum we obtain
\begin{eqnarray}
\Omega=\frac{\Delta^2}{U}+\sum_{\bf k}\left[\xi_{\bf k}-{\cal
W}(E_{\bf k}^+)-{\cal W}(E_{\bf k}^-)\right],
\end{eqnarray}
where ${\cal W}(E)=E/2+T\ln(1+e^{- E/T})$. Note that  the term
$\sum_{\bf k}\xi_{\bf k}\equiv\frac{1}{2}\sum_{\bf k}(\xi_{\bf
k}^++\xi_{\bf k}^-)$ is added to recover the correct ground state
energy for the normal state ($\Delta=0$).

\subsection{Ground-state energy}
At zero temperature, the ground-state energy $E_{\rm
G}\equiv\Omega(T=0)$ is $E_{\rm G}=\Delta^2/U+(1/2)\sum_{\bf
k}(2\xi_{\bf k}-E_{\bf k}^+-E_{\bf k}^-)$. Using the fact that the
binding energy $E_{\rm B}$ satisfies the equation
\begin{eqnarray}
\frac{1}{U}=\frac{1}{2}\sum_{\alpha=\pm}\int_0^\infty
\frac{k^2dk}{2\pi^2}\frac{1}{k^2+2\alpha \lambda k+E_{\text B}},
\end{eqnarray}
we can express the ground-state energy in terms of $E_{\text B}$ as
\begin{eqnarray}
E_{\rm G}=\frac{1}{2}\sum_{\alpha=\pm}\int_0^\infty
\frac{k^2dk}{2\pi^2}\left(\frac{\Delta^2}{k^2+2\alpha \lambda
k+E_{\text B}}-E_k^\alpha+\xi_k^\alpha\right).
\end{eqnarray}
Since the integral is convergent, we can use the trick $k^2\pm
2\lambda k=(k\pm \lambda)^2-\lambda^2$ and convert the integration
variables to $k\pm\lambda$. Then, we obtain
\begin{eqnarray}
E_{\rm G}=\int_0^\infty
\frac{dk}{2\pi^2}(k^2+\lambda^2)\left(\frac{\Delta^2}{k^2+E_{\text
B}-\lambda^2}-\tilde{E}_k+\tilde{\xi}_k\right),
\end{eqnarray}
where
\begin{eqnarray}
\tilde{\xi}_k=\epsilon_k-\tilde{\mu},\ \ \ \ \ \
\tilde{E}_k=\sqrt{\left(\epsilon_k-\tilde{\mu}\right)^2+\Delta^2}
\end{eqnarray}
with $\tilde{\mu}=\mu+\lambda^2/2$.

Using the above expression for $E_{\text G}$, the gap $\Delta$ and
the chemical potential $\mu$ can be determined by $\partial E_{\rm
G}/\partial \Delta=0$ and $\partial E_{\rm G}/\partial \mu=-n$,
i.e.,
\begin{eqnarray}
&&\int_0^\infty dk(k^2+\lambda^2)\left[\frac{1}{k^2+E_{\text
B}-\lambda^2}-\frac{1}{2\sqrt{\left(\epsilon_k-\tilde{\mu}\right)^2+\Delta^2}}\right]=0,\nonumber\\
&&\int_0^\infty
dk(k^2+\lambda^2)\left[1-\frac{\epsilon_k-\tilde{\mu}}{\sqrt{\left(\epsilon_k-\tilde{\mu}\right)^2+\Delta^2}}\right]=2\pi^2n
\end{eqnarray}
We notice that the above expressions for the gap and number
equations can be analytically evaluated using the elliptic
functions, such as the analytical treatment for the gap and number
equations in the absence of SOC \cite{ANAGAP}.

\subsection{Fermion Green's function}
The explicit form of the fermion Green's function ${\cal
G}(i\omega_n,{\bf k})$ can be evaluated using the formula for block
matrix. In the Nambu-Gor'kov space, it takes the form
\begin{eqnarray}
{\cal G}(i\omega_n,{\bf k})=\left(\begin{array}{cc}{\cal G}_{11}(i\omega_n,{\bf k})&{\cal G}_{12}(i\omega_n,{\bf k})\\
{\cal G}_{21}(i\omega_n,{\bf k})& {\cal G}_{22}(i\omega_n,{\bf
k})\end{array}\right).
\end{eqnarray}
The matrix elements can be expressed as
\begin{eqnarray}
&&{\cal G}_{11}(i\omega_n,{\bf k})={\cal A}_{11}(i\omega_n,{\bf
k})\hat{I}+{\cal
B}_{11}(i\omega_n,{\bf k})\hat{M},\nonumber\\
&&{\cal
G}_{22}(i\omega_n,{\bf k})={\cal A}_{22}(i\omega_n,{\bf k})\hat{I}+{\cal B}_{22}(i\omega_n,{\bf k})\hat{M}^*,\nonumber\\
&&{\cal G}_{12}(i\omega_n,{\bf k})=-i\sigma_y\left[{\cal
A}_{12}(i\omega_n,{\bf k})\hat{I}+{\cal
B}_{12}(i\omega_n,{\bf k})\hat{M}^*\right],\nonumber\\
&&{\cal G}_{21}(i\omega_n,{\bf k})=i\sigma_y\left[{\cal
A}_{21}(i\omega_n,{\bf k})\hat{I}+{\cal B}_{21}(i\omega_n,{\bf
k})\hat{M}\right],
\end{eqnarray}
where $\hat{I}$ is the identity operator in the spin space and the
operators $\hat{M}$ and $\hat{M}^*$ are defined as
\begin{eqnarray}
\hat{M}=\frac{\mbox{\boldmath{$\sigma$}}\cdot{\bf k}}{|{\bf k}|},\ \
\ \  \hat{M}^*&=&\frac{\mbox{\boldmath{$\sigma$}}^*\cdot{\bf
k}}{|{\bf k}|}.
\end{eqnarray}
The explicit forms of the quantities ${\cal A}_{ij}$ and ${\cal
B}_{ij}$ are given by
\begin{eqnarray}
&&{\cal A}_{11}(i\omega_n,{\bf
k})=\frac{1}{2}\left[\frac{i\omega_n+\xi_{\bf
k}^+}{(i\omega_n)^2-(E_{\bf k}^+)^2}+\frac{i\omega_n+\xi_{\bf
k}^-}{(i\omega_n)^2-(E_{\bf k}^-)^2}\right],\nonumber\\
&&{\cal A}_{22}(i\omega_n,{\bf
k})=\frac{1}{2}\left[\frac{i\omega_n-\xi_{\bf
k}^+}{(i\omega_n)^2-(E_{\bf k}^+)^2}+\frac{i\omega_n-\xi_{\bf
k}^-}{(i\omega_n)^2-(E_{\bf k}^-)^2}\right],\nonumber\\
&&{\cal A}_{12}(i\omega_n,{\bf
k})=\frac{1}{2}\left[\frac{\Delta}{(i\omega_n)^2-(E_{\bf
k}^+)^2}+\frac{\Delta}{(i\omega_n)^2-(E_{\bf
k}^-)^2}\right],\nonumber\\
&&{\cal A}_{21}(i\omega_n,{\bf k})={\cal A}_{12}(i\omega_n,{\bf k}),
\end{eqnarray}
and
\begin{eqnarray}
&&{\cal B}_{11}(i\omega_n,{\bf
k})=\frac{1}{2}\left[\frac{i\omega_n+\xi_{\bf
k}^+}{(i\omega_n)^2-(E_{\bf k}^+)^2}-\frac{i\omega_n+\xi_{\bf
k}^-}{(i\omega_n)^2-(E_{\bf k}^-)^2}\right],\nonumber\\
&&{\cal B}_{22}(i\omega_n,{\bf
k})=-\frac{1}{2}\left[\frac{i\omega_n-\xi_{\bf
k}^+}{(i\omega_n)^2-(E_{\bf k}^+)^2}-\frac{i\omega_n-\xi_{\bf
k}^-}{(i\omega_n)^2-(E_{\bf k}^-)^2}\right],\nonumber\\
&&{\cal B}_{12}(i\omega_n,{\bf
k})=-\frac{1}{2}\left[\frac{\Delta}{(i\omega_n)^2-(E_{\bf
k}^+)^2}-\frac{\Delta}{(i\omega_n)^2-(E_{\bf
k}^-)^2}\right],\nonumber\\
&&{\cal B}_{21}(i\omega_n,{\bf k})=-{\cal B}_{12}(i\omega_n,{\bf
k}).
\end{eqnarray}

Using the matrix elements of the Green's function, we can calculate
various quantities. First, the momentum distributions
$n_\uparrow({\bf k})$ and $n_\downarrow({\bf k})$ for the two spin
components can be evaluated as
\begin{eqnarray}
&&n_\uparrow({\bf k})\equiv\langle\bar{\psi}_{{\bf
k}\uparrow}\psi_{{\bf k}\uparrow}\rangle\nonumber\\
&=&\frac{1}{\beta}\sum_n\left[{\cal A}_{11}(i\omega_n,{\bf
k})+\frac{k_z}{|{\bf k}|}{\cal
B}_{11}(i\omega_n,{\bf k})\right]e^{i\omega_n0^+},\nonumber\\
&&n_\downarrow({\bf k})\equiv\langle\bar{\psi}_{{\bf
k}\downarrow}\psi_{{\bf k}\downarrow}\rangle\nonumber\\
&=&\frac{1}{\beta}\sum_n\left[{\cal A}_{11}(i\omega_n,{\bf
k})-\frac{k_z}{|{\bf k}|}{\cal B}_{11}(i\omega_n,{\bf
k})\right]e^{i\omega_n0^+}.
\end{eqnarray}
Second, the singlet and triplet pairing amplitudes can be expressed
as
\begin{eqnarray}
&&\phi_{\uparrow\downarrow}({\bf k})\equiv\langle\psi_{{\bf
k}\uparrow}\psi_{-{\bf k}\downarrow}\rangle\nonumber\\
&=&\frac{1}{\beta}\sum_n\left[-{\cal A}_{21}(i\omega_n,{\bf
k})+\frac{k_z}{|{\bf k}|}{\cal B}_{21}(i\omega_n,{\bf
k})\right],\nonumber\\
&&\phi_{\downarrow\uparrow}({\bf k})\equiv\langle\psi_{{\bf
k}\downarrow}\psi_{-{\bf k}\uparrow}\rangle\nonumber\\
&=&\frac{1}{\beta}\sum_n\left[{\cal A}_{21}(i\omega_n,{\bf
k})+\frac{k_z}{|{\bf k}|}{\cal B}_{21}(i\omega_n,{\bf
k})\right],\nonumber\\
&&\phi_{\uparrow\uparrow}({\bf k})\equiv\langle\psi_{{\bf
k}\uparrow}\psi_{-{\bf k}\uparrow}\rangle\nonumber\\
&=&-\frac{k_x-ik_y}{k}\frac{1}{\beta}\sum_n{\cal
B}_{21}(i\omega_n,{\bf
k}),\nonumber\\
&&\phi_{\downarrow\downarrow}({\bf k})\equiv\langle\psi_{{\bf
k}\downarrow}\psi_{-{\bf k}\downarrow}\rangle\nonumber\\
&=&\frac{k_x+ik_y}{k}\frac{1}{\beta}\sum_n{\cal
B}_{21}(i\omega_n,{\bf k}).
\end{eqnarray}
Third, the gap equation for $\Delta$ can be expressed
as
\begin{eqnarray}
\Delta=-U\frac{1}{\beta}\sum_n\sum_{\bf k}{\cal
A}_{12}(i\omega_n,{\bf k}).
\end{eqnarray}

\subsection{Gap and chemical potential}
Using the ground state energy $E_{\rm G}$, the original forms of the
gap and number equations at $T=0$ are
\begin{eqnarray}
&&\frac{1}{U}=\frac{1}{2}\sum_{\bf k}\left(\frac{1}{2E_{\bf
k}^+}+\frac{1}{2E_{\bf k}^-}\right),\nonumber\\
&&n=\sum_{\bf k}\left(1-\frac{\xi_{\bf k}^+}{2E_{\bf
k}^+}-\frac{\xi_{\bf k}^-}{2E_{\bf k}^-}\right).
\end{eqnarray}
The pairing gap $\Delta$ and the chemical potential $\mu$ can be
numerically solved for given values of $1/(k_{\rm F}a_s)$ and
$\lambda/k_{\rm F}$. From now on, we denote the saddle point
solution for the gap at zero temperature as $\Delta_0$. We also
notice the relation
\begin{eqnarray}
\frac{1}{\lambda a_s}=\frac{1}{k_{\text
F}a_s}\left(\frac{\lambda}{k_{\rm F}}\right)^{-1}.
\end{eqnarray}

{\bf (A) Analytical Results for Large SOC.} We first obtain the
analytical solution at large SOC, $\lambda/k_{\rm F}\gg 1$. For
large SOC, we expect $\mu<0$ and $\Delta_0\ll|\mu|$. Therefore, we
can expand the equations in powers of $\Delta_0/|\mu|$ and keep only
the leading order terms. The gap equation becomes
\begin{eqnarray}
\frac{1}{U}=\frac{1}{2}\sum_{\alpha=\pm}\int_0^\infty
\frac{k^2dk}{2\pi^2}\frac{1}{k^2+2\alpha \lambda
k-2\mu}+O\left(\frac{\Delta_0^2}{|\mu|^2}\right),
\end{eqnarray}
Comparing with the two-body problem, we obtain
\begin{eqnarray}
\mu\simeq-\frac{E_{\text B}}{2}.
\end{eqnarray}
Substituting this into the number equation, we obtain
\begin{eqnarray}
n&=&\frac{\Delta_0^2}{8\pi^2}\int_0^\infty
k^2dk\left[\frac{1}{(\xi_k^+)^2}+\frac{1}{(\xi_k^-)^2}\right]+O\left(\frac{\Delta_0^2}{|\mu|^2}\right)\nonumber\\
&\simeq&\frac{\Delta_0^2}{\pi^2}\int_0^\infty k^2dk
\frac{(k^2+E_{\text B})^2+4\lambda^2k^2}{\left[(k^2+E_{\text
B})^2-4\lambda^2k^2\right]^2}.
\end{eqnarray}
We notice that this integral also appears in Eq. (25). Completing
the integral analytically, we obtain
\begin{eqnarray}
\Delta_0^2&\simeq&4\pi\lambda n\frac{({\cal J}-1)^{3/2}}{{\cal J}}\nonumber\\
 &=&\frac{4\lambda\left[2\epsilon_{\rm
F}({\cal J}-1)\right]^{3/2}}{3\pi{\cal J}}.
\end{eqnarray}
Therefore we have
\begin{eqnarray}
\frac{\Delta_0}{\epsilon_{\text F}}\simeq
\sqrt{\frac{16}{3\pi}\frac{({\cal J}-1)^{3/2}}{{\cal
J}}\frac{\lambda}{k_{\text F}}}.
\end{eqnarray}
In the limit $\lambda a_s\rightarrow\infty$, we have ${\cal J}=2$
and therefore
\begin{eqnarray}
\Delta_0^2(\lambda a_s\rightarrow\infty)\simeq2\pi\lambda n
=\frac{2\lambda(2\epsilon_{\rm F})^{3/2}}{3\pi}.
\end{eqnarray}
It can be written as another interesting form
\begin{eqnarray}
\frac{\Delta_0(\lambda a_s\rightarrow\infty)}{\epsilon_{\text
F}}\simeq \sqrt{\frac{8}{3\pi}\frac{\lambda}{k_{\text F}}}.
\end{eqnarray}
Therefore, for very large SOC, the gap $\Delta_0$ increases as
$\Delta_0\sim\sqrt{\lambda}$.

\begin{figure}[!htb]
\begin{center}
\includegraphics[width=7.2cm]{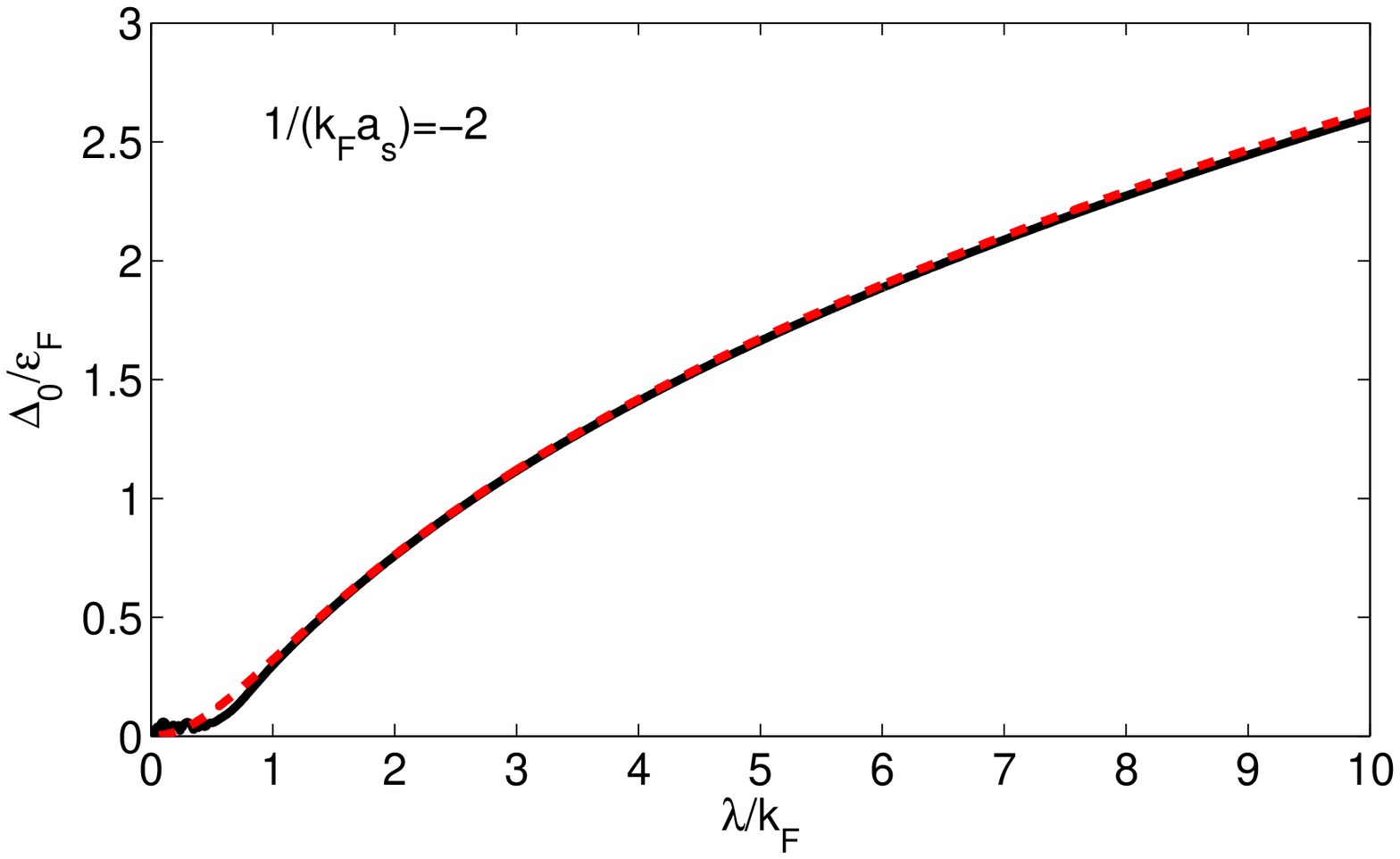}
\includegraphics[width=7.2cm]{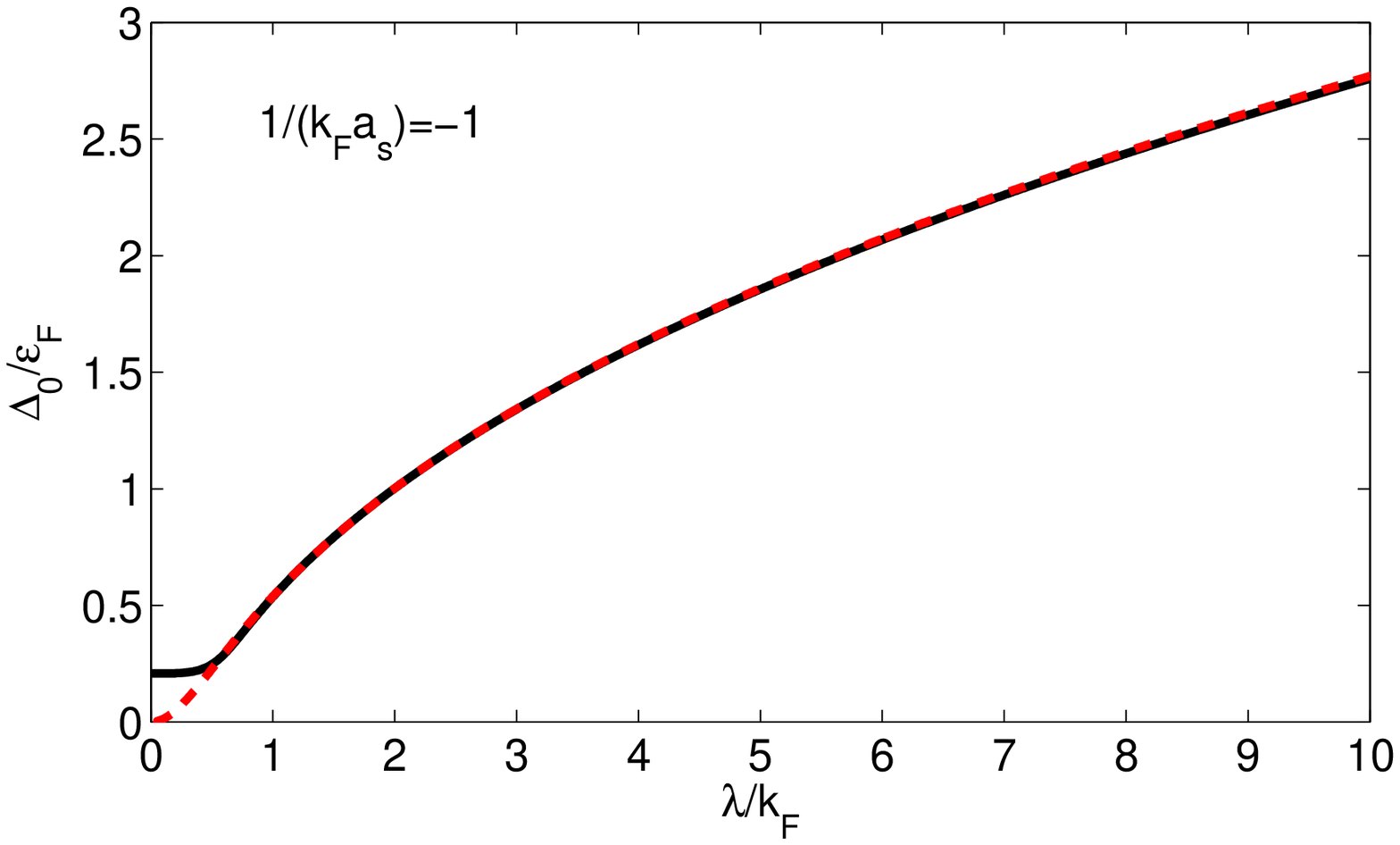}
\includegraphics[width=7.2cm]{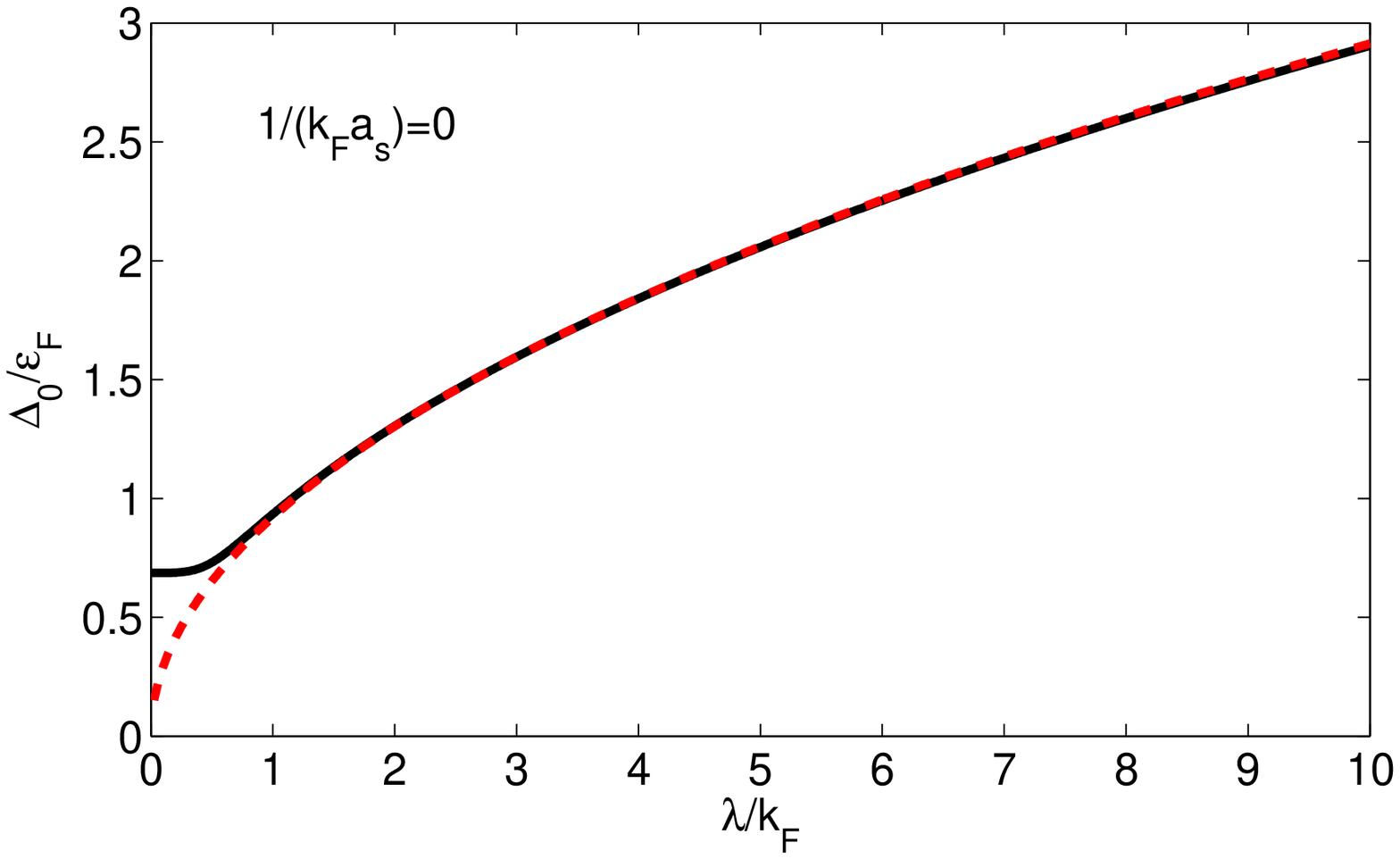}
\includegraphics[width=7.2cm]{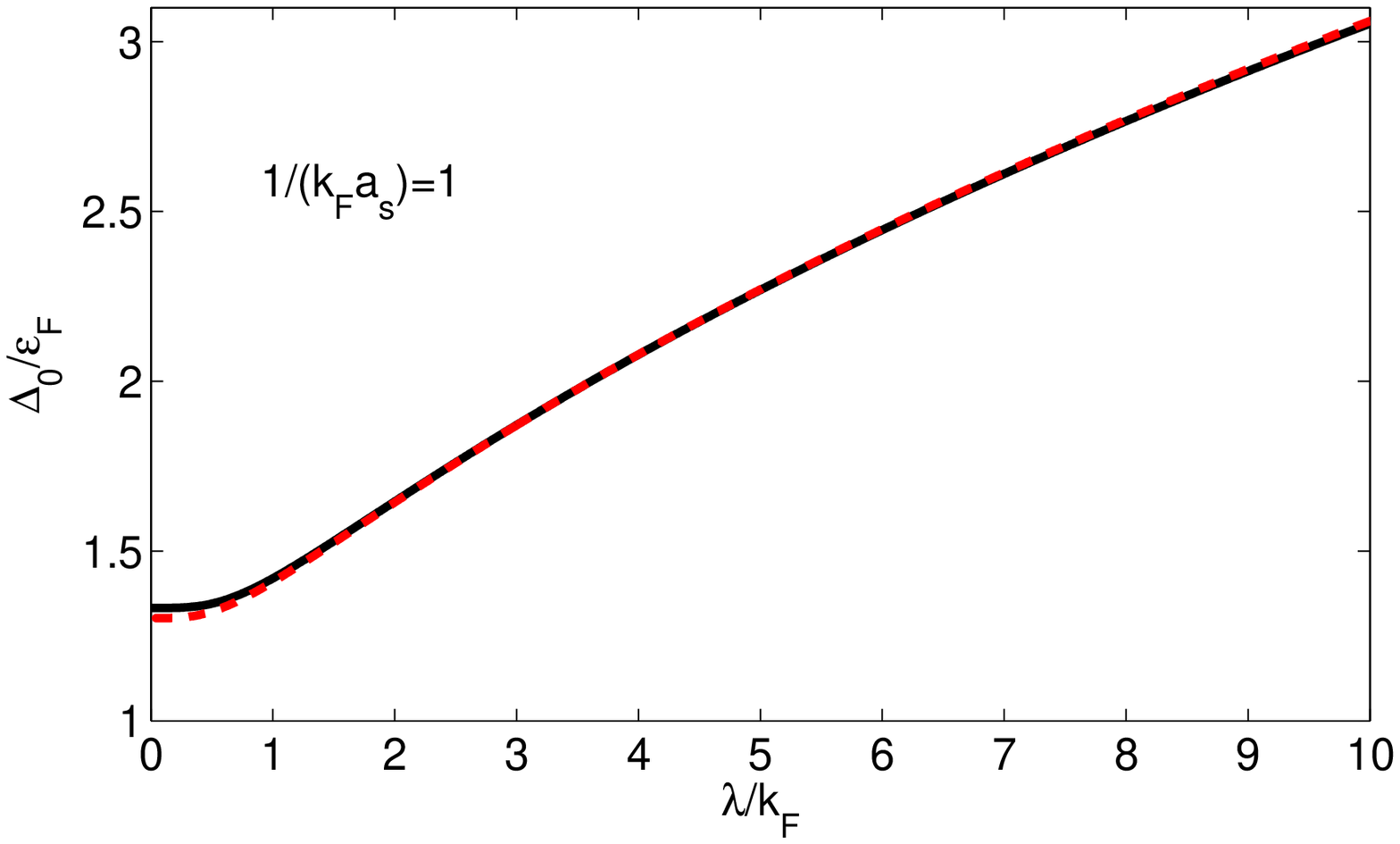}
\includegraphics[width=7.2cm]{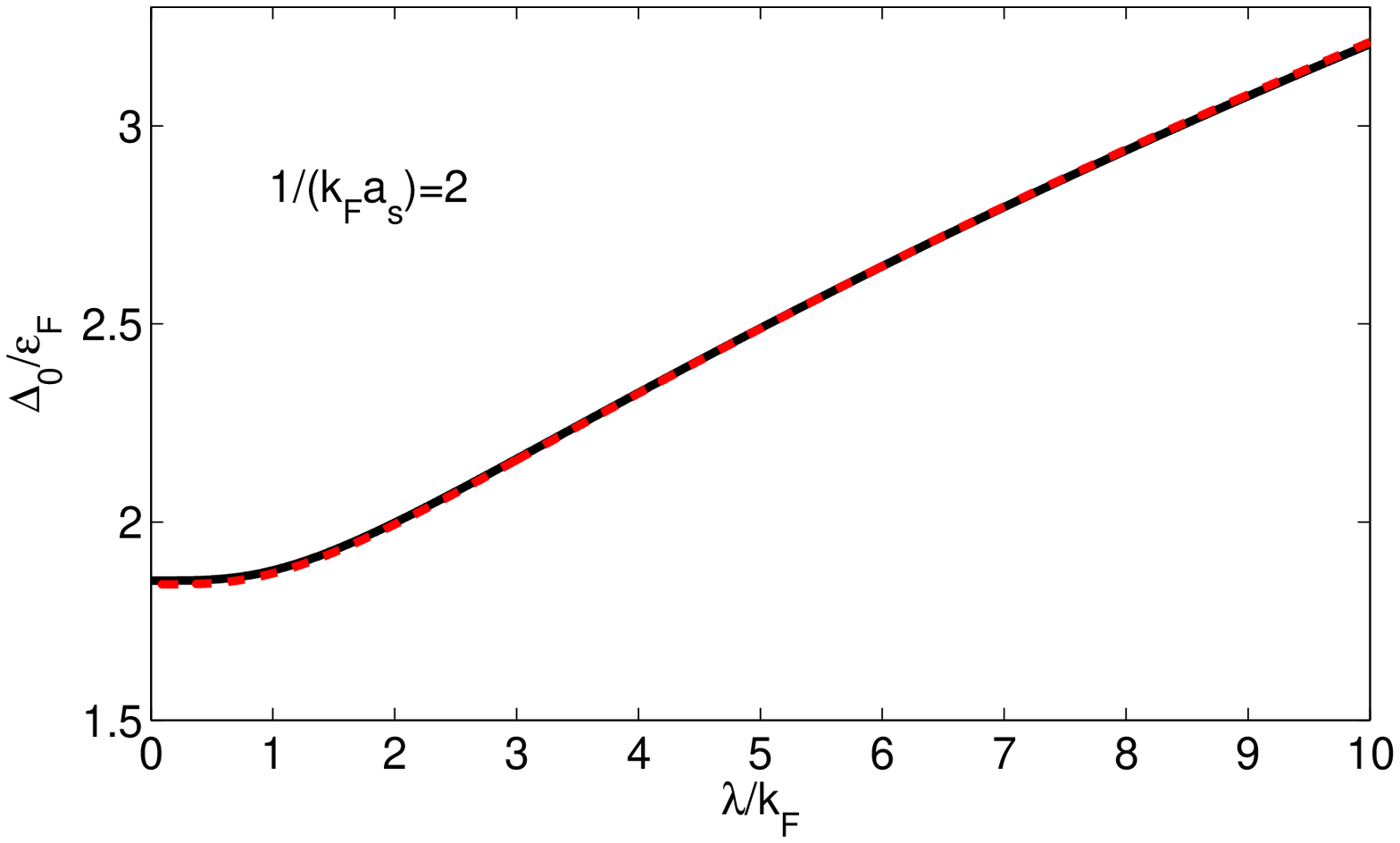}
\caption{The pairing gap $\Delta_0$ (divided by $\epsilon_{\rm F}$)
as a function of $\lambda/k_{\rm F}$. The red dashed line shows the
analytical result (54) or (60).
 \label{fig3}}
\end{center}
\end{figure}

\begin{figure}[!htb]
\begin{center}
\includegraphics[width=7.2cm]{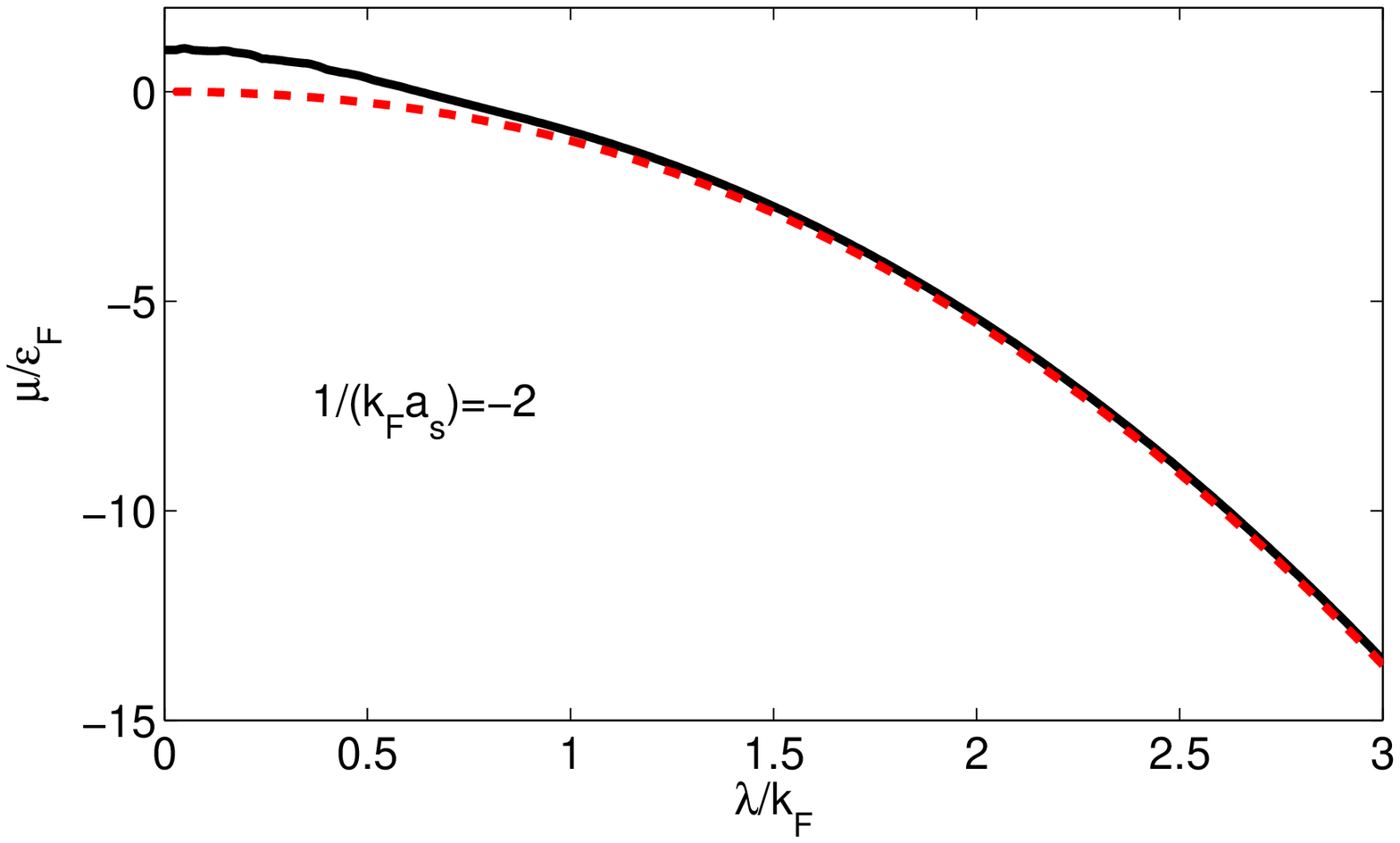}
\includegraphics[width=7.2cm]{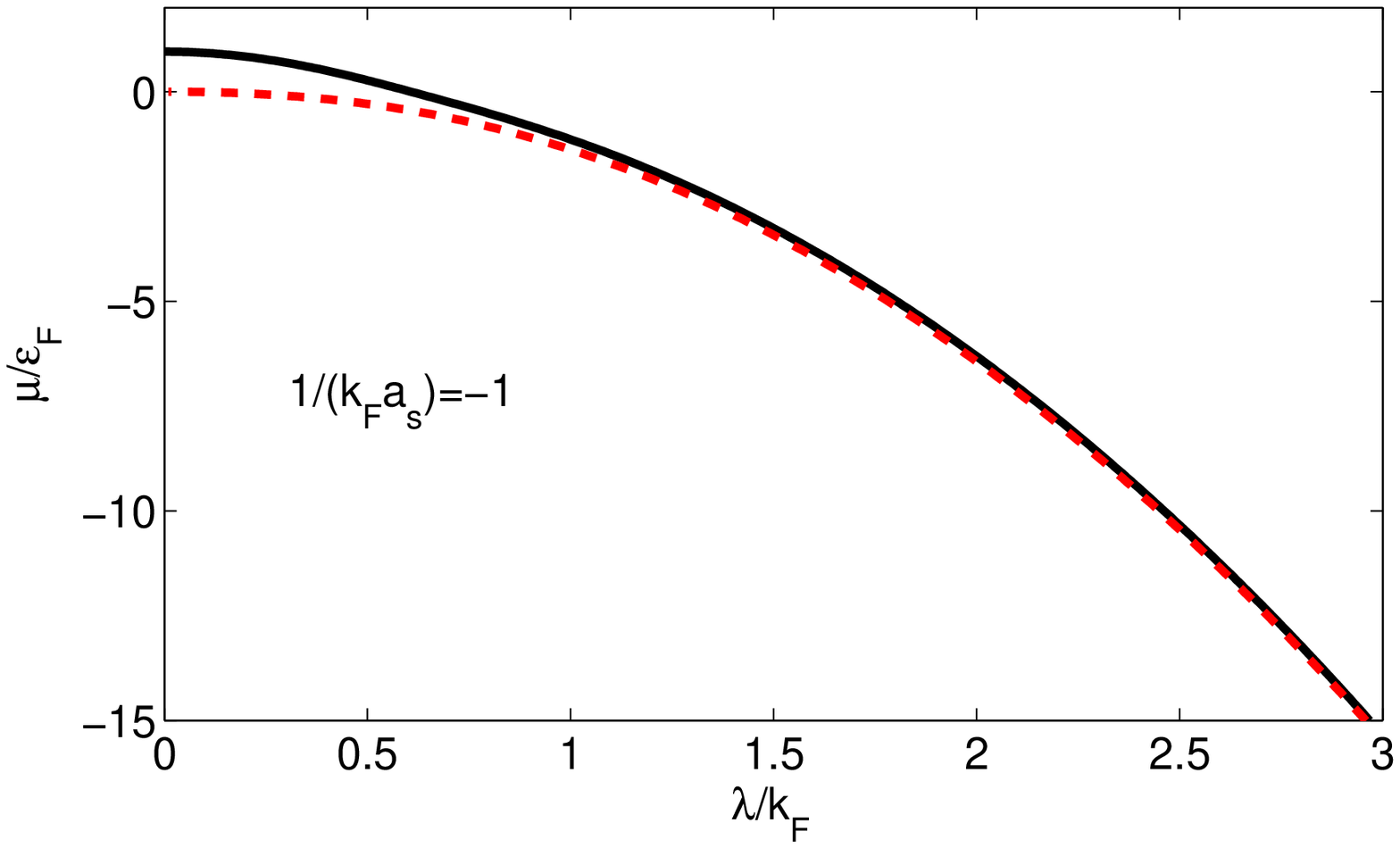}
\includegraphics[width=7.2cm]{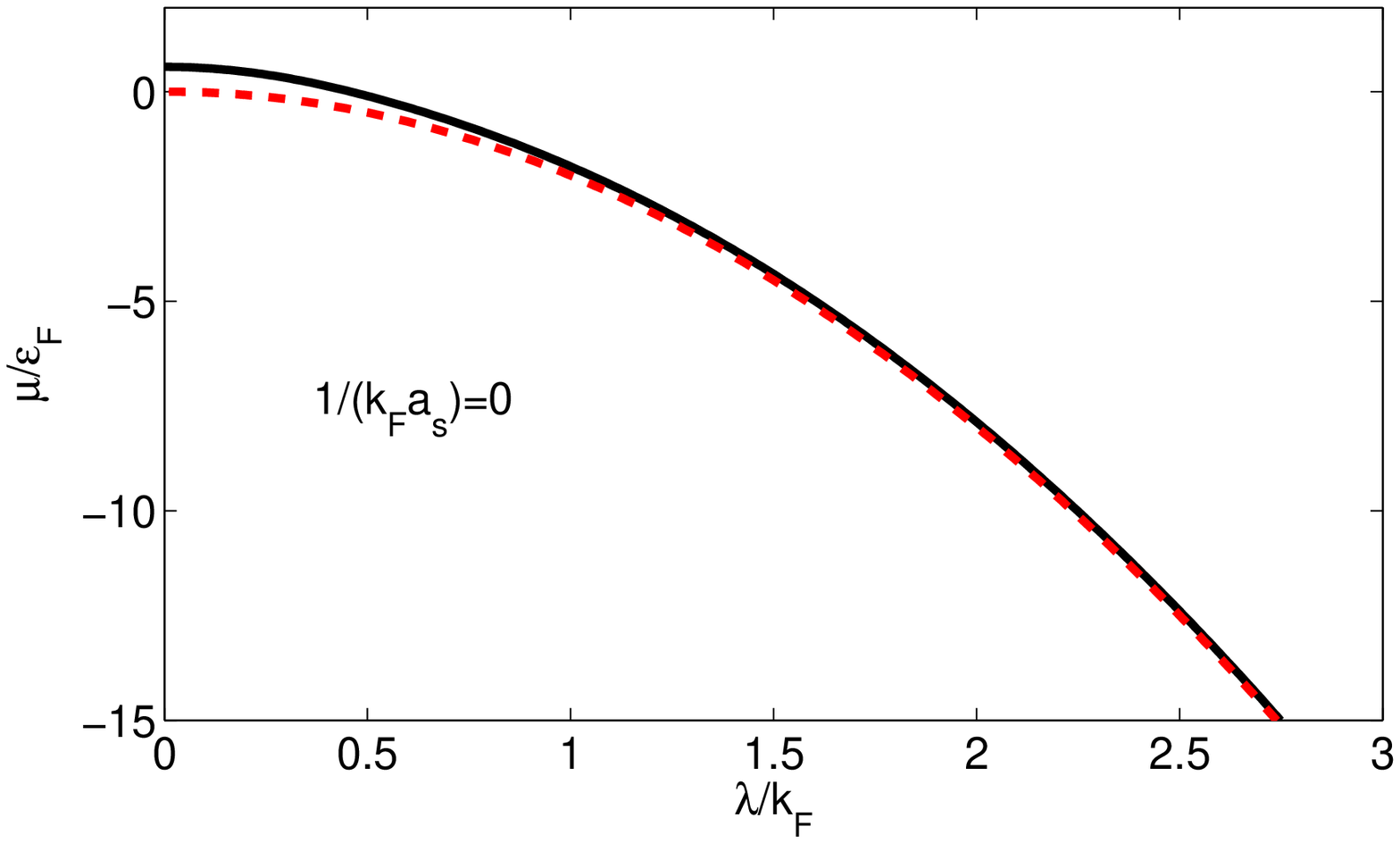}
\includegraphics[width=7.2cm]{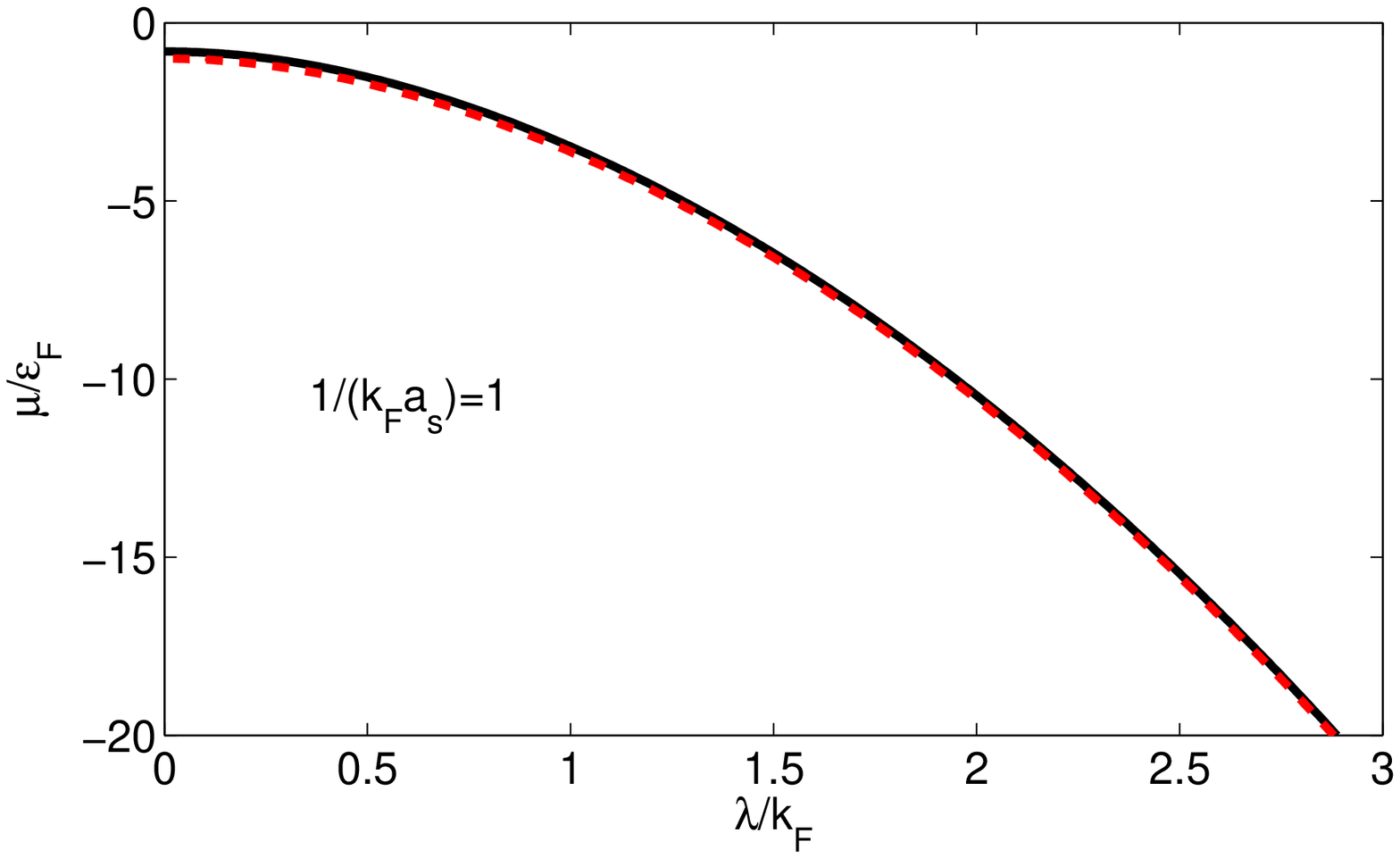}
\includegraphics[width=7.2cm]{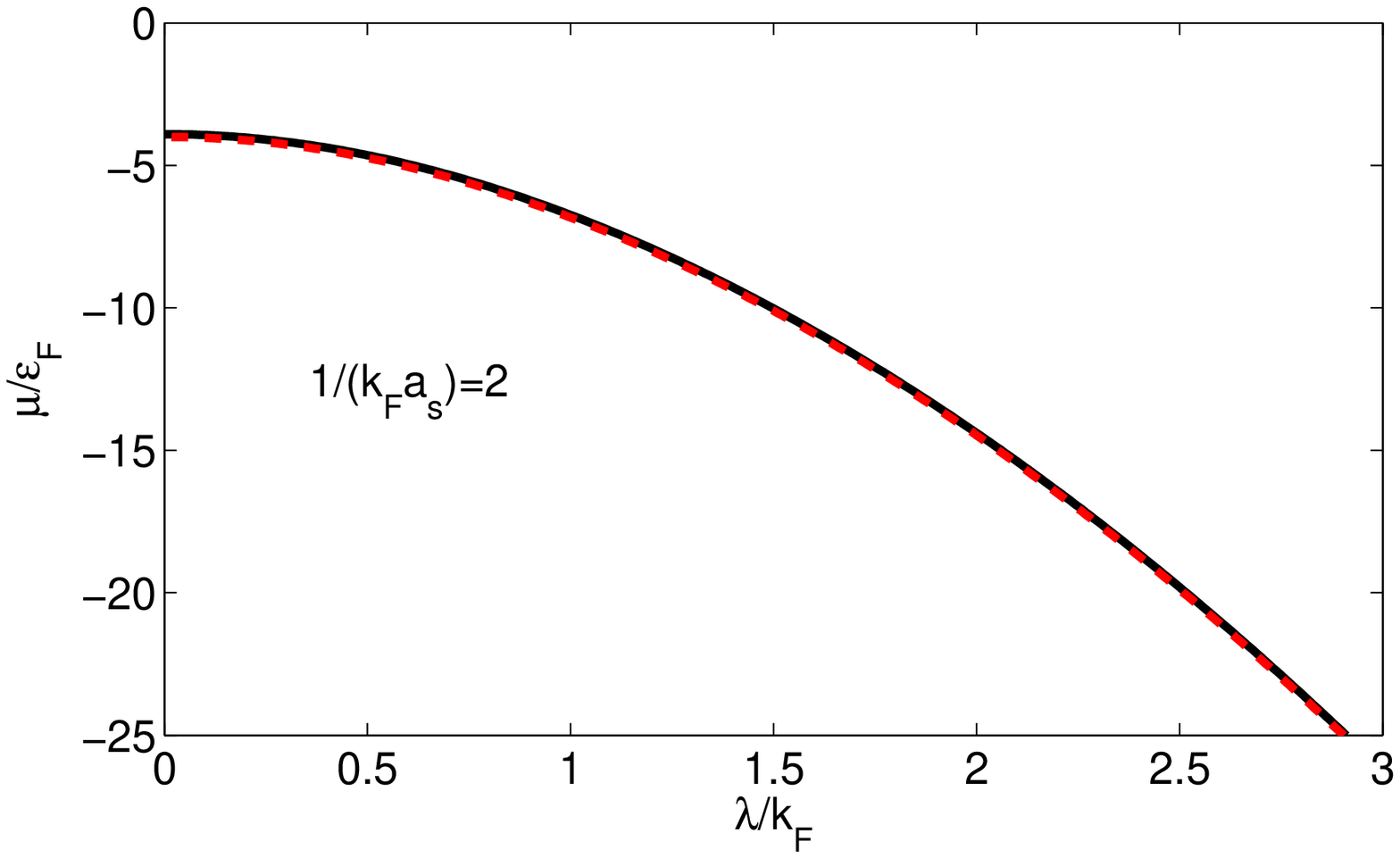}
\caption{The chemical potential $\mu$ (divided by $\epsilon_{\rm
F}$) as a function of $\lambda/k_{\rm F}$. The red dashed line shows
the analytical result $\mu\simeq-\frac{E_{\rm B}}{2}$ for large SOC.
 \label{fig4}}
\end{center}
\end{figure}

Beyond the leading order, we can write the chemical potential $\mu$
as
\begin{eqnarray}
\mu=-\frac{E_{\text B}}{2}+\frac{\mu_{\text B}}{2},
\end{eqnarray}
where $\mu_{\text B}=2\mu+E_{\rm B}\ll E_{\rm B}$ is referred to as
the effective chemical potential for bosons (rashbons). We will give
an explicit expression for $\mu_{\text B}$ in Section V.

{\bf (B) Numerical Results.} The gap and number equations (39) and
(48) are equivalent. For numerical calculations, it is convenient to
employ Eq. (39). If we define the following dimensionless quantities
\begin{eqnarray}
g_1=\frac{\lambda}{k_{\rm F}}, \ \ \ g_2=\frac{1}{k_{\rm F}a_s},\ \
\ x_1=\frac{\mu}{\epsilon_{\rm F}},\ \ \
x_2=\frac{\Delta_0}{\epsilon_{\rm F}},
\end{eqnarray}
the gap and number equations can be written as the following
dimensionless form
\begin{eqnarray}
&&\int_0^\infty dz(z^2+g_1^2)\Bigg[\frac{1}{z^2+g_1^2{\cal
J}(g_2/g_1)-g_1^2}\nonumber\\
&&\ \ \ \ \ \ \ \ \ \ \ \ \ \ \ \ \ \ \ \ \ \ \ -\frac{1}{\sqrt{(z^2-g_1^2-x_1)^2+x_2^2}}\Bigg]=0,\nonumber\\
&&\int_0^\infty
dz(z^2+g_1^2)\Bigg[1-\frac{z^2-g_1^2-x_1}{\sqrt{(z^2-g_1^2-x_1)^2+x_2^2}}\Bigg]=\frac{2}{3}.
\end{eqnarray}
The integrals in the above equations can be analytically evaluated
using elliptic functions \cite{ANAGAP}. For given values of $g_1$
and $g_2$, these two equations determine $x_1$ and $x_2$.

The numerical results are shown in Fig. \ref{fig3} and Fig.
\ref{fig4}. The red dashed lines correspond to the analytical
results for large SOC,
\begin{eqnarray}
&&x_1=-g_1^2{\cal J}(g_2/g_1),\nonumber\\
&&x_2=\sqrt{\frac{16g_1}{3\pi}\frac{[{\cal
J}(g_2/g_1)-1]^{3/2}}{{\cal J}(g_2/g_1)}}.
\end{eqnarray}
We find that the pairing gap generally increases with increased
$\lambda/k_{\rm F}$. The numerical results become in good agreement
with the analytical results when $\lambda/k_{\rm F}\gtrsim 1$.
Therefore, the system enters the rashbon BEC regime at
$\lambda/k_{\text F}\sim1$. For large positive value of $1/(k_{\rm
F}a_s)$, the analytical results are in good agreement with the
numerical results even for small values of $\lambda/k_{\rm F}$. For
very large $\lambda$, we find the numerical results fit very well
with the following scaling behavior
\begin{eqnarray}
\frac{\Delta_0}{\epsilon_{\rm
F}}\simeq\sqrt{\frac{8}{3\pi}}\sqrt{\frac{\lambda}{k_{\rm F}}},\ \ \
\ \frac{\mu}{\epsilon_{\rm F}}\simeq-2\left(\frac{\lambda}{k_{\rm
F}}\right)^2,
\end{eqnarray}
for both negative and positive values of $1/(k_{\rm F}a_s)$.

\begin{figure}[!htb]
\begin{center}
\includegraphics[width=7.2cm]{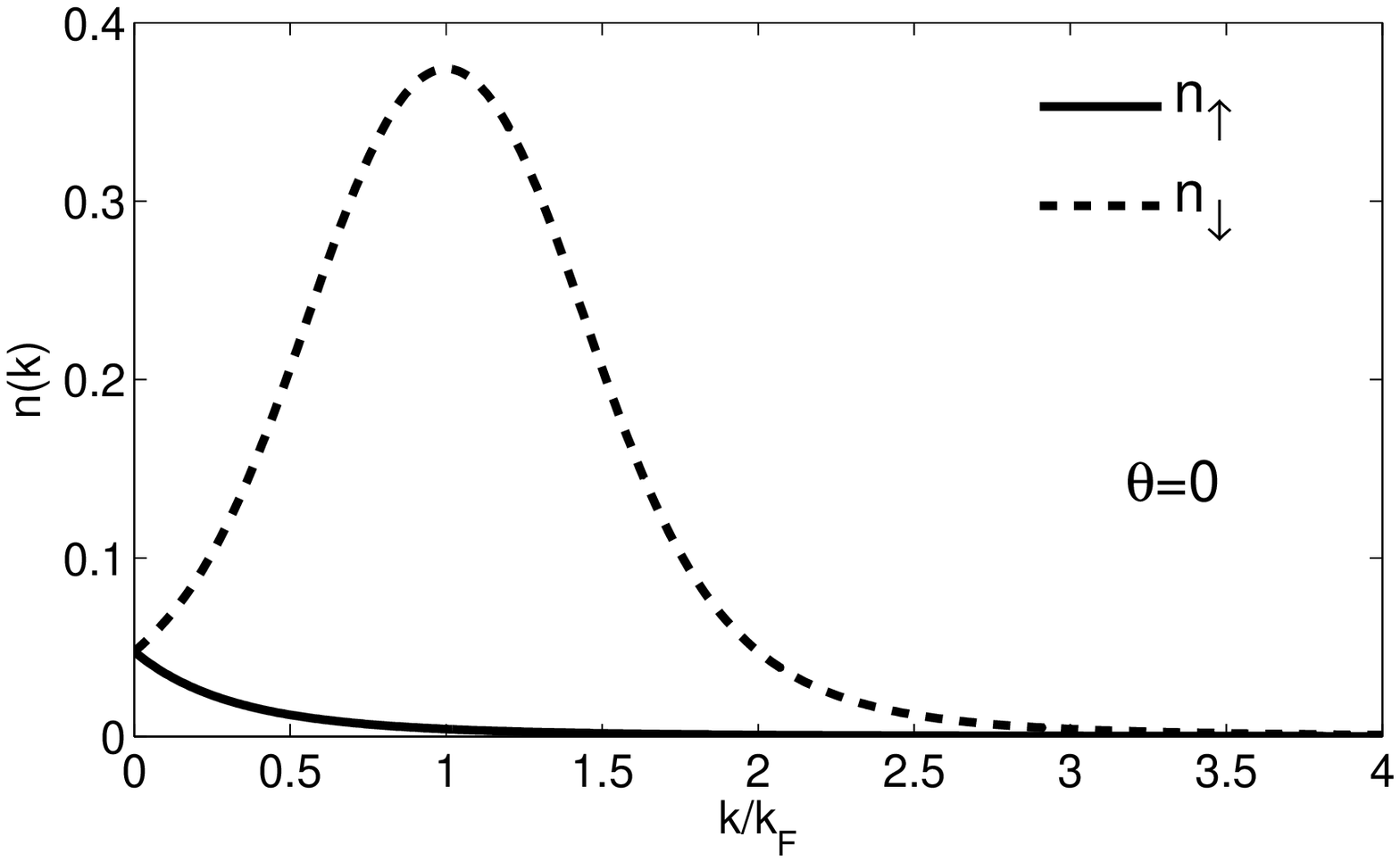}
\includegraphics[width=7.2cm]{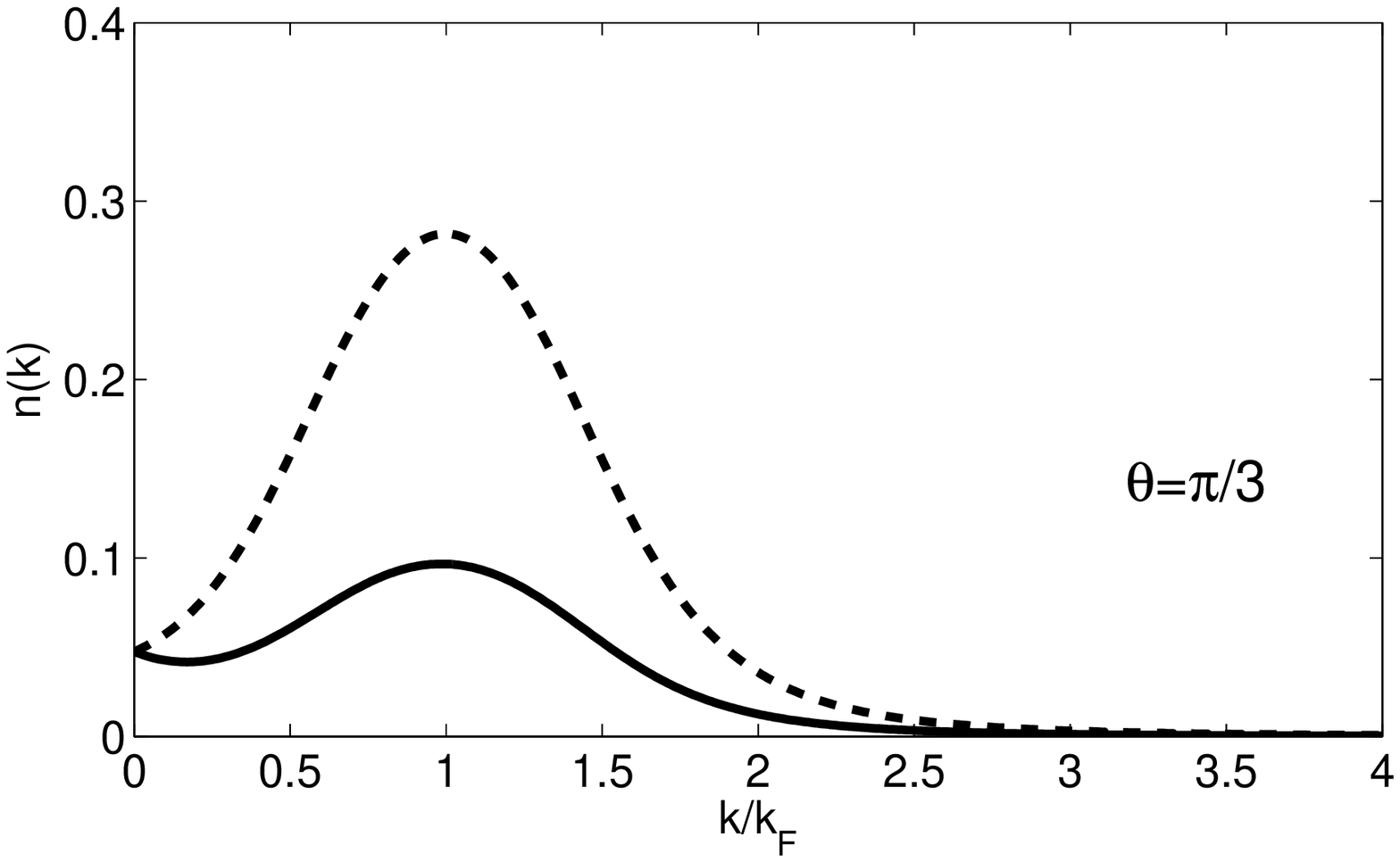}
\includegraphics[width=7.2cm]{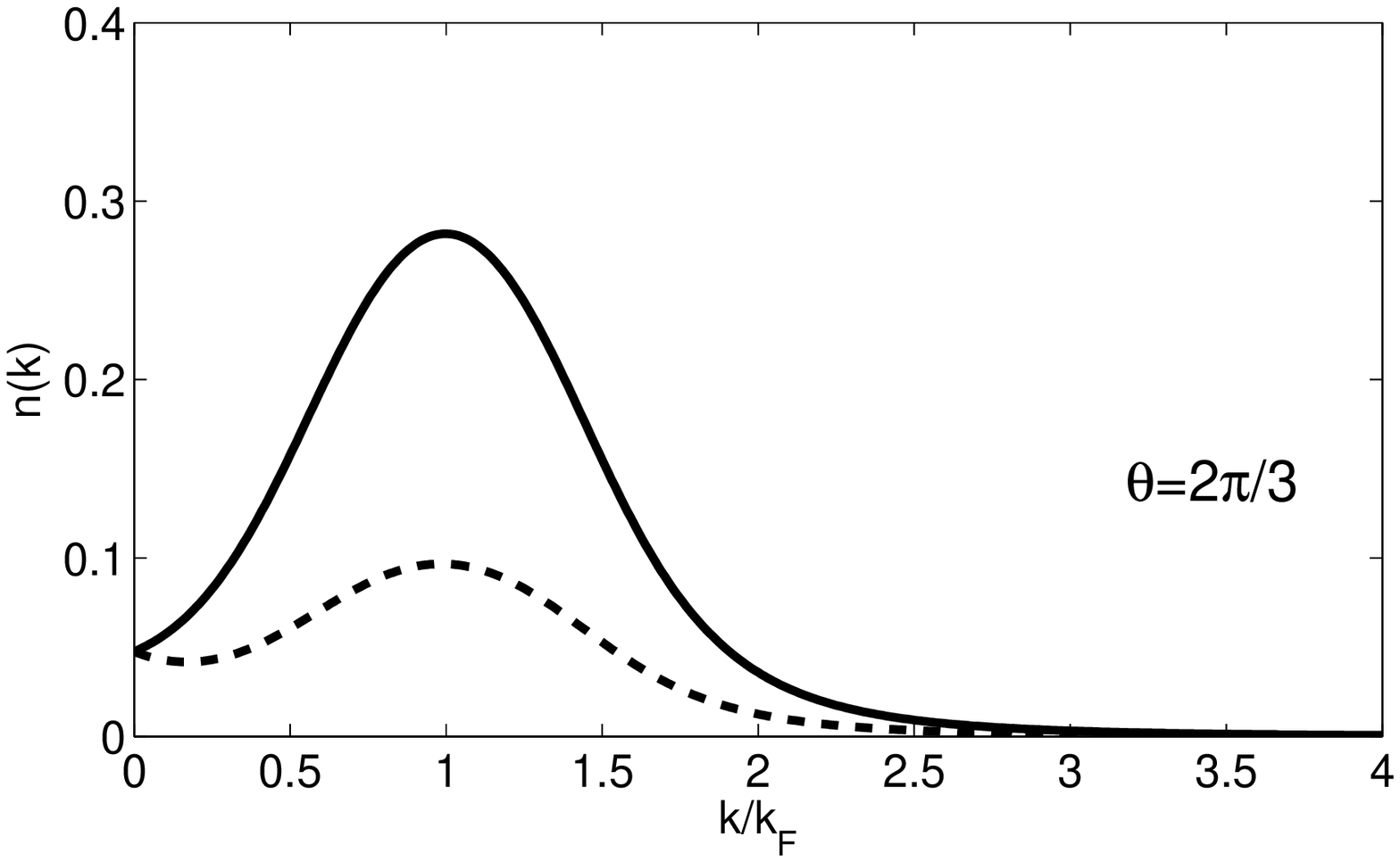}
\includegraphics[width=7.2cm]{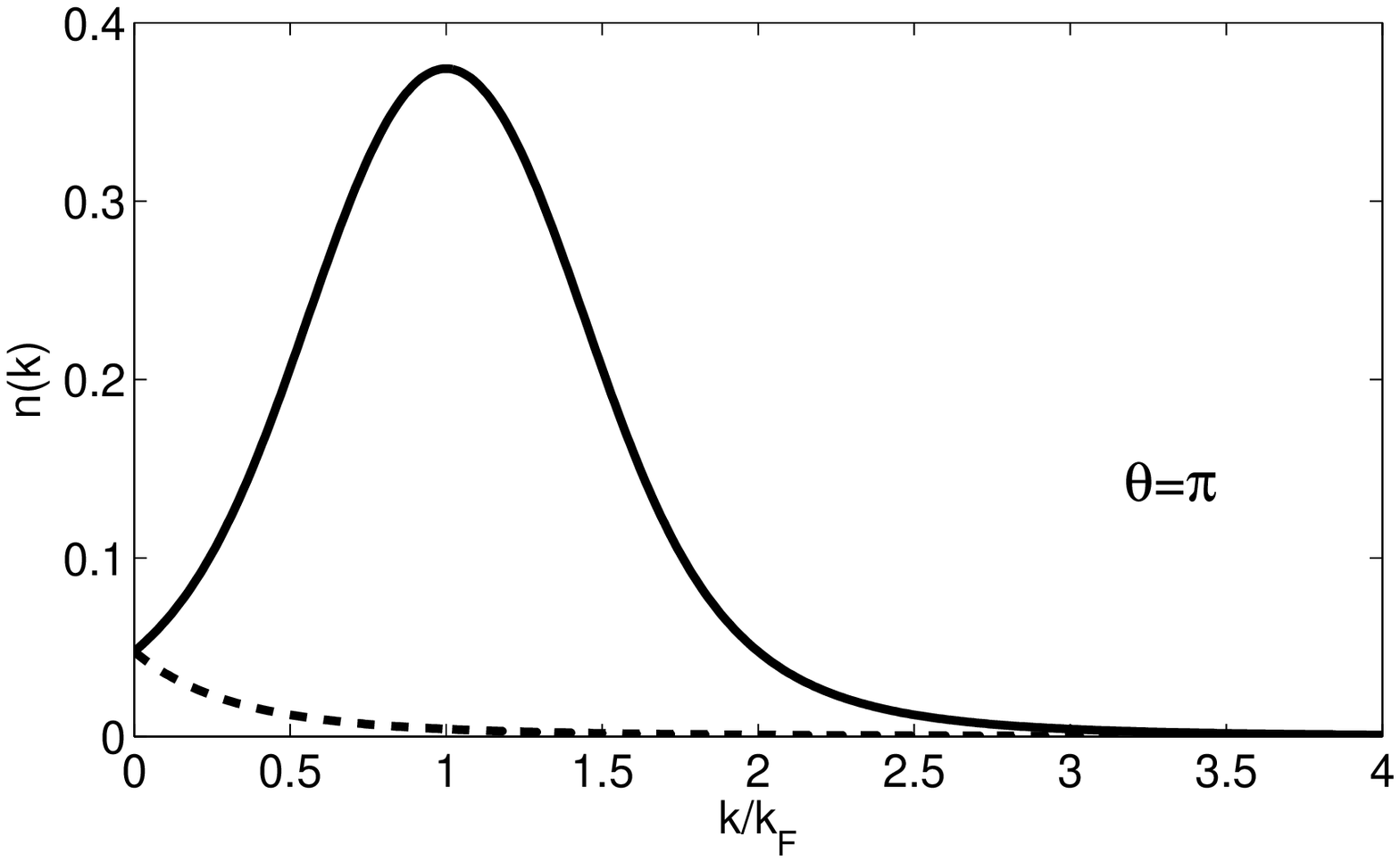}
\caption{The fermion momentum distributions $n_\uparrow(k)$ and
$n_\downarrow(k)$ for various values of the polar angle $\theta$. We
set $1/(k_{\rm F}a_s)=-1$ and $\lambda/k_{\rm F}=1$ in this
calculation.
 \label{fig5}}
\end{center}
\end{figure}

\subsection{Fermion momentum distribution}
From the matrix elements of the fermion Green's function ${\cal
G}(i\omega_n,{\bf k})$, we can obtain the momentum distributions
$n_\uparrow({\bf k})$ and $n_\downarrow({\bf k})$ for the two spin
components. The density of each component reads $n_\sigma=\sum_{\bf
k}n_\sigma({\bf k}). $We find that even though the density of the
two components are the same, $n_\uparrow=n_\downarrow$, their
distributions in the momentum space are different. At zero
temperature, their explicit expressions are given by
\begin{eqnarray}
n_\uparrow(k,\theta)&=&\frac{1}{4}\sum_\alpha\left(1-\frac{\xi_k^\alpha}{E_k^\alpha}\right)
+\frac{\cos\theta}{4}\sum_\alpha\alpha\left(1-\frac{\xi_k^\alpha}{E_k^\alpha}\right),\nonumber\\
n_\downarrow(k,\theta)&=&\frac{1}{4}\sum_\alpha\left(1-\frac{\xi_k^\alpha}{E_k^\alpha}\right)
-\frac{\cos\theta}{4}\sum_\alpha\alpha\left(1-\frac{\xi_k^\alpha}{E_k^\alpha}\right),
\end{eqnarray}
where $\theta$ is the polar angle in the momentum space. We find
that $n_\uparrow({\bf k})=n_\downarrow({\bf k})$ only for
$\theta=\pi/2$. We have $n_\uparrow({\bf k})<n_\downarrow({\bf k})$
for $0<\theta<\pi/2$ and $n_\uparrow({\bf k})>n_\downarrow({\bf k})$
for $\pi/2<\theta<\pi$. The reason of $n_\uparrow({\bf k})\neq
n_\downarrow({\bf k})$ can be understood from the fact that the
inversion symmetry ($z\rightarrow -z$) does not hold due to the
presence of SOC. Meanwhile, we have $n_\uparrow({\bf
k})=n_\downarrow(-{\bf k})$ due to the time reversal symmetry.

In general, with increased SOC, the distribution broadens, which
indicates a BCS-BEC crossover. A numerical example for $1/(k_{\rm
F}a_s)=-1$ and $\lambda/k_{\rm F}=1$ is shown in Fig. \ref{fig5}.
The new feature here is that the distributions generally display
non-monotonous behavior due to the SOC effect. We note that the
peaks in the distributions are just located at $k=\lambda$.

\subsection{Condensate density}

According to Leggett's definition~\cite{FC}, the condensate number
of fermion pairs is given by
\begin{eqnarray}
N_0=\frac{1}{2}\sum_{\sigma,\sigma^\prime=\uparrow,\downarrow}\int\int
d^3{\bf r}d^3{\bf r}^\prime|\langle\psi_{\sigma}({\bf
r})\psi_{\sigma^\prime}({\bf r}^\prime)\rangle|^2. \label{A18}
\end{eqnarray}
For systems with only singlet pairing, this recovers the usual
result $N_0=\int\int d^3{\bf r}d^3{\bf
r}^\prime|\langle\psi_{\uparrow}({\bf r})\psi_{\downarrow}({\bf
r}^\prime)\rangle|^2$ \cite{FCFM}. Converting this to the momentum
space, we find that the condensate density $n_0=N_0/V$ is a sum of
all absolute squares of the pairing amplitudes,
\begin{eqnarray}
n_0=\frac{1}{2}\sum_{\bf k}\left[|\phi_{\uparrow\downarrow}({\bf
k})|^2+|\phi_{\downarrow\uparrow}({\bf
k})|^2+|\phi_{\uparrow\uparrow}({\bf
k})|^2+|\phi_{\downarrow\downarrow}({\bf k})|^2\right]\nonumber\\
=\sum_{\bf k}\left\{\left[\frac{1}{\beta}\sum_n{\cal
A}_{21}(i\omega_n,{\bf k})\right]^2+\left[\frac{1}{\beta}\sum_n{\cal
B}_{21}(i\omega_n,{\bf k})\right]^2\right\}.
\end{eqnarray}

Completing the Matsubara frequency summation and taking the zero
temperature limit, we obtain the explicit expression for $T=0$,
\begin{eqnarray}
n_0&=&\frac{\Delta_0^2}{16\pi^2}\int_0^\infty
k^2dk\left[\frac{1}{(E_k^+)^2}+\frac{1}{(E_k^-)^2}\right]\nonumber\\
&=&\frac{\Delta_0^2}{8\pi^2}\int_0^\infty
dk\frac{k^2+\lambda^2}{(\epsilon_k-\tilde{\mu})^2+\Delta_0^2}.
\end{eqnarray}
Generally, we can show that $n_0<n/2$. For large SOC and/or
attraction, we have $\Delta_0\ll|\mu|$. Using the number equation
(39) or (48) and expanding all terms in powers of $\Delta_0/|\mu|$,
we find that
\begin{eqnarray}
n_0=\frac{n}{2}-O\left(\frac{\Delta_0^4}{|\mu|^4}\right).
\end{eqnarray}
Therefore, the condensate fraction $2N_0/N$ approaches unity at
large SOC and/or attraction, indicating the fact that the ground
state at large SOC is a Bose condensate of weakly interacting
rashbons.

\begin{figure}[!htb]
\begin{center}
\includegraphics[width=8cm]{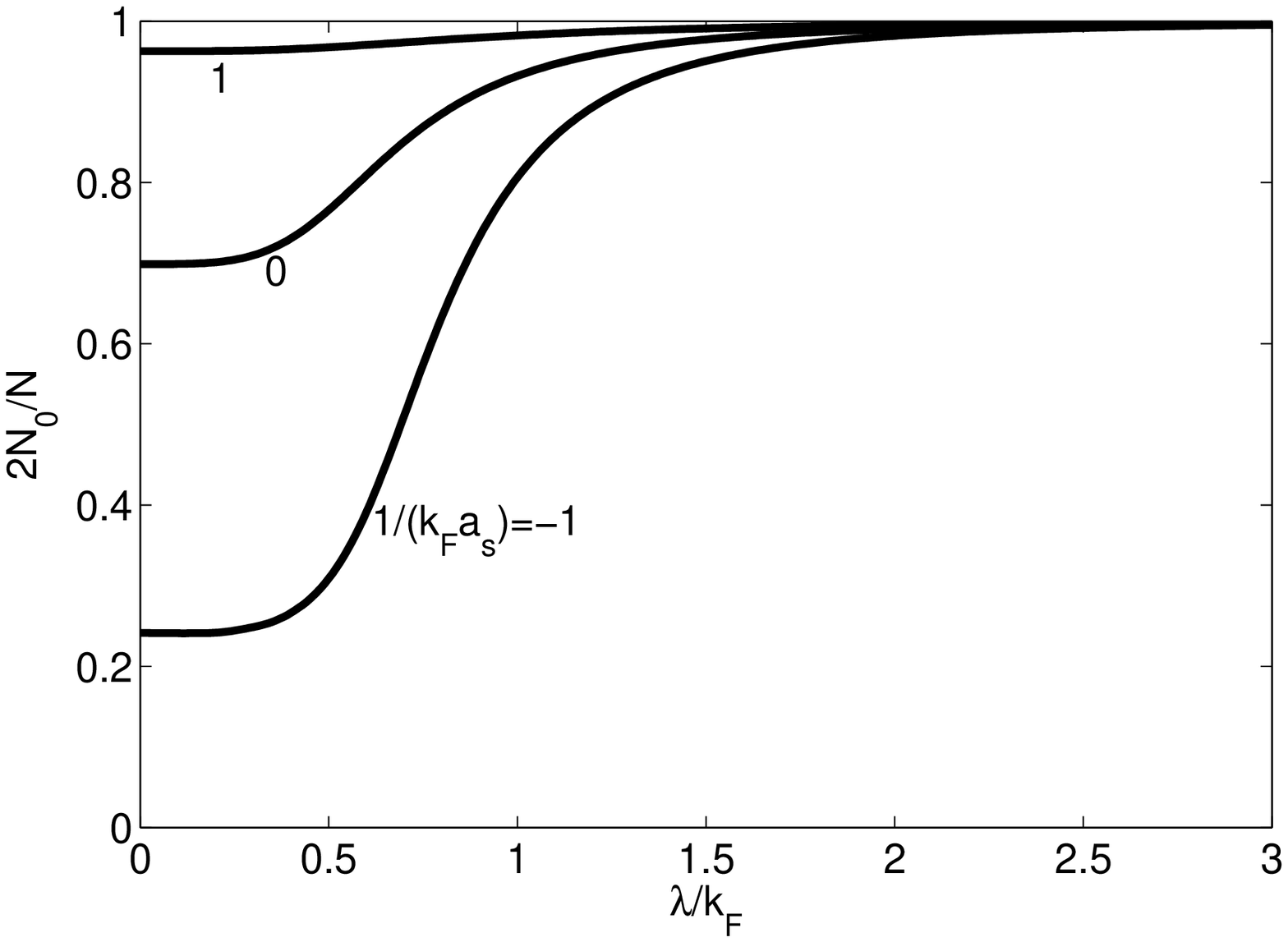}
\caption{The condensate fraction $2N_0/N$ as a function of
$\lambda/k_{\rm F}$ for various values of $1/(k_{\rm F}a_s)$.
 \label{fig6}}
\end{center}
\end{figure}

In general, the condensate fraction $2N_0/N$ can be expressed as
\begin{eqnarray}
\frac{2N_0}{N}=\frac{3x_2^2}{4}\int_0^\infty dz
\frac{z^2+g_1^2}{(z^2-g_1^2-x_1)^2+x_2^2}.
\end{eqnarray}
It can be numerically obtained using the solutions of $x_1$ and
$x_2$ from the gap and number equations. The numerical results are
shown in Fig. \ref{fig6}. We find that, even for negative values of
$1/(k_{\rm F}a_s)$, the condensate fraction approaches unity around
$\lambda/k_{\rm F}\sim 2$. This is consistent with the observation
from the solutions of the gap and number equations that the system
enters the rashbon BEC regime at $\lambda/k_{\rm F}\simeq 1$ for
negative and small positive values of $1/(k_{\rm F}a_s)$.

\subsection{Superfluid density}
To evaluate the superfluid density $n_s$, we can employ the standard
definition \cite{NS01,NS02}. When the superfluid moves with a
uniform velocity
$\mbox{\boldmath{$\upsilon$}}_s=(\upsilon_x,\upsilon_y,\upsilon_z)$,
the superfluid order parameter transforms like $\Phi\rightarrow \Phi
e^{2i{\bf q}_s\cdot {\bf r}}$ and $\Phi^*\rightarrow \Phi^*
e^{-2i{\bf q}_s\cdot {\bf r}}$, where ${\bf
q}_s=m\mbox{\boldmath{$\upsilon$}}_s$ ($m=1$ in our units). The
superfluid density $n_s$ is defined as the response of the
thermodynamic potential $\Omega$ to an infinitesimal velocity
velocity $\mbox{\boldmath{$\upsilon$}}_s$, i.e.,
\begin{eqnarray}
\Omega({\bf q}_s)=\Omega({\bf 0})+\frac{1}{2}n_s{\bf q}_s^2+O({\bf
q}_s^4).
\end{eqnarray}
The thermodynamic potential in the presence of a velocity
$\mbox{\boldmath{$\upsilon$}}_s$ can be evaluated by a gauge
transformation for the fermion field $\psi\rightarrow\psi e^{-i{\bf
q}_s\cdot {\bf r}}$. We have
\begin{eqnarray}
\Omega({\bf
q}_s)=\frac{\Delta^2}{U}-\frac{1}{2}\frac{1}{\beta}\sum_n\sum_{\bf
k}\text{lndet}{\cal G}_s^{-1}(i\omega_n,{\bf k}),
\end{eqnarray}
where the inverse fermion Green function in the presence of
$\mbox{\boldmath{$\upsilon$}}_s$ reads
\begin{eqnarray}
{\cal G}_s^{-1}(i\omega_n,{\bf k})={\cal G}^{-1}(i\omega_n,{\bf
k})-\Sigma({\bf q}_s).
\end{eqnarray}
Here the velocity-dependent part $\Sigma({\bf q}_s)$ includes three
parts, $\Sigma({\bf q}_s)=\Sigma_1({\bf q}_s) +\Sigma_2({\bf
q}_s)+\Sigma_3({\bf q}_s)$, where
\begin{eqnarray}
&&\Sigma_1({\bf q}_s)=\frac{1}{2}{\bf q}_s^2\tau_3,\nonumber\\
&&\Sigma_2({\bf q}_s)={\bf k}\cdot{\bf q}_s\tau_0,\nonumber\\
&&\Sigma_3({\bf
q}_s)=\lambda(\sigma_xq_x\tau_3+\sigma_yq_y\tau_0+\sigma_zq_z\tau_3).
\end{eqnarray}
Here $\tau_i$ ($i=1,2,3$) and $\tau_0$ are the Pauli matrices and
the identity matrix in the Nambu-Gor'kov space, respectively. We
note that the term $\Sigma_3({\bf q}_s)$ is purely due to the
presence of SOC.

{\bf (A) Derivation of the Superfluid Density.} The superfluid
density $n_s$ can be obtained by the method of derivative expansion
for $\Omega({\bf q}_s)$, i.e.,
\begin{eqnarray}
\Omega({\bf q}_s)=\Omega({\bf
0})+\frac{1}{2}\sum_n\frac{1}{n}\text{Tr}\left[{\cal G}\Sigma({\bf
q}_s)\right]^n.
\end{eqnarray}
We find that there are four types of nonzero contributions at the
order $O({\bf q}_s^2)$:
\begin{eqnarray}
&&\Omega_1\sim\text{Tr}({\cal G}\Sigma_1),\ \ \ \ \ \ \ \ \ \
\Omega_2\sim
\text{Tr}({\cal G}\Sigma_2{\cal G}\Sigma_2),\nonumber\\
&&\Omega_3\sim \text{Tr}({\cal G}\Sigma_3{\cal G}\Sigma_3),\ \ \
\Omega_4\sim \text{Tr}({\cal G}\Sigma_2{\cal G}\Sigma_3).
\end{eqnarray}
Since the superfluid state is isotropic, the superfluid density
tensor should also be isotropic. We have carefully checked that all
anisotropic terms vanish exactly. Completing the trace in the
Nambu-Gor'kov and spin spaces, we finally obtain the following
expressions for the four types of contributions:
\begin{widetext}
\begin{eqnarray}
&&\Omega_1=\frac{{\bf q}_s^2}{2}\frac{1}{\beta}\sum_n\sum_{\bf
k}\frac{1}{2}\left({\cal A}_{11}e^{i\omega0^+}-{\cal
A}_{22}e^{-i\omega_n0^+}\right)\nonumber\\
&&\Omega_2=\frac{{\bf q}_s^2}{2}\frac{1}{\beta}\sum_n\sum_{\bf
k}\frac{{\bf k}^2}{3}\left({\cal A}_{11}^2+{\cal B}_{11}^2+{\cal
A}_{22}^2+{\cal B}_{22}^2+2{\cal A}_{21}^2+2{\cal
B}_{21}^2\right),\nonumber\\
&&\Omega_3=\frac{{\bf q}_s^2}{2}\frac{1}{\beta}\sum_n\sum_{\bf
k}\lambda^2\left[\left({\cal A}_{11}^2+{\cal A}_{22}^2+2{\cal
A}_{21}^2\right)-\frac{1}{3}\left({\cal B}_{11}^2+{\cal
B}_{22}^2+2{\cal
B}_{21}^2\right)\right],\nonumber\\
&&\Omega_4=\frac{{\bf q}_s^2}{2}\frac{1}{\beta}\sum_n\sum_{\bf
k}\frac{4\lambda|{\bf k}|}{3}\left({\cal A}_{11}{\cal B}_{11}-{\cal
A}_{22}{\cal B}_{22}+2{\cal A}_{21}{\cal B}_{21}\right).
\end{eqnarray}
Note that the first contribution is just from the total particle
density $n$, $\Omega_1=\frac{1}{2}n{\bf q}_s^2$. Collecting all
terms, the superfluid density $n_s$ is given by
\begin{eqnarray}
&&n_s=n+\frac{1}{\beta}\sum_n\sum_{\bf k}\Bigg[\frac{{\bf
k}^2}{3}\left({\cal A}_{11}^2+{\cal B}_{11}^2+{\cal A}_{22}^2+{\cal
B}_{22}^2+2{\cal A}_{21}^2+2{\cal B}_{21}^2\right)+\frac{4\lambda
|{\bf k}|}{3}\left({\cal A}_{11}{\cal B}_{11}-{\cal A}_{22}{\cal
B}_{22}+2{\cal A}_{21}{\cal
B}_{21}\right)\nonumber\\
&&\ \ \ \ \ +\lambda^2\left({\cal A}_{11}^2+{\cal A}_{22}^2+2{\cal
A}_{21}^2\right)-\frac{\lambda^2}{3}\left({\cal B}_{11}^2+{\cal
B}_{22}^2+2{\cal B}_{21}^2\right)\Bigg].
\end{eqnarray}
Completing the Matsubara frequency sum, we obtain the
finite-temperature expression
\begin{eqnarray}
n_s&=&n-\int_0^\infty\frac{k^2dk}{2\pi^2}\left[\frac{(k+\lambda)^2}{6}\frac{1}{2T}\frac{1}{\cosh^2\left(\frac{E_k^+}{2T}\right)}+\frac{(k-\lambda)^2}{6}\frac{1}{2T}\frac{1}{\cosh^2\left(\frac{E_k^-}{2T}\right)}\right]\nonumber\\
&-&\frac{\lambda}{3}\int_0^\infty\frac{k
dk}{2\pi^2}\left[\left(\xi_k^++\frac{\Delta^2}{\xi_k}\right)\frac{1-2f(E_k^+)}{E_k^+}-\left(\xi_k^-+\frac{\Delta^2}{\xi_k}\right)\frac{1-2f(E_k^-)}{E_k^-}\right].
\end{eqnarray}
\end{widetext}
We have checked that this expression is consistent with the result
for ordinary fermionic superfluids in the absence of SOC
\cite{NS01,NS02}. Also, setting $\Delta=0$, we find that
$n_s(\Delta=0)$ vanishes exactly.

We are interested in the zero temperature case. At zero temperature,
the superfluid density reduces to
\begin{eqnarray}
n_s=n-n_\lambda,
\end{eqnarray}
where $n_\lambda$ is given by
\begin{eqnarray}
n_\lambda=\frac{\lambda}{6\pi^2}\int_0^\infty
kdk\left[\left(\xi_k^++\frac{\Delta_0^2}{\xi_k}\right)\frac{1}{E_k^+}-\left(\xi_k^-+\frac{\Delta_0^2}{\xi_k}\right)\frac{1}{E_k^-}\right].
\end{eqnarray}
We notice that $n_\lambda$ vanishes in the absence of SOC and we
recover the usual result $n_s=n$ at $T=0$ for ordinary fermionic
superfluids \cite{NS01,NS02}. However, for nonzero SOC, $n_\lambda$
is always positive and we have $n_s(\lambda\neq0)<n$. Therefore, the
SOC leads to suppression of the superfluid density.

{\bf (B) Analytical Result for Large SOC.} To understand this
interesting phenomenon, we first take a look at the large SOC limit.
In this case we have $\mu\simeq -E_{\text{B}}/2$ and
$\Delta\ll|\mu|$. Therefore, we can expand the expression in powers
of $\Delta/|\mu|$ and keep only the leading order terms. Doing so,
we obtain
\begin{eqnarray}
n&\simeq&\frac{\Delta_0^2}{8\pi^2}\int_0^\infty
k^2dk\left[\frac{1}{(\xi_k^+)^2}+\frac{1}{(\xi_k^-)^2}\right]\nonumber\\
&\simeq&\frac{\Delta_0^2}{\pi^2}\int_0^\infty k^2dk
\frac{(k^2+E_{\text B})^2+4\lambda^2k^2}{\left[(k^2+E_{\text
B})^2-4\lambda^2k^2\right]^2}
\end{eqnarray}
and
\begin{eqnarray}
n_\lambda&\simeq&\frac{\lambda\Delta_0^2}{6\pi^2}\int_0^\infty
kdk\left\{\frac{1}{\xi_k}\left(\frac{1}{\xi_k^+}-\frac{1}{\xi_k^-}\right)-\frac{1}{2}\left[\frac{1}{(\xi_k^+)^2}-\frac{1}{(\xi_k^-)^2}\right]\right\}\nonumber\\
&\simeq&\frac{\Delta_0^2}{\pi^2}\frac{4}{3}\int_0^\infty k^2dk
\frac{8\lambda^4k^2}{(k^2+E_{\text B})\left[(k^2+E_{\text
B})^2-4\lambda^2k^2\right]^2}.
\end{eqnarray}
Comparing the above results with the equation for the molecule
effective mass, we find that $n_\lambda/n\simeq1-2m/m_{\text B}$.
Therefore, at large SOC, the superfluid density is suppressed by a
factor $2m/m_{\rm B}$, i.e.,
\begin{eqnarray}
n_s\simeq\frac{2m}{m_{\text B}}n.
\end{eqnarray}
For $\lambda\rightarrow\infty$, using the result for $2m/m_{\rm B}$
at $\kappa=0$, we find that the ratio $n_s/n$ approaches a universal
value,
\begin{eqnarray}
\frac{n_s}{n}(\lambda/k_{\rm
F}\rightarrow\infty)\rightarrow\frac{14}{3(4+\sqrt{2})}=0.862.
\end{eqnarray}

\begin{figure}[!htb]
\begin{center}
\includegraphics[width=7.2cm]{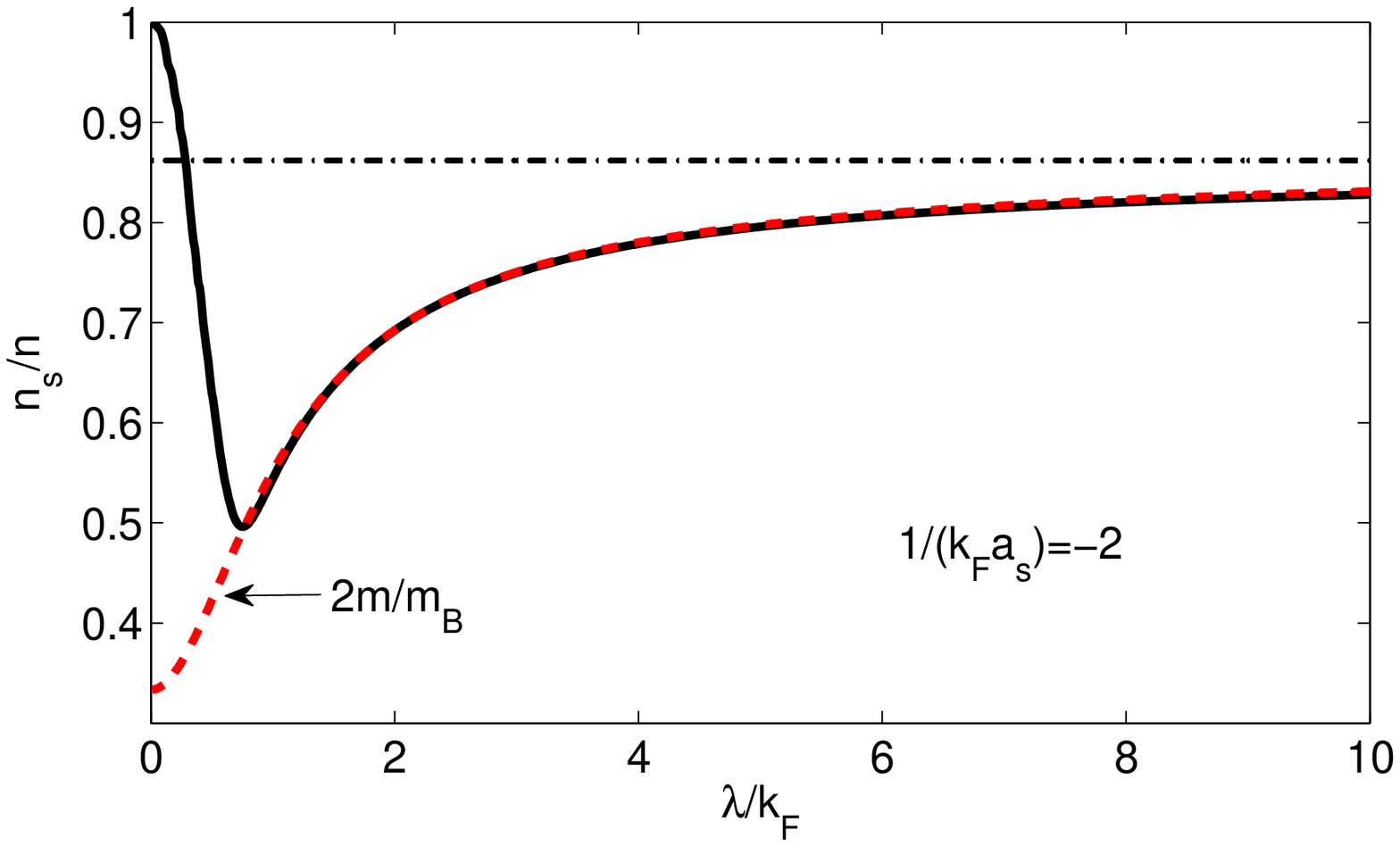}
\includegraphics[width=7.2cm]{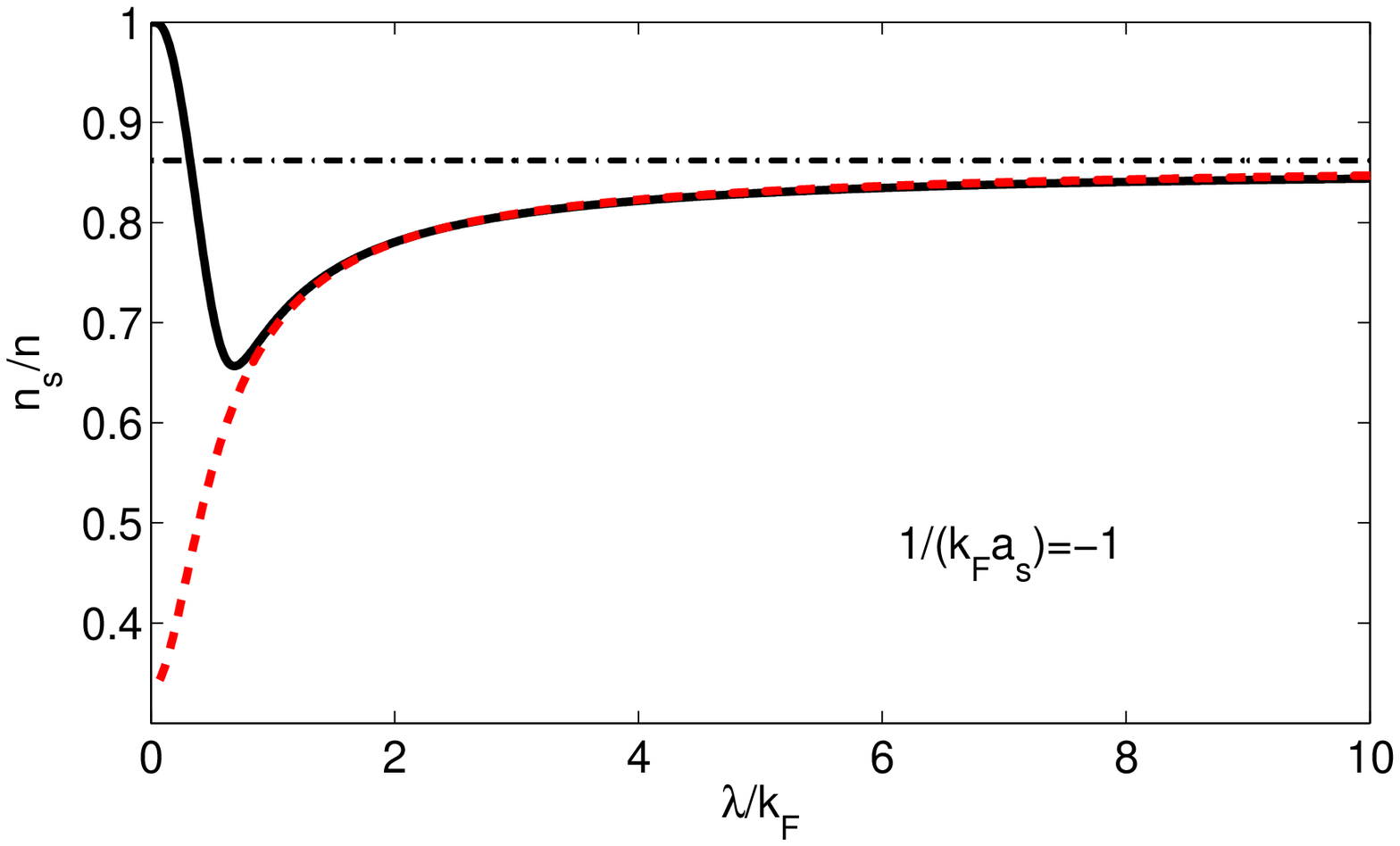}
\includegraphics[width=7.2cm]{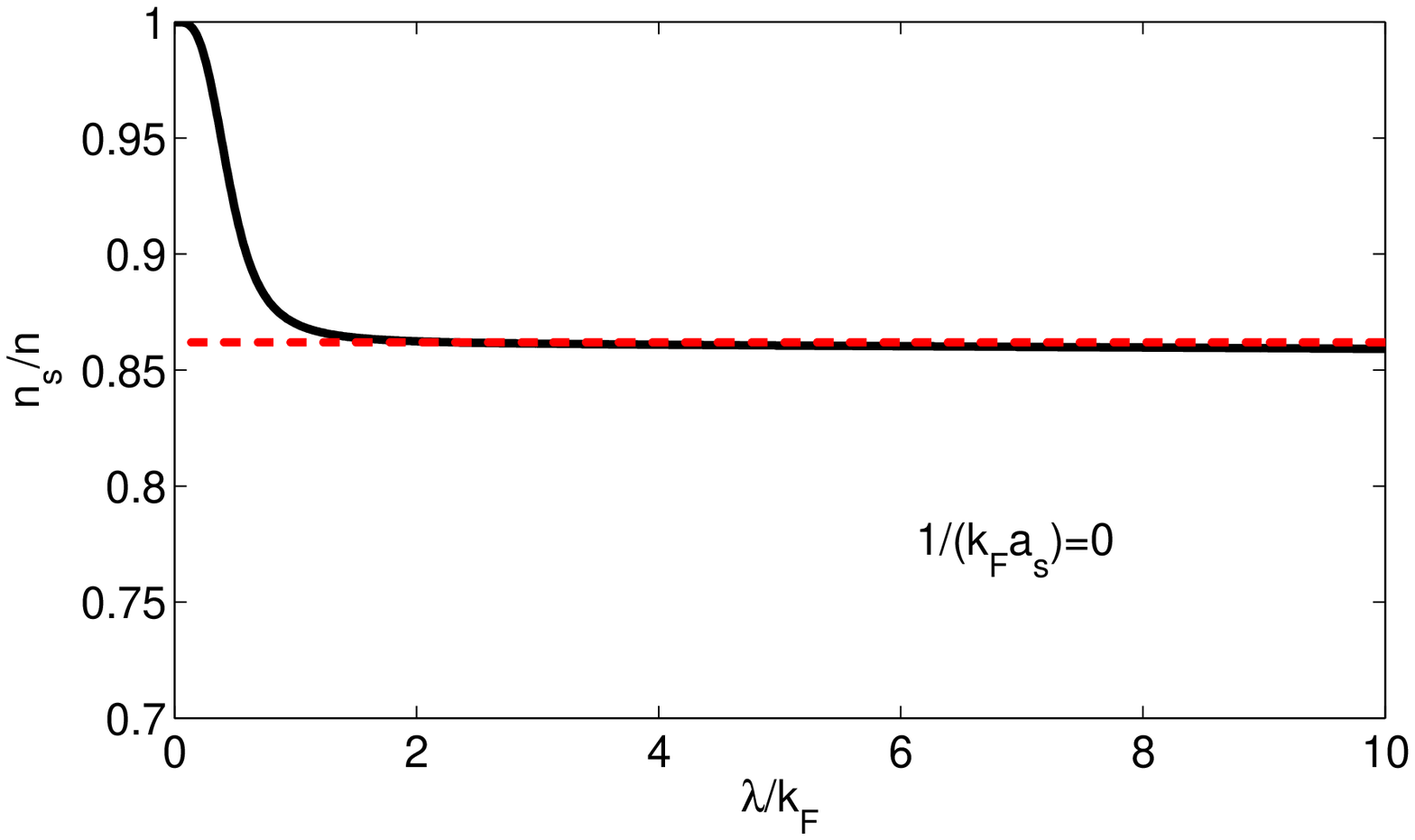}
\includegraphics[width=7.2cm]{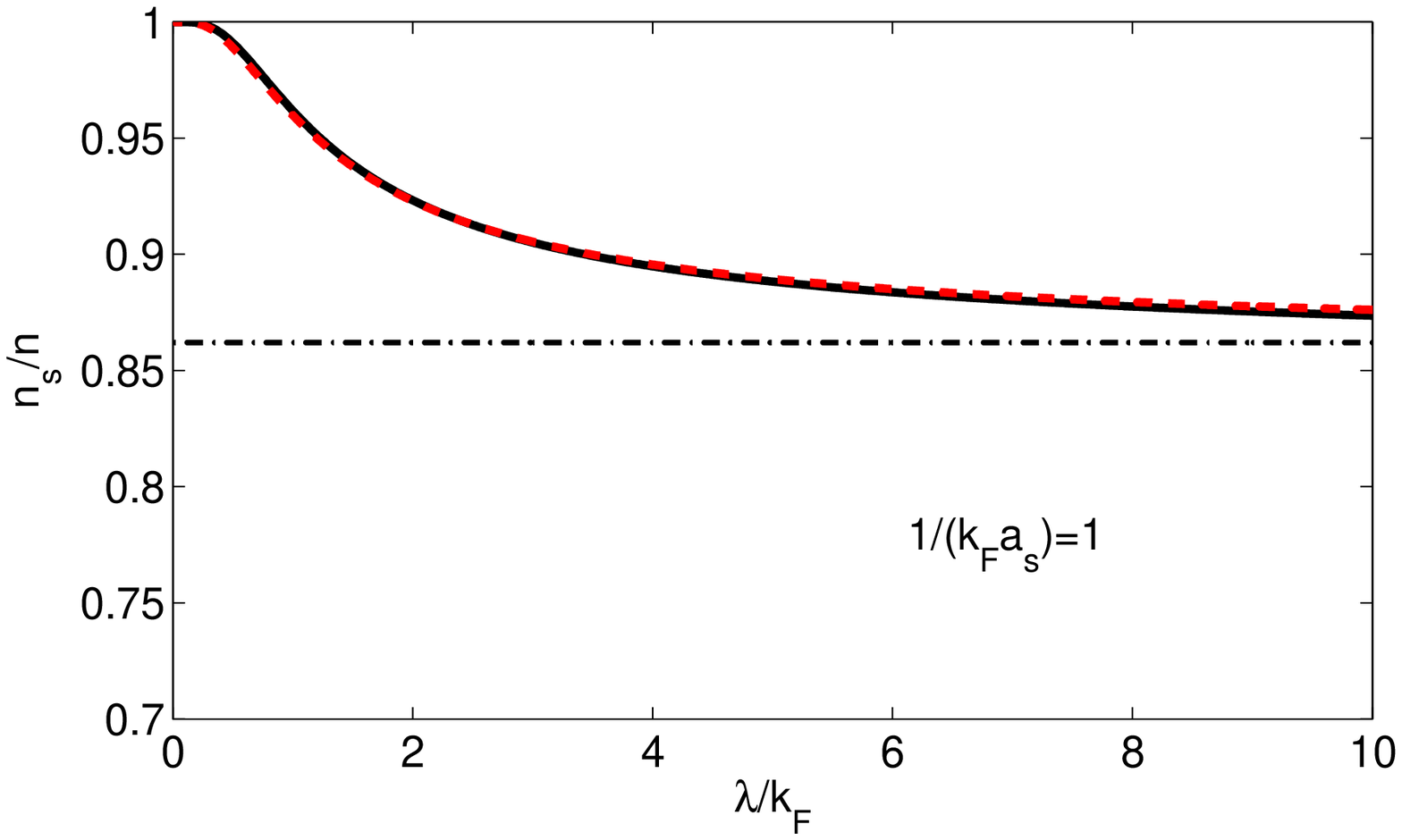}
\includegraphics[width=7.2cm]{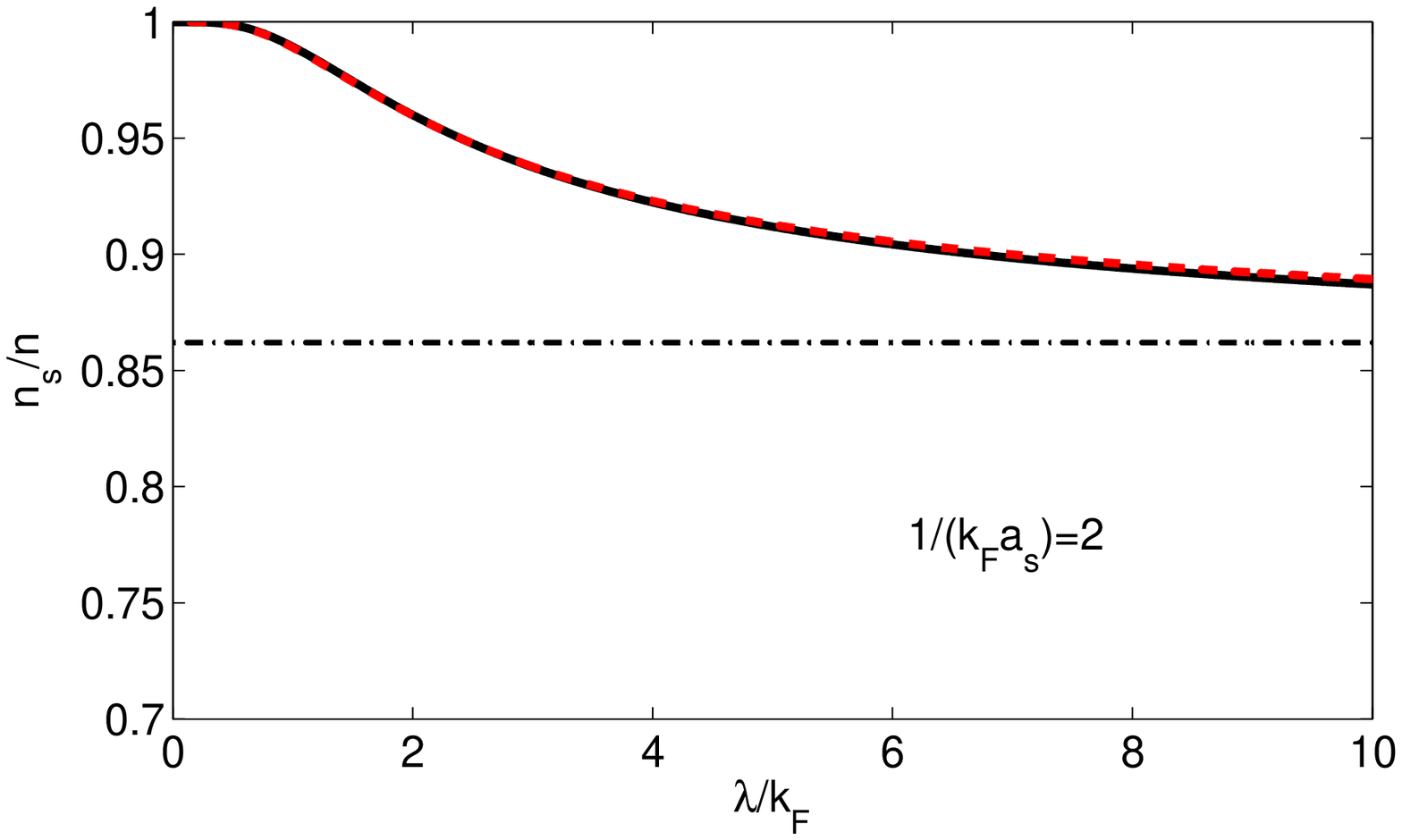}
\caption{The superfluid density $n_s$ (divided by $n$) as a function
of $\lambda/k_{\rm F}$ for various values of $1/(k_{\rm F}a_s)$. The
red dashed line shows the analytical result $2m/m_{\text B}$. The
dot-dashed line corresponds to the value $0.862$ for $\lambda
a_s\rightarrow \infty$.
 \label{fig7}}
\end{center}
\end{figure}

To further understand this result, we consider the effective action
for the phase field $\theta(x)$. To this end, we write the order
parameter as $\Phi(x)=\Delta(x)e^{i\theta(x)}$. In the static limit,
we can obtain the effective Hamiltonian for the phase field,
$H_{\text{eff}}=(J_s/2)\int d^3{\bf r}[\nabla\theta(\bf r)]^2$,
where the superfluid phase stiffness $J_s$ is related to the
superfluid density $n_s$ by $J_s=n_s/(4m)$. Therefore, at large SOC,
we have
\begin{eqnarray}
J_s\simeq\frac{2m}{m_{\text B}}\frac{n}{4m}=\frac{n_{\text
B}}{m_{\text B}},
\end{eqnarray}
where $n_{\text B}=n/2$ is the density of bosons (rashbons). This
means that, at large SOC, the superfluid phase stiffness
self-consistently recovers that for a rashbon gas with a non-trivial
effective mass $m_{\rm B}$. We emphasize that this interesting
result was first observed by us in 2D Fermi gases with Rashba
spin-orbit coupling \cite{2Dhe}.

This result also indicates that the Galilean invariance, which is
explicitly broken in the original fermion Hamiltonian, can be viewed
as a low-energy emergent symmetry at large SOC. This is due to the
fact that at large SOC the system becomes a weakly interacting
Bose-Einstein condensate of non-relativistic rashbons which have a
non-trivial effective mass $m_{\rm B}$. We will show this conclusion
explicitly in the next section by deriving the Gross-Pitaevskii free
energy for the dilute rashbon condensate at large SOC.

{\bf (C) Numerical Results.} The superfluid density at zero
temperature can be expressed in terms of the dimensionless
parameters as
\begin{eqnarray}
\frac{n_s}{n}=1-\frac{g_1}{2}\sum_{\alpha=\pm}\alpha\int_0^\infty
zdz\frac{z^2+2\alpha
g_1z-x_1+\frac{x_2^2}{z^2-x_1}}{\sqrt{(z^2+2\alpha
g_1z-x_1)^2+x_2^2}}.
\end{eqnarray}
It can be numerically obtained using the solutions of $x_1$ and
$x_2$ from the gap and number equations.

The numerical results for $n_s/n$ as a function of $\lambda/k_{\rm
F}$ for different values of $1/(k_{\rm F}a_s)$ are shown in Fig.
\ref{fig7}. For negative values or small positive values of
$1/(k_{\rm F}a_s)$, the numerical result becomes in good agreement
with the analytical result $n_s/n\simeq 2m/m_{\text B}$ when
$\lambda/k_{\rm F}>1$, which is consistent with the observation that
the system enters the rashbon BEC regime at $\lambda/k_{\rm F}\sim
1$. For large positive values of $1/(k_{\rm F}a_s)$ (in fact even
for $1/(k_{\rm F}a_s)=1$), the numerical results are always in good
agreement with the analytical result for all values of
$\lambda/k_{\rm F}$.

For both negative and positive values of $1/(k_{\rm F}a_s)$, we find
that $n/n_s$ approaches a universal value $0.862$ when
$\lambda/k_{\rm F}\rightarrow\infty$, as indicated from the
analytical observation.

\subsection{Spin susceptibility}

Since the superfluid ground state exhibits spin-triplet pairing, the
spin susceptibility $\chi$ can be nonzero even at zero temperature
\cite{spin}, in contrast to the case of vanishing SOC. The spin
susceptibility is defined as the response of the system to an
infinitesimal ``magnetic field" ${\bf H}$, which induces an
additional term $\mbox{\boldmath{$\sigma$}}\cdot{\bf H}$ in the
Hamiltonian. Since the ground state is rotationally symmetric, the
spin susceptibility is also isotropic. It can be evaluated by the
definition
\begin{eqnarray}
\Omega({\bf H})=\Omega({\bf 0})-\frac{1}{2}\chi{\bf H}^2+\cdots.
\end{eqnarray}
Using the derivative expansion, the spin susceptibility can be
evaluated as
\begin{eqnarray}
\chi&=&-\frac{1}{\beta}\sum_n\sum_{\bf k}\left({\cal A}_{11}^2+{\cal
A}_{22}^2+2{\cal A}_{21}^2\right)\nonumber\\
&+&\frac{1}{3}\frac{1}{\beta}\sum_n\sum_{\bf k}\left({\cal
B}_{11}^2+{\cal B}_{22}^2+2{\cal B}_{21}^2\right).
\end{eqnarray}

\begin{figure}[!htb]
\begin{center}
\includegraphics[width=8.2cm]{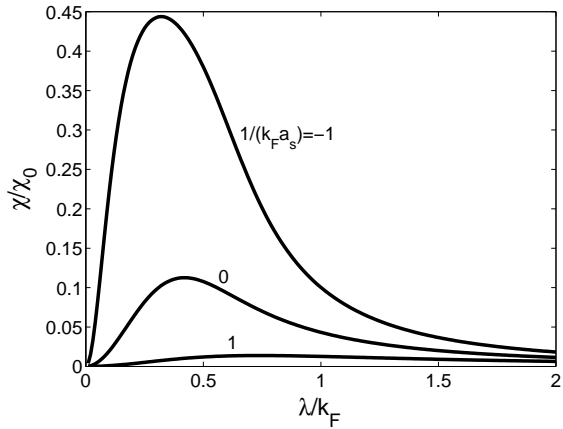}
\caption{The spin susceptibility $\chi$ (divided by $\chi_0$) as a
function of $\lambda/k_{\rm F}$ for various values of $1/(k_{\rm
F}a_s)$.
 \label{figchi}}
\end{center}
\end{figure}

At zero temperature, the spin susceptibility reads
\begin{eqnarray}
\chi=\frac{1}{6\pi^2\lambda}\int_0^\infty
kdk\left[\left(\xi_k^++\frac{\Delta_0^2}{\xi_k}\right)\frac{1}{E_k^+}-\left(\xi_k^-+\frac{\Delta_0^2}{\xi_k}\right)\frac{1}{E_k^-}\right].
\end{eqnarray}
This result shows explicitly that $\chi\neq0$ for nonzero SOC. An
interesting relation is that $\chi$ is proportional to the normal
fluid density $n_\lambda=n-n_s$. We have
\begin{eqnarray}
\chi=\frac{n-n_s}{\lambda^2}.
\end{eqnarray}
Using the result $\chi_0=3n/(2\epsilon_{\rm F})$ for non-interacting
Fermi gases in the absence of SOC, we obtain
\begin{eqnarray}
\frac{\chi}{\chi_0}=\frac{1}{3}\left(\frac{\lambda}{k_{\text
F}}\right)^{-2}\left(1-\frac{n_s}{n}\right).
\end{eqnarray}
Therefore, at large SOC, the spin susceptibility behaves as
$\chi\sim(\lambda/k_{\rm F})^{-2}$. The numerical results are shown
in Fig. \ref{figchi}. In general, increasing the attractive strength
suppresses the magnitude of $\chi$.

\section{Bose-Einstein Condensation of Weakly Interacting Rashbons}
As we have shown in the last section, the superfluid state in the
large SOC limit is a Bose-Einstein condensation of rashbons. We are
interested in the interactions among the rashbons. In this section,
we will derive the Gross-Pitaevskii free energy for a dilute rashbon
condensate, which allow us to extract the rashbon-rashbon scattering
length. Another goal of this section is to show that the Galilean
invariance, which is explicitly broken in the original fermion
Hamiltonian, can be effectively recovered at the boson (rashbon)
level at large SOC.

To this end, we consider the mean field theory where the auxiliary
boson field $\Phi(x)$ is replaced by its expectation value
$\langle\Phi(x)\rangle=\Delta(x)$. In the large SOC limit
$\lambda\rightarrow\infty$, the fermion chemical potential $\mu$
approaches $-E_{\text B}/2$. Since the pairing gap
$|\Delta|\ll|\mu|$, we can expand the effective action in powers of
$|\Delta|$ (as well as in powers of its space-time derivatives),
which results in a Ginzburg-Landau free energy functional
\begin{eqnarray}
&&V_{\text{GL}}[\Delta(x)]=\int dx\Bigg[\Delta^\dagger(x)\left(
a\frac{\partial}{\partial\tau}-b\mbox{\boldmath{$\nabla$}}^2\right)\Delta(x)\nonumber\\
&&\ \ \ \ \ \ \ \ \ \ \ \ \ \ \ \ \ +\ \
c|\Delta(x)|^2+\frac{1}{2}d|\Delta(x)|^4\Bigg].
\end{eqnarray}

\subsection{Calculation of the Ginzburg-Landau coefficients}

The coefficients $c$ and $d$ of the potential terms can be obtained
from the mean field thermodynamic potential $\Omega_0=(T/V){\cal
S}_{\text{eff}}[\Delta^\dagger,\Delta]$ which can be evaluated as
\begin{eqnarray}
\Omega=-\frac{|\Delta|^2}{4\pi a_s}-\sum_{\bf k}\left(\frac{E_{\bf
k}^++E_{\bf k}^-}{2}-\xi_{\bf k}-\frac{|\Delta|^2}{2\epsilon_{\bf
k}}\right).
\end{eqnarray}
We have
\begin{eqnarray}
c=\frac{\partial\Omega}{\partial|\Delta|^2}\bigg|_{\Delta=0},\ \ \ \
d=\frac{\partial^2\Omega}{\partial(|\Delta|^2)^2}\bigg|_{\Delta=0}.
\end{eqnarray}
After a simple algebra, the coefficients $\alpha$ and $\beta$ can be
evaluated as
\begin{eqnarray}
&&c=\frac{1}{4\pi}\left(\frac{-2\mu-2\lambda^2}{\sqrt{-2\mu-\lambda^2}}-\frac{1}{a_s}\right),\nonumber\\
&&d=\frac{1}{16\pi}\frac{-2\mu+2\lambda^2}{(-2\mu-\lambda^2)^{5/2}}.
\end{eqnarray}
From the expression of $c$, we find that a quantum phase transition
from vacuum to Bose condensation takes place at $\mu=-E_{\rm B}/2$.
Thus near the phase transition, $c$ can be simplified as
\begin{eqnarray}
c\simeq-\frac{1}{8\pi}\frac{E_{\text B}}{\left(E_{\rm
B}-\lambda^2\right)^{3/2}}\mu_{\text B},
\end{eqnarray}
where $\mu_{\text B}=2\mu+E_{\rm B}\ll E_{\text B}$ is the boson
chemical potential. Further, setting $\mu= -E_{\rm B}/2$, $d$ can be
reduced to
\begin{eqnarray}
d\simeq\frac{1}{16\pi}\frac{E_{\rm B}+2\lambda^2}{(E_{\rm
B}-\lambda^2)^{5/2}}.
\end{eqnarray}

The coefficients $a$ and $b$ of the kinetic terms can be obtained
from the inverse boson propagator ${\cal D}^{-1}(Q)$ with
$\Delta=0$. It can be evaluated as
\begin{eqnarray}
{\cal
D}^{-1}(Q)=\frac{1}{U}-\frac{1}{4}\sum_{\alpha,\gamma=\pm}\sum_{{\bf
k}}\frac{1+\alpha\gamma {\cal T}_{\bf{kq}}}{\xi_{{\bf k}+{\bf
q}/2}^\alpha+\xi_{{\bf k}-{\bf q}/2}^\gamma-i\nu_n}
\end{eqnarray}
In the large SOC limit, the coefficients $a$ and $b$ can be obtained
by the small momentum expansion for ${\cal D}^{-1}(Q)$. We have
\begin{eqnarray}
{\cal D}^{-1}(Q)\simeq-a\left(i\nu_n+\mu_{\text B}-\frac{{\bf
q}^2}{2m_{\rm B}}\right),
\end{eqnarray}
where $m_{\text B}$ is the rashbon effective mass determined by
(28), and $a$ is given by
\begin{eqnarray}
a=\frac{1}{8\pi}\frac{E_{\text B}}{\left(E_{\rm
B}-\lambda^2\right)^{3/2}}.
\end{eqnarray}
We observe the relation $c={\cal D}^{-1}(0)=-a\mu_{\text B}$.

\subsection{Gross-Pitaevskii free energy}

According to the above results for the Ginzburg-Landau coefficients,
if we define the new condensate wave function $\psiup(x)$ by
\begin{eqnarray}
\psiup(x)=\sqrt{a}\Delta(x),
\end{eqnarray}
the Ginzgurg-Landau free energy can be reduced to the
Gross-Pitaevskii free energy of a dilute Bose gas,
\begin{eqnarray}
&&V_{\text{GP}}[\psiup(x)]=\int dx\Bigg[\psiup^\dagger(x)\left(
\frac{\partial}{\partial\tau}-\frac{\mbox{\boldmath{$\nabla$}}^2}{2m_{\text B}}\right)\psiup(x)\nonumber\\
&&\ \ \ \ \ \ \ \ \ \ \ \ \ \ \ \ \ -\ \ \mu_{\text
B}|\psiup(x)|^2+\frac{1}{2}\frac{4\pi a_{\text{BB}}}{m_{\text
B}}|\psiup(x)|^4\Bigg],
\end{eqnarray}
where $a_{\text{BB}}$ is the boson-boson scattering length. Its
explicit expression is
\begin{eqnarray}
a_{\text{BB}}=m_{\text B}\frac{E_{\rm B}+2\lambda^2}{E_{\rm
B}^2}\sqrt{E_{\rm B}-\lambda^2}.
\end{eqnarray}
Note that $m=1$ in our units. For $\lambda=0$ and $a_s>0$, using the
result $m_{\text B}=2$ and $E_{\rm B}=1/a_s^2$, we recover the
well-known result $a_{\text{BB}}=2a_s$ \cite{BCSBEC3}. One remark
here is that this result is the mean field result which is not
exact. In the absence of SOC, exact four-body calculation shows that
$a_{\text{BB}}\simeq 0.6a_s$ \cite{fourbody}. Therefore, it is
interesting to explore the exact rashbon-rashbon scattering length
in the future studies. Another theoretical framework to obtain more
exact $a_{\text{BB}}$ is to include the Gaussian fluctuations
\cite{FL}.

The Gross-Pitaevskii free energy explicitly shows that the Galilean
invariance, which is explicitly broken in the original fermion
Hamiltonian, can be effectively viewed as a low-energy emergent
symmetry at large SOC.

\begin{figure}[!htb]
\begin{center}
\includegraphics[width=8cm]{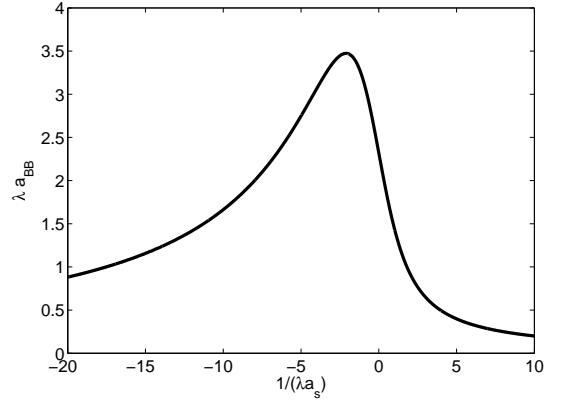}
\caption{The molecule scattering length $a_{\text{BB}}$ (divided by
$1/\lambda$) as a function of the dimensionless parameter
$\kappa=1/(\lambda a_s)$.
 \label{fig8}}
\end{center}
\end{figure}

\subsection{Rashbon-rashbon scattering length}

Using the expressions for the binding energy $E_{\rm B}$ and the
effective mass $m_{\text B}$, we obtain
\begin{eqnarray}
a_{\text{BB}}=\frac{1}{\lambda}\frac{2({\cal J}+2)\sqrt{{\cal
J}-1}}{{\cal J}^2\left[\frac{7}{3}-\frac{4}{3}\left(\frac{{\cal
J}-1}{{\cal J}}\right)^{3/2}-\frac{2}{{\cal J}}\right]}.
\end{eqnarray}
We find that the quantity $\lambda a_{\text{BB}}$ depends only on
the dimensionless parameter $\kappa=1/(\lambda a_s)$. For the case
$\lambda a_s\rightarrow\infty$ or $\kappa=0$, we have ${\cal J}=2$
and, therefore,
\begin{eqnarray}
a_{\text{BB}}(\lambda a_s\rightarrow
\infty)=\frac{1}{\lambda}\frac{3(4+\sqrt{2})}{7}=\frac{2.32}{\lambda}.
\end{eqnarray}
The numerical result for the scattering length $a_{\text{BB}}$ is
shown in Fig. \ref{fig8}. We find that the quantity $\lambda
a_{\text{BB}}$ has a maximum near the point $\kappa=0$, at
$\kappa=-2.11$.

\subsection{Rashbon chemical potential}

For a uniform system, the expectation value of the condensate
$\psiup(x)$ should be determined by minimizing the Gross-Pitaevskii
free energy. We find that the minimum is given by
\begin{eqnarray}
|\psiup_0|^2=\frac{\mu_{\rm B}}{g_0},
\end{eqnarray}
where $g_0=4\pi a_{\text{BB}}/m_{\text B}$. The total density of the
bosons is $n_{\text B}=n/2=|\psiup_0|^2=a\Delta_0^2$. Therefore, the
boson chemical potential can be given by
\begin{eqnarray}
\mu_{\text B}=\frac{2\pi a_{\text{BB}}}{m_{\text B}}n.
\end{eqnarray}
For the case $\lambda a_s\rightarrow\infty$, using the result for
$m_{\text B}$ and $a_{\text{BB}}$, we obtain
\begin{eqnarray}
\mu_{\text B}(\lambda a_s\rightarrow\infty)=\frac{2\pi n}{\lambda}.
\end{eqnarray}

\section{Gaussian Fluctuation and Collective Excitations}

To study the collective excitations, we consider the fluctuations
around the mean field. Making the field shift
$\Phi(x)\rightarrow\Delta_0+\phi(x)$, we can expand the effective
action ${\cal S}_{\text{eff}}$ in powers of the fluctuations. The
zeroth order term ${\cal S}_{\text{eff}}^{(0)}$ is just the mean
field result, and the linear terms vanish automatically guaranteed
by the saddle point condition for $\Delta_0$. The quadratic terms,
corresponding to Gaussian fluctuations, can be evaluated as
\begin{eqnarray}
{\cal
S}_{\text{eff}}^{(2)}[\phi,\phi^\dag]=\frac{1}{2}\sum_Q\left(\begin{array}{cc}
\phi^\dagger(Q) & \phi(-Q)\end{array}\right){\bf M}(Q)\left(\begin{array}{cc} \phi(Q)\\
\phi^\dagger(-Q)
\end{array}\right),
\end{eqnarray}
where the inverse boson propagator ${\bf M}$ takes the form
\begin{eqnarray}
{\bf M}(Q)=\left(\begin{array}{cc}{\bf M}_{11}(Q)&{\bf M}_{12}(Q)\\
{\bf M}_{21}(Q)& {\bf M}_{22}(Q)\end{array}\right)
\end{eqnarray}
with the relations ${\bf M}_{11}(Q)={\bf M}_{22}(-Q)$ and ${\bf
M}_{12}(Q)={\bf M}_{21}(Q)$. The matrix elements of ${\bf M}(Q)$ can
be expressed in terms of the fermion propagator ${\cal G}(K)$. We
have
\begin{eqnarray}
&&{\bf M}_{11}(Q)=\frac{1}{U}+\frac{1}{2}\sum_K\text{Tr}\left[{\cal
G}_{11}(K+Q)\sigma_y{\cal G}_{22}(K)\sigma_y\right],\nonumber\\
&&{\bf M}_{12}(Q)=-\frac{1}{2}\sum_K\text{Tr}\left[{\cal
G}_{12}(K+Q)\sigma_y{\cal G}_{12}(K)\sigma_y\right].
\end{eqnarray}
At zero temperature, the explicit form of ${\bf M}(Q)$ can be
evaluated as
\begin{widetext}
\begin{eqnarray}
{\bf M}_{11}(Q)
=\frac{1}{U}+\frac{1}{4}\sum_{\alpha,\gamma=\pm}\sum_{\bf
k}\left[\frac{\left(u_{{\bf k}+{\bf
q}/2}^\alpha\right)^2\left(u_{{\bf k}-{\bf
q}/2}^\gamma\right)^2}{i\nu_n-E_{{\bf k}+{\bf q}/2}^\alpha-E_{{\bf
k}-{\bf q}/2}^\gamma}-\frac{\left(v_{{\bf k}+{\bf
q}/2}^\alpha\right)^2\left(v_{{\bf k}-{\bf
q}/2}^\gamma\right)^2}{i\nu_n+E_{{\bf k}+{\bf q}/2}^\alpha+E_{{\bf
k}-{\bf q}/2}^\gamma} \right]\left(1+\alpha\gamma{\cal T}_{\bf
kq}\right)
\end{eqnarray}
and
\begin{eqnarray}
{\bf M}_{12}(Q) =\frac{1}{4}\sum_{\alpha,\gamma=\pm}\sum_{\bf
k}\left[\frac{u_{{\bf k}+{\bf q}/2}^\alpha v_{{\bf k}+{\bf
q}/2}^\alpha u_{{\bf k}-{\bf q}/2}^\gamma v_{{\bf k}-{\bf
q}/2}^\gamma}{i\nu_n+E_{{\bf k}+{\bf q}/2}^\alpha+E_{{\bf k}-{\bf
q}/2}^\gamma}-\frac{u_{{\bf k}+{\bf q}/2}^\alpha v_{{\bf k}+{\bf
q}/2}^\alpha u_{{\bf k}-{\bf q}/2}^\gamma v_{{\bf k}-{\bf
q}/2}^\gamma}{i\nu_n-E_{{\bf k}+{\bf q}/2}^\alpha-E_{{\bf k}-{\bf
q}/2}^\gamma}\right]\left(1+\alpha\gamma{\cal T}_{\bf kq}\right).
\end{eqnarray}
\end{widetext}
Here the BCS distribution functions are defined as $(v_{\bf
k}^\alpha)^2=(1-\xi_{\bf k}^\alpha/E_{\bf k}^\alpha)/2$ and $(u_{\bf
k}^\alpha)^2=(1+\xi_{\bf k}^\alpha/E_{\bf k}^\alpha)/2$. In the
absence of SOC, $\lambda=0$, the expressions for ${\bf M}_{11}(Q)$
and ${\bf M}_{12}(Q)$ recover the results obtained in
Ref. \cite{BCSBEC3}.

\subsection{Bogoliubov excitation in the rashbon condensate}
At large SOC and/or attraction, the superfluid state is a
Bose-Einstein condensation of weakly interacting Bose gas. Thus, we
expect that the low-energy collective excitation in this case
recover the well-known Bogoliubov excitation spectrum in a weakly
interacting Bose condensate \cite{Naoto}. In this part, we will give
an explicit proof for this.

In the large SOC and/or strong-coupling limit, the chemical
potential reads $\mu\simeq -E_{\rm B}/2$ and we have
$\Delta_0\ll|\mu|$. In this case, we can expand the matrix elements
of ${\bf M}$ in powers of $\Delta_0/|\mu|$ and keep only the
leading-order terms. Following this spirit, we obtain
\begin{eqnarray}
&&{\bf M}_{11}(Q)\simeq{\cal D}^{-1}(Q)+X\Delta_0^2,\nonumber\\
&&{\bf M}_{12}(Q)\simeq Y\Delta_0^2,
\end{eqnarray}
where the coefficients $X$ and $Y$ are given by
\begin{eqnarray}
X=2Y=\frac{1}{4}\sum_{\bf k}\left[\frac{1}{(\xi_{\bf
k}^+)^3}+\frac{1}{(\xi_{\bf k}^-)^3}\right]=2d.
\end{eqnarray}
Further, taking the small momentum expansion for ${\cal D}^{-1}(Q)$,
we obtain
\begin{eqnarray}
{\bf M}_{11}(Q)\simeq-a\left(i\nu_n+\mu_{\text B}-\frac{{\bf
q}^2}{2m_{\rm B}}\right)+2d\Delta_0^2.
\end{eqnarray}
Therefore, in the large SOC and/or strong coupling limit, the boson
propagator ${\bf M}(Q)$ can be well approximated by
\begin{eqnarray}
&&{\bf M}_{11}(Q)\simeq a\left(-i\nu_n+\frac{{\bf q}^2}{2m_{\text B}}-\mu_{\text B}+2g_0|\psiup_0|^2\right)\nonumber\\
&&{\bf M}_{22}(Q)\simeq a\left(i\nu_n+\frac{{\bf q}^2}{2m_{\text B}}-\mu_{\text B}+2g_0|\psiup_0|^2\right)\nonumber\\
&&{\bf M}_{12}(Q)={\bf M}_{21}(Q)\simeq ag_0|\psiup_0|^2,
\end{eqnarray}
where $g_0=4\pi a_{\text{BB}}/m_{\text B}$ and
$|\psiup_0|^2=\mu_{\text B}/g_0$ is the minimum of the
Gross-Pitaevskii free energy (corresponding to the saddle point
$\Delta_0$ of the effective potential). From the Gross-Pitaevskii
free energy, the boson density reads $n_{\text B}=n/2=|\psiup_0|^2$.
Utilizing these results, we obtain
\begin{eqnarray}
{\bf M}(Q)\simeq a\left(\begin{array}{cc} -i\nu_n+\frac{{\bf q}^2}{2m_{\text B}}+g_0n_{\rm B}&g_0n_{\rm B}\\
g_0n_{\rm B}&i\nu_n+\frac{{\bf q}^2}{2m_{\text B}}+g_0n_{\rm
B}\end{array}\right).
\end{eqnarray}

By taking the analytical continuation $i\nu_n\rightarrow\omega+i0^+$,
the dispersion $\omega=\omega({\bf q})$ of the collective mode is
obtained by solving the equation
\begin{equation}
\det{{\bf M}[{\bf q},\omega({\bf q})]}=0.
\end{equation}
Therefore, the Goldstone mode takes a dispersion relation given by
\begin{eqnarray}
\omega({\bf q})=\sqrt{\frac{{\bf q}^2}{2m_{\text B}}\left(\frac{{\bf
q}^2}{2m_{\text B}}+\frac{8\pi a_{\text{BB}}n_{\text B}}{m_{\text
B}}\right)}.
\end{eqnarray}
This is just the Bogoliubov excitation spectrum in a dilute Bose
condensate where the bosons possess a mass $m_{\text{B}}$ and a
two-body scattering length $a_{\text{BB}}$.

\subsection{Collective modes in the BCS-BEC crossover}
The dispersions of the collective modes are, in principle, determined
by the equation $\det{{\bf M}[{\bf q},\omega({\bf q})]}=0$. To make
the result more physical, we decompose the complex fluctuation field
$\phi(x)$ into its amplitude mode $\lambda(x)$ and phase mode
$\theta(x)$, $\phi(x)=\lambda(x)+i\theta(x)$. Then, the fluctuation
part of the effective action takes the form
\begin{equation}
{\cal S}_{\text{eff}}^{(2)}=\frac{1}{2}\sum_Q\left(\begin{array}{cc} \lambda^*(Q)&\theta^*(Q)\end{array}\right){\bf N}(Q)\left(\begin{array}{c} \lambda(Q)\\
\theta(Q)\end{array}\right),
\end{equation}
where the matrix ${\bf N}(Q)$ is defined as
\begin{equation}
{\bf N}(Q)=2\left(\begin{array}{cc} {\bf M}_{11}^++{\bf M}_{12}&i{\bf M}_{11}^-\\
-i{\bf M}_{11}^- & {\bf M}_{11}^+-{\bf M}_{12}\end{array}\right).
\end{equation}
Here the quantities ${\bf M}_{11}^\pm$ are defined
as
\begin{equation}
{\bf M}_{11}^\pm({\bf q},\omega)=\frac{1}{2}\left[{\bf M}_{11}({\bf
q},\omega)\pm {\bf M}_{11}({\bf q},-\omega)\right].
\end{equation}
We notice that ${\bf M}_{11}^+$ and ${\bf M}_{11}^-$ are even and
odd functions of $\omega$, respectively.

From the explicit form of ${\bf M}_{11}(Q)$, we have ${\bf
M}_{11}^-({\bf q},0)=0$. Therefore, the amplitude and phase modes
decouple completely at $\omega=0$. Furthermore, using the saddle-point 
condition for the order parameter $\Delta_0$, we find ${\bf
M}_{11}^+({\bf 0},0)={\bf M}_{12}({\bf 0},0)$, which ensures that
the phase mode at ${\bf q}=0$ is gapless, i.e., the Goldstone mode.

We now determine the velocity $c_s$ of the Goldstone mode,
$\omega({\bf q})=c_s|{\bf q}|$ for $\omega,|{\bf q}|\ll
\text{min}_{\bf k}\{E_{\bf k}^\pm\}$. For this purpose, we make a
small ${\bf q}$ and $\omega$ expansion of ${\bf N}(Q)$,
\begin{eqnarray}
{\bf M}_{11}^++{\bf M}_{12} &=& A+C|{\bf q}|^2-D\omega^2+\cdots,\nonumber\\
{\bf M}_{11}^+-{\bf M}_{12} &=& Q|{\bf q}|^2-R\omega^2+\cdots,\nonumber\\
{\bf M}_{11}^- &=& -B\omega+\cdots.
\end{eqnarray}
Here we note that the coefficient $Q$ is proportional to the
superfluid density $n_s$ and the superfluid phase stiffness $J_s$.
The explicit form of $A,\ B,\ D,\ R$ and $Q$ can be calculated as
\begin{eqnarray}
A &=& \frac{1}{4}\sum_{\alpha=\pm}\sum_{\bf k}\frac{\Delta_0^2}{(E_{\bf k}^\alpha)^3} ,\nonumber\\
B &=& \frac{1}{8}\sum_{\alpha=\pm}\sum_{\bf k}\frac{\xi_{\bf k}^\alpha}{(E_{\bf k}^\alpha)^3},\nonumber\\
D &=& \frac{1}{16}\sum_{\alpha=\pm}\sum_{\bf k}\left[\frac{1}{(E_{\bf k}^\alpha)^3}-\frac{\Delta_0^2}{(E_{\bf k}^\alpha)^5}\right] ,\nonumber\\
R &=& \frac{1}{16}\sum_{\alpha=\pm}\sum_{\bf k}\frac{1}{(E_{\bf k}^\alpha)^3},\nonumber\\
Q &=& \frac{J_s}{2\Delta_0^2}=\frac{n_s}{8m\Delta_0^2}.
\end{eqnarray}

The Goldstone mode velocity or the so-called sound velocity in the
superfluid state is given by
\begin{equation}
c_s=\sqrt{Q\over B^2/A+R}.
\end{equation}
The corresponding eigenvector of ${\bf N}$ is
$(\lambda,\theta)=(-ic|{\bf q}|B/A,1)$, which is a pure phase mode
at ${\bf q}=0$ but has an admixture of the amplitude mode controlled
by $B$ at finite ${\bf q}$. Another massive mode, or the so-called
Anderson-Higgs mode, has a mass gap
\begin{eqnarray}
M_g=\sqrt{\frac{B^2+AR}{DR}}.
\end{eqnarray}

{\bf (A) Analytical Results for Large SOC.} In the rashbon BEC limit
$\lambda/k_{\text F}\gg1$, we have $\mu\simeq E_{\text B}/2$ and
$\Delta_0\ll|\mu|$. Therefore, the coefficients $A,\ B,\ D,\ R$ and
$Q$ can be well approximated as
\begin{eqnarray}
A &\simeq& \frac{\Delta_0^2}{4}\sum_{\alpha=\pm}\sum_{\bf k}\frac{1}{(\xi_{\bf k}^\alpha)^3}=2\Delta_0^2d ,\nonumber\\
B &\simeq& \frac{1}{8}\sum_{\alpha=\pm}\sum_{\bf k}\frac{1}{(\xi_{\bf k}^\alpha)^2}=a,\nonumber\\
D &\simeq& \frac{1}{16}\sum_{\alpha=\pm}\sum_{\bf k}\frac{1}{(\xi_{\bf k}^\alpha)^3}=\frac{d}{2} ,\nonumber\\
R &\simeq& \frac{1}{16}\sum_{\alpha=\pm}\sum_{\bf k}\frac{1}{(\xi_{\bf k}^\alpha)^3}=\frac{d}{2},\nonumber\\
Q &\simeq& \frac{1}{2\Delta_0^2}\frac{n_{\rm B}}{m_{\text B}}.
\end{eqnarray}
In this case, we find that $B^2/A\gg R$ and therefore the amplitude
and phase modes are strongly coupled.

\begin{figure}[!htb]
\begin{center}
\includegraphics[width=8cm]{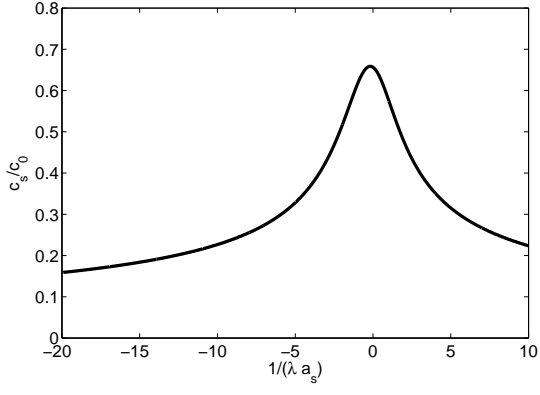}
\caption{The sound velocity $c_s$ of the Goldstone mode (divided by
$c_0$) in the RBEC regime as a function of the dimensionless
parameter $\kappa=1/(\lambda a_s)$.
 \label{fig9}}
\end{center}
\end{figure}

The sound velocity $c_s$ and the mass gap $M_g$ read
\begin{eqnarray}
&&c_s=\sqrt{AQ\over B^2}\simeq\sqrt{\frac{d}{a^2}\frac{n_{\rm B}}{m_{\rm B}}},\nonumber\\
&&M_g=\sqrt{\frac{B^2}{DR}}\simeq\frac{2a}{d}.
\end{eqnarray}
Using the relation $d/a^2=4\pi a_{\text{BB}}/m_{\text B}$, the sound
velocity recovers the result for a weakly interacting rashbon gas,
\begin{eqnarray}
c_s=\sqrt{\frac{4\pi a _{\text{BB}}n_{\rm B}}{m_{\rm
B}^2}}=\sqrt{\frac{\mu_{\rm B}}{m_{\rm B}}}.
\end{eqnarray}
Therefore, in the BEC limit, the quantity $c_s/c_0$ depends only on
the dimensionless parameter $\kappa=1/(\lambda a_s)$, where
$c_0=\sqrt{2\pi n/\lambda}$. Using the results for $m_{\rm B}$ and
$a_{\text{BB}}$, we obtain
\begin{eqnarray}
c_s=c_0\sqrt{\frac{({\cal J}+2)\sqrt{{\cal J}-1}}{2{\cal
J}^2}\left[\frac{7}{3}-\frac{4}{3}\left(\frac{{\cal J}-1}{{\cal
J}}\right)^{3/2}-\frac{2}{{\cal J}}\right]}.
\end{eqnarray}
The numerical result for the quantity $c/c_0$ is shown in Fig.
\ref{fig9}. We find that it has a maximum near the point $\kappa=0$,
at $\kappa=-0.18$. For the case $\lambda a_s\rightarrow\infty$, we
have
\begin{eqnarray}
c_s(\lambda
a_s\rightarrow\infty)=c_0\sqrt{\frac{7}{3(4+\sqrt{2})}}=0.66c_0.
\end{eqnarray}

\begin{figure}[!htb]
\begin{center}
\includegraphics[width=8cm]{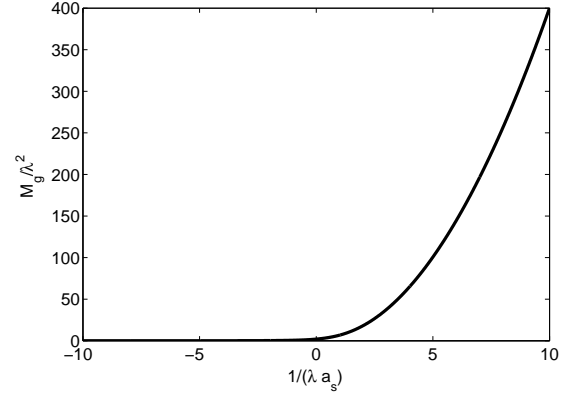}
\caption{The mass gap $M_g$ of the Anderson-Higgs mode (divided by
$\lambda^2$) in the RBEC regime as a function of the dimensionless
parameter $\kappa=1/(\lambda a_s)$.
 \label{fig10}}
\end{center}
\end{figure}

Using the expressions for $a$ and $d$, we obtain the explicit form
of the mass gap $M_g$,
\begin{eqnarray}
M_g=\frac{4E_{\text B}(E_{\text B}-\lambda^2)}{E_{\text
B}+2\lambda^2}=\lambda^2\frac{4{\cal J}({\cal J}-1)}{{\cal J}+2}.
\end{eqnarray}
Therefore, in the BEC limit, the quantity $M_g/\lambda^2$ depends
only on the dimensionless parameter $\kappa=1/(\lambda a_s)$. The
numerical result is shown in Fig. \ref{fig10}. We find that it is
very small in the limit $\kappa\rightarrow -\infty$, and increases
rapidly in the regime $\kappa>0$. For the case $\lambda
a_s\rightarrow\infty$, we have $E_{\text B}=2\lambda^2$ and
therefore
\begin{eqnarray}
M_g(\lambda a_s\rightarrow\infty)=2\lambda^2.
\end{eqnarray}

\begin{figure}[!htb]
\begin{center}
\includegraphics[width=7.5cm]{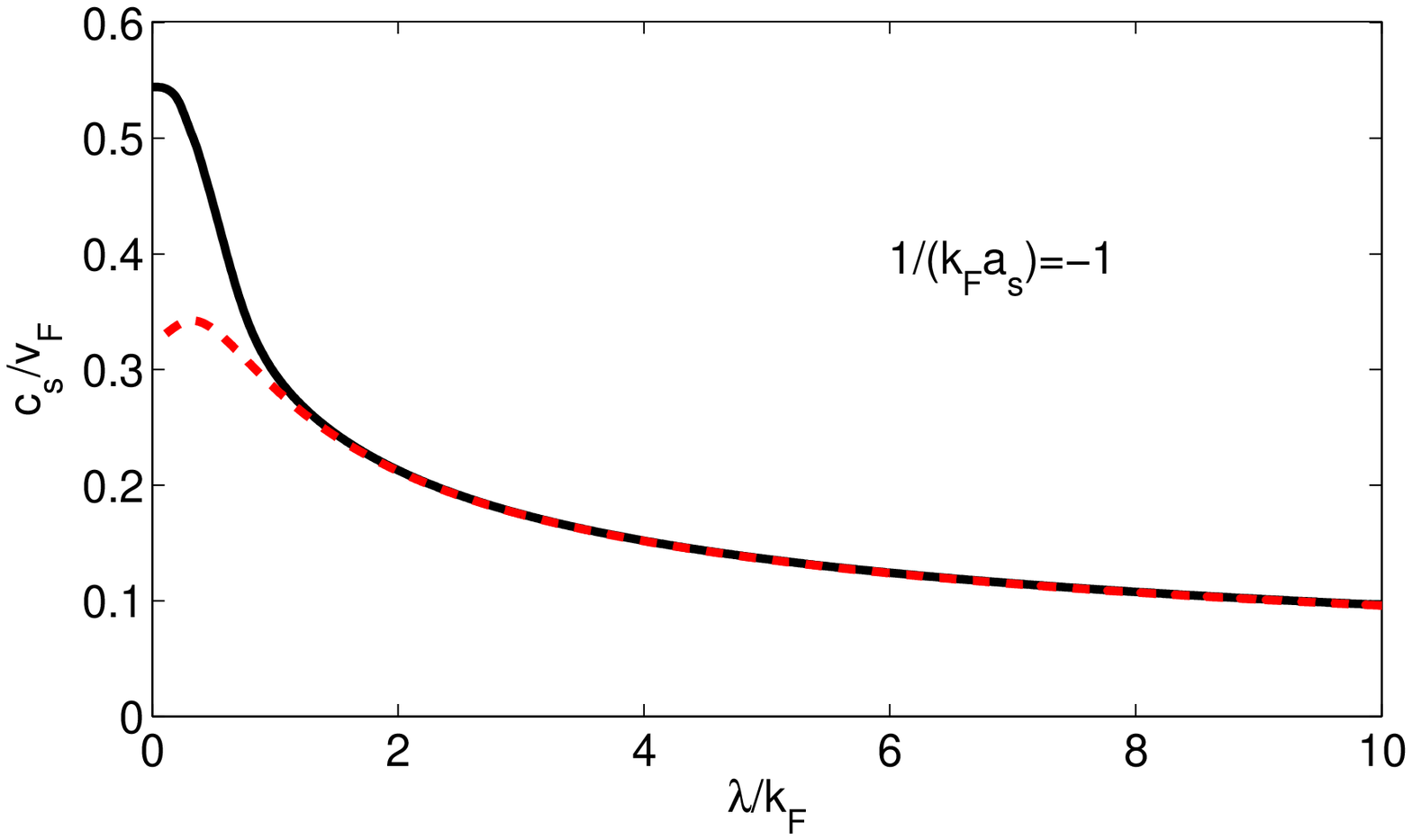}
\includegraphics[width=7.5cm]{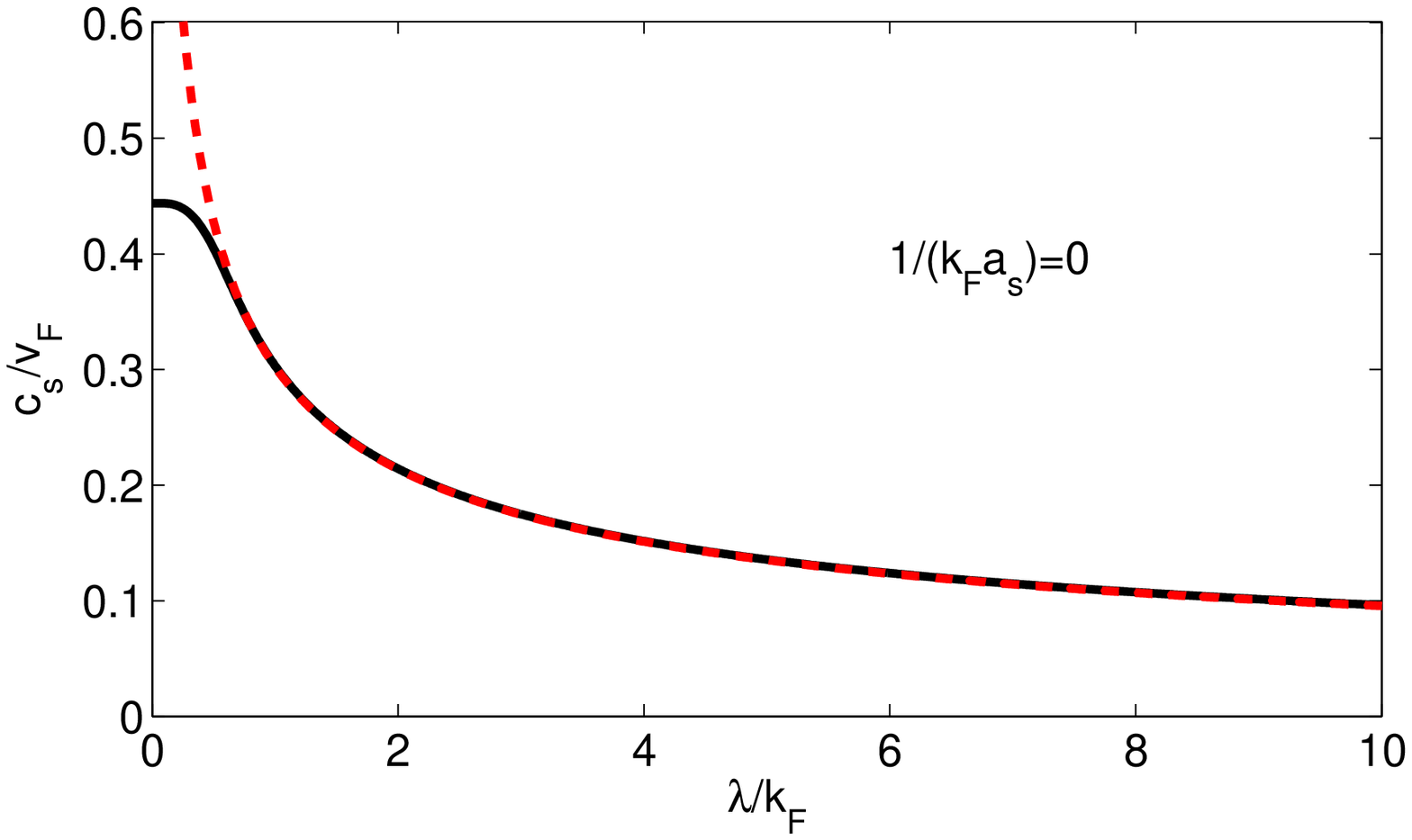}
\includegraphics[width=7.5cm]{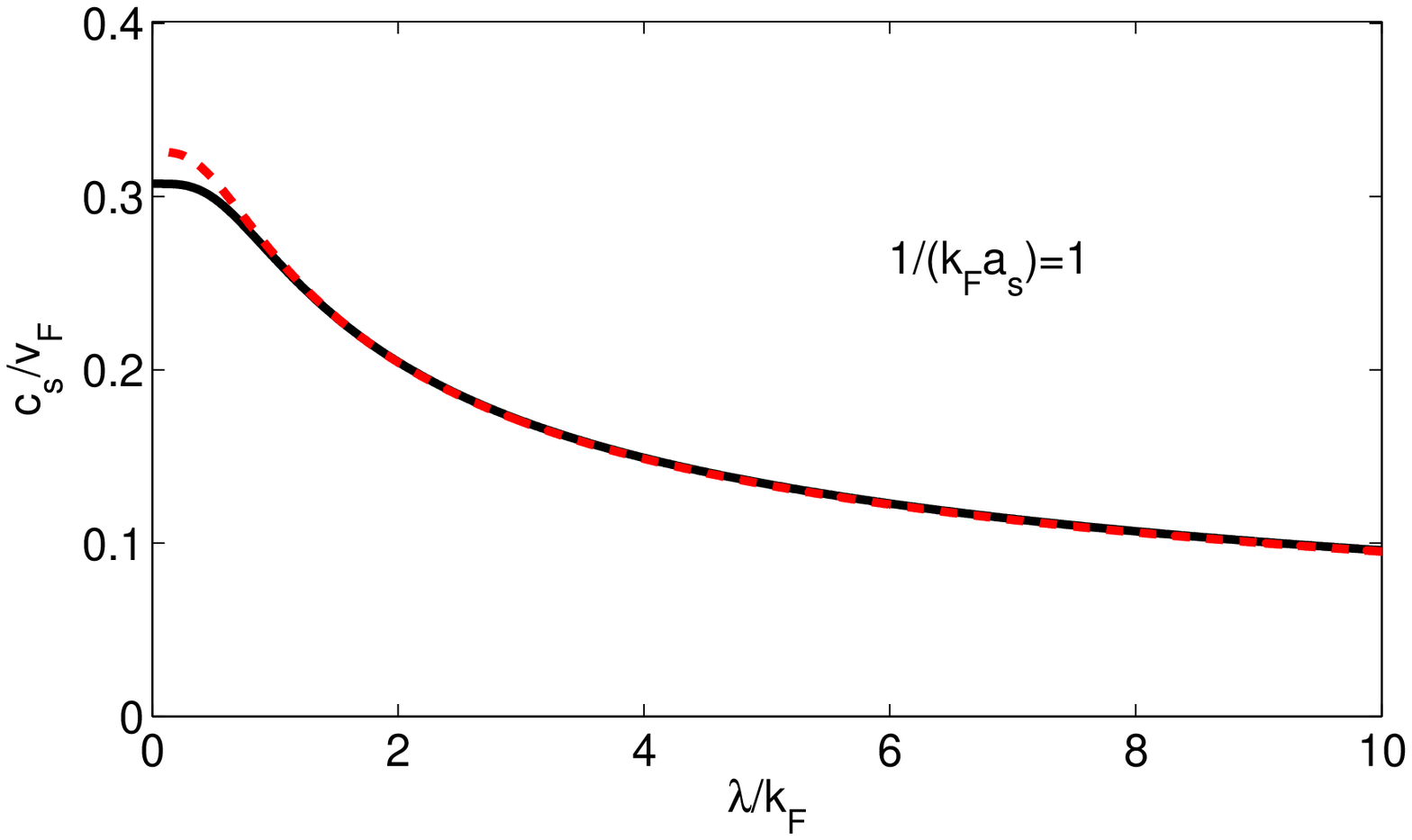}
\includegraphics[width=7.5cm]{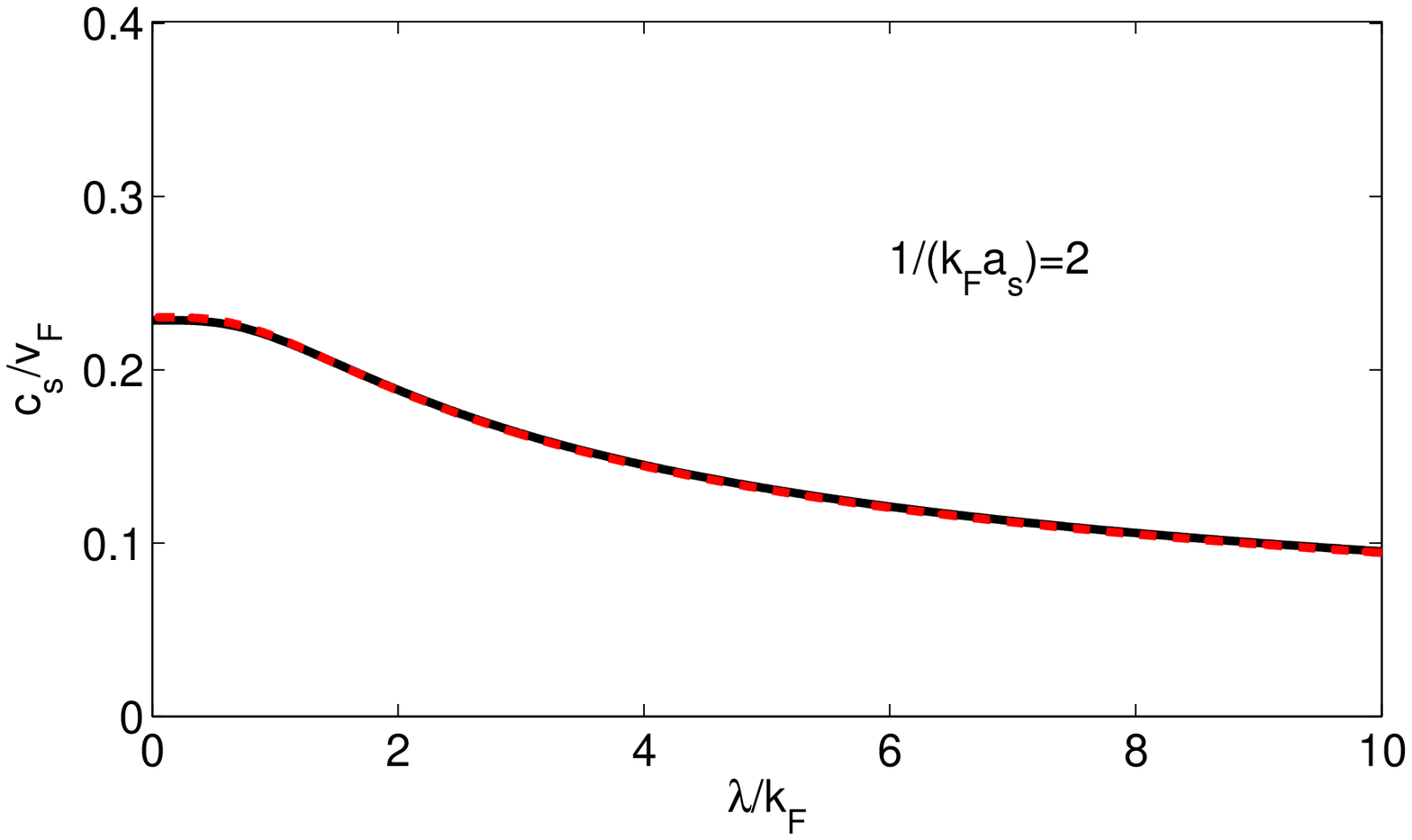}
\caption{The velocity of the Goldstone mode (sound velocity) $c_s$
(divided by $\upsilon_{\rm F}$) as a function of $\lambda/k_{\rm F}$
for various values of $1/k_{\rm F}a_s$. The red dashed lines
corresponds to the analytical result (134).
 \label{fig11}}
\end{center}
\end{figure}

\begin{figure}[!htb]
\begin{center}
\includegraphics[width=7.5cm]{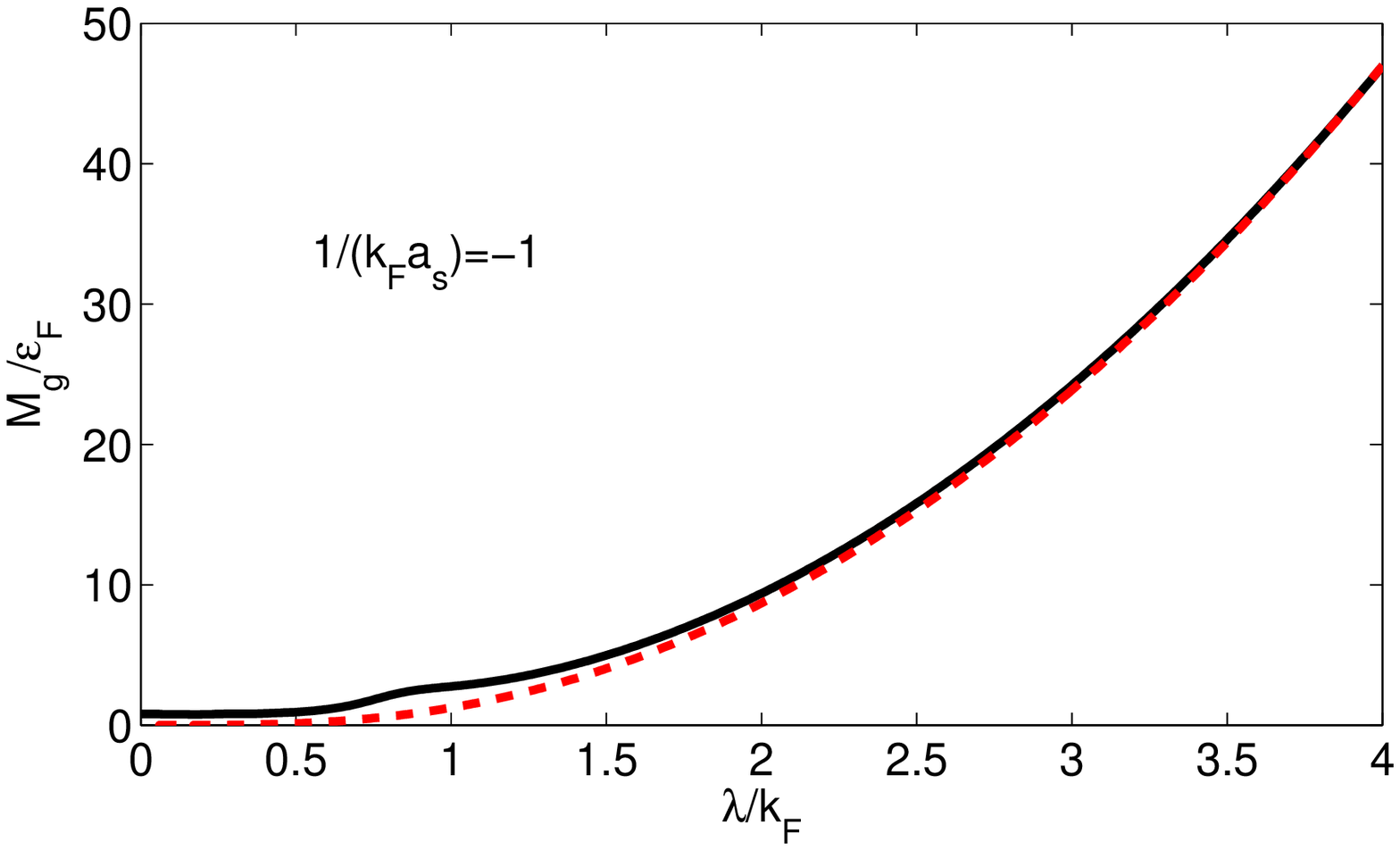}
\includegraphics[width=7.5cm]{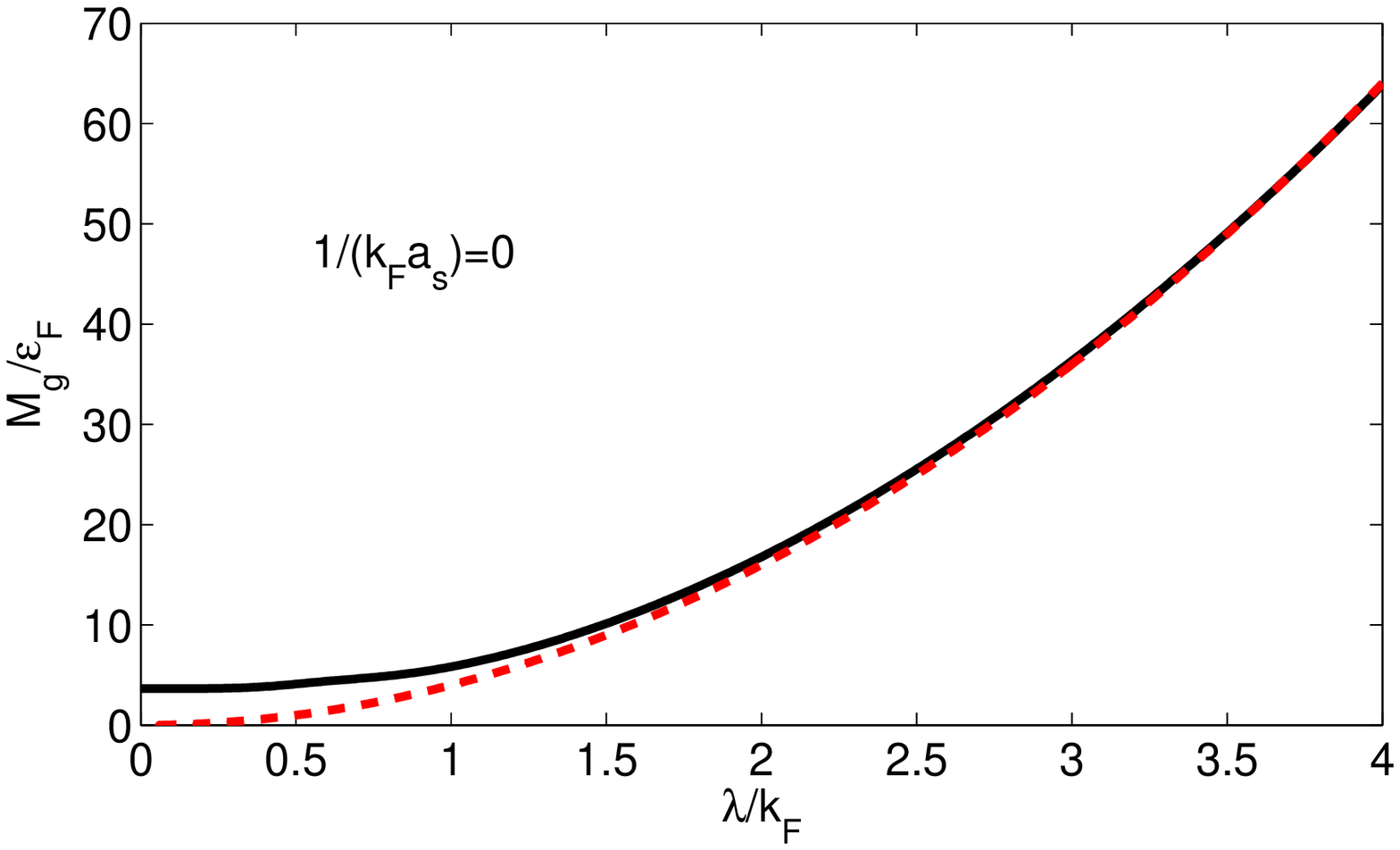}
\includegraphics[width=7.5cm]{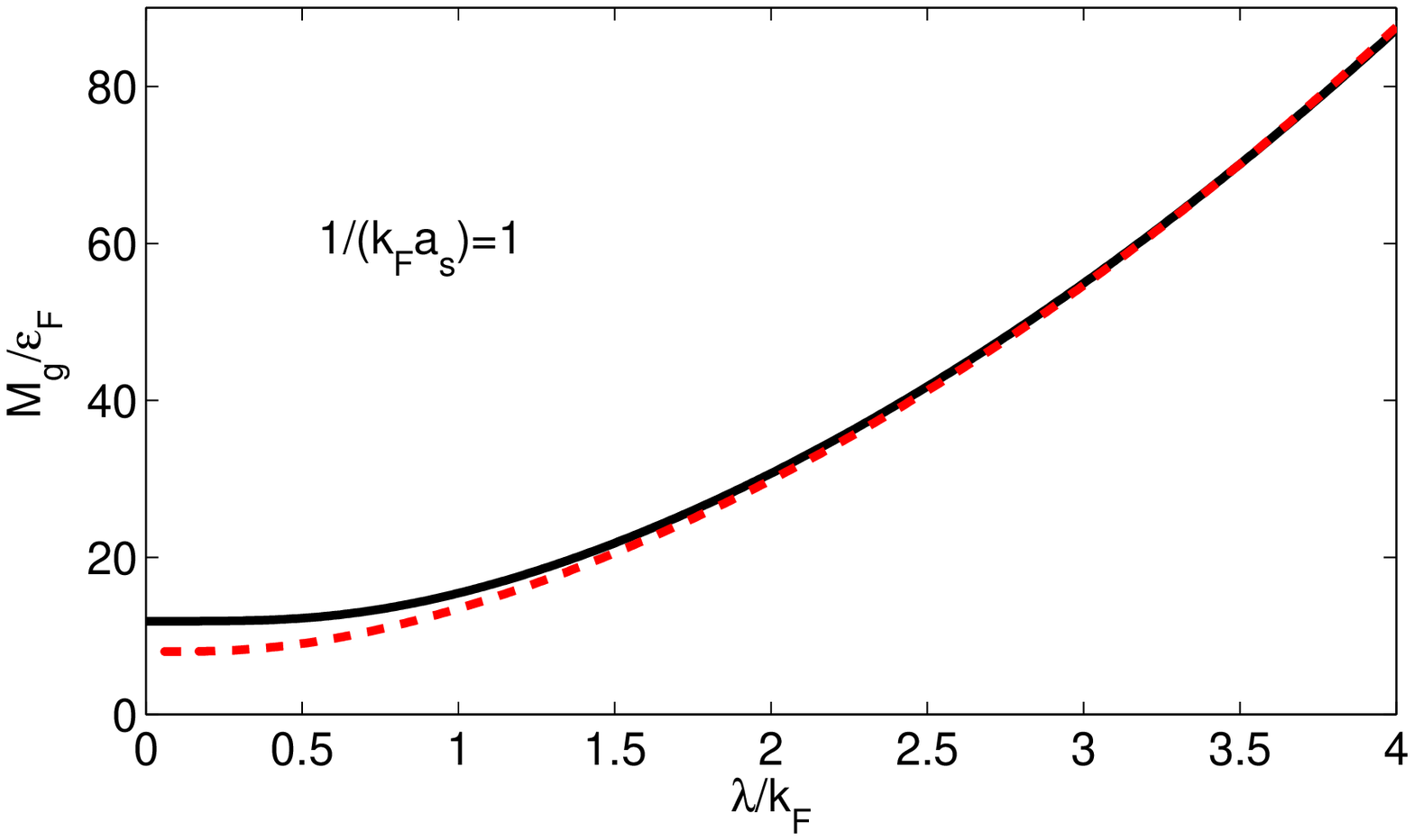}
\includegraphics[width=7.5cm]{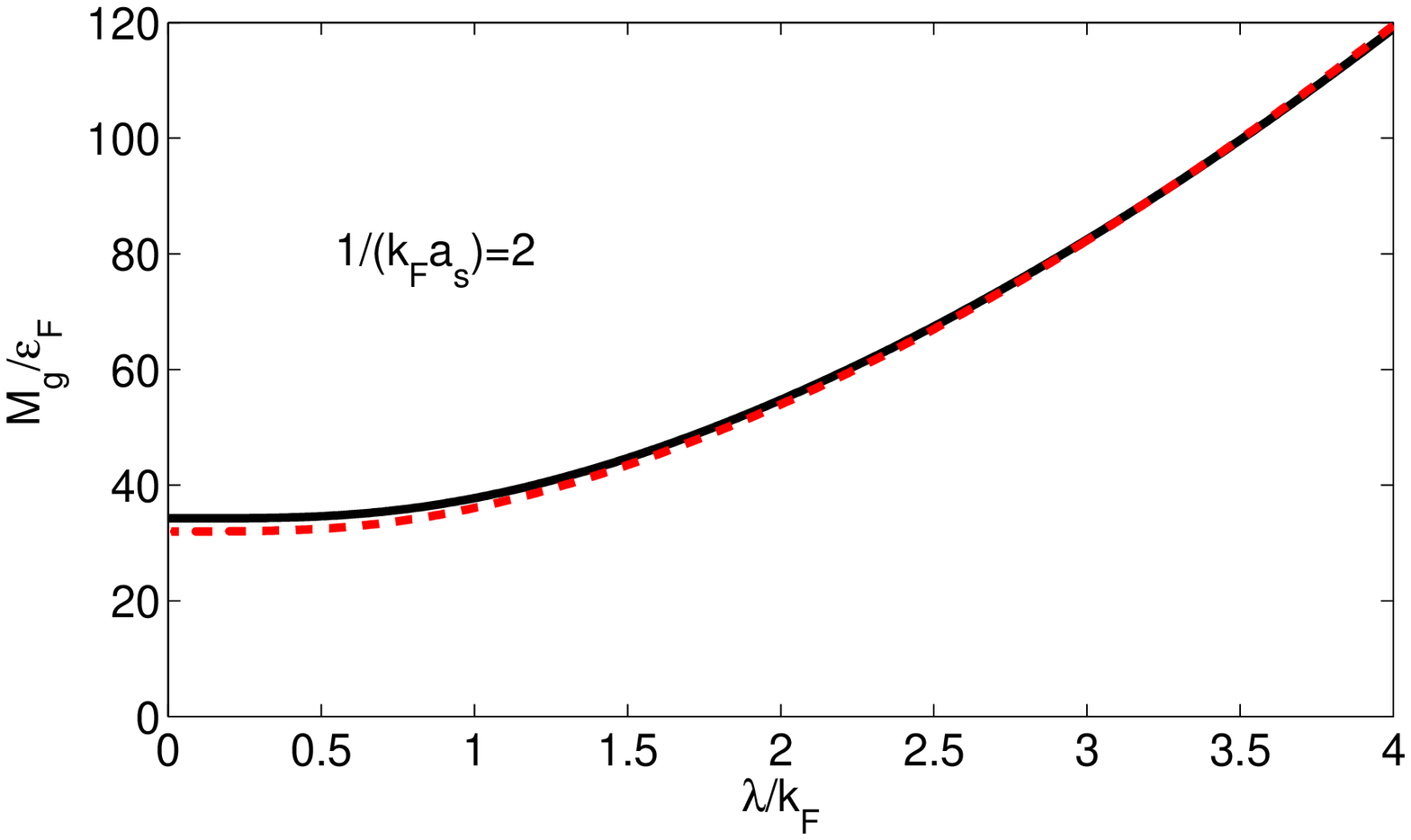}
\caption{The mass gap $M_g$ of the Anderson-Higgs mode (divided by
$\epsilon_{\rm F}$) as a function of $\lambda/k_{\rm F}$ for various
va;lues of $1/(k_{\rm F} a_s)$. The red dashed lines corresponds to
the analytical result (136).
 \label{fig12}}
\end{center}
\end{figure}

{\bf (B) Numerical Results.} Using the same trick in Section IV, we
obtain
\begin{eqnarray}
A &=& \frac{\Delta_0^2}{4\pi^2}\int_0^\infty dk \frac{k^2+\lambda^2}{[(\epsilon_k-\tilde{\mu})^2+\Delta_0^2]^{3/2}}
\equiv\frac{k_{\text F}}{2\pi^2}\tilde{A},\nonumber\\
B &=& \frac{1}{8\pi^2}\int_0^\infty dk
\frac{(k^2+\lambda^2)(\epsilon_k-\tilde{\mu})}{[(\epsilon_k-\tilde{\mu})^2+\Delta_0^2]^{3/2}}
\equiv\frac{1}{2\pi^2k_{\text F}}\tilde{B},\nonumber\\
D &=& \frac{1}{16\pi^2}\int_0^\infty dk
\frac{(k^2+\lambda^2)(\epsilon_k-\tilde{\mu})^2}{[(\epsilon_k-\tilde{\mu})^2+\Delta_0^2]^{5/2}}
\equiv\frac{1}{2\pi^2k_{\text F}^3}\tilde{D},\nonumber\\
R &=& \frac{1}{16\pi^2}\int_0^\infty dk
\frac{k^2+\lambda^2}{[(\epsilon_k-\tilde{\mu})^2+\Delta_0^2]^{3/2}}
\equiv\frac{1}{2\pi^2k_{\text F}^3}\tilde{R},\nonumber\\
Q &\equiv&\frac{1}{2\pi^2k_{\text F}}\tilde{Q},
\end{eqnarray}
where the dimensionless quantities
$\tilde{A},\tilde{B},\tilde{D},\tilde{R}$ and $\tilde{Q}$ are
defined as
\begin{eqnarray}
&&\tilde{A} = x_2^2\int_0^\infty dz\frac{z^2+g_1^2}{[(z^2-g_1^2-x_1)^2+x_2^2]^{3/2}},\nonumber\\
&&\tilde{B} = \int_0^\infty dz (z^2+g_1^2)\frac{z^2-g_1^2-x_1}{[(z^2-g_1^2-x_1)^2+x_2^2]^{3/2}},\nonumber\\
&&\tilde{D} = \int_0^\infty dz (z^2+g_1^2)\frac{(z^2-g_1^2-x_1)^2}{[(z^2-g_1^2-x_1)^2+x_2^2]^{5/2}},\nonumber\\
&&\tilde{R} = \int_0^\infty dz\frac{z^2+g_1^2}{[(z^2-g_1^2-x_1)^2+x_2^2]^{3/2}},\nonumber\\
&&\tilde{Q} = \frac{1}{3x_2^2}-\frac{
g_1}{6x_2^2}\sum_{\alpha=\pm}\alpha\int_0^\infty
zdz\frac{z^2+2\alpha
g_1z-x_1+\frac{x_2^2}{z^2-x_1}}{\sqrt{(z^2+2\alpha
g_1z-x_1)^2+x_2^2}}.\nonumber\\
\end{eqnarray}
Therefore, we have
\begin{equation}
\frac{c_s}{\upsilon_{\rm F}}=\sqrt{\tilde{Q}\over
\tilde{B}^2/\tilde{A}+\tilde{R}}
\end{equation}
and
\begin{eqnarray}
\frac{M_g}{\epsilon_{\rm
F}}=2\sqrt{\frac{\tilde{B}^2+\tilde{A}\tilde{R}}{\tilde{D}\tilde{R}}},
\end{eqnarray}
where $\upsilon_{\rm F}=k_{\text F}/m$ $(m=1)$ is the Fermi velocity
for the non-interacting Fermi gas in the absence of SOC.

Using the solutions of $x_1$ and $x_2$ from the gap and number
equations, we can calculate the quantity $c_s/\upsilon_{\rm F}$ and
$M_g/\epsilon_{\rm F}$ for given values of $1/(k_{\rm F}a_s)$ and
$\lambda/k_{\text F}$. The numerical results are shown in Figs.
\ref{fig11} and  \ref{fig12}. For large negative values of
$1/(k_{\rm F}a_s)$ and $\lambda/k_{\rm F}\rightarrow0$, we recover
the well-known result $c_s=\upsilon_{\rm F}/\sqrt{3}$ for weak
coupling fermionic superfluids \cite{BCSBEC3}. For negative values
or small positive values of $1/(k_{\rm F}a_s)$, the numerical result
becomes already in good agreement with the analytical results (124)
and (126) at $\lambda/k_{\rm F}\sim1$, which is consistent with the
observation that the system enters the rashbon BEC regime at
$\lambda/k_{\rm F}\sim 1$. For large positive values of $1/(k_{\rm
F}a_s)$, the numerical results are in good agreement with the
analytical results for all values of $\lambda/k_{\rm F}$.

For very large $\lambda/k_{\rm F}$, we find that the numerical
results fit very well with the following scaling behavior
\begin{eqnarray}
\frac{c_s}{\upsilon_{\rm
F}}=0.66\sqrt{\frac{2\pi}{3}}\left(\frac{\lambda}{k_{\rm
F}}\right)^{-1/2},\ \ \ \ \frac{M_g}{\epsilon_{\rm
F}}=4\left(\frac{\lambda}{k_{\rm F}}\right)^2,
\end{eqnarray}
for both negative and positive values of $1/(k_{\rm F}a_s)$, as
indicated from the analytical observations.

\section{Summary}
In summary, we have presented a comprehensive study of the BCS-BEC
crossover problem in 3D Fermi gases with a spherical spin-orbit
coupling which can be realized by a 3D symmetrical configuration of
the synthetic SU(2) gauge field. The two-body problem, the
superfluid ground-state properties, and the behaviors of collective
excitations are studied. Analytical results and interesting
universal behaviors for various physical quantities at large SOC are
obtained. We notice that there has been experimental proposal for
the realization of 3D spherical spin-orbit coupling in cold
fermionic atoms \cite{3DSOC}. Therefore, it is interesting to test
our theoretical predictions in future experiments of cold Fermi
gases with 3D spherical spin-orbit coupling.

\emph{Acknowledgments} --- L. He and X. -G. Huang acknowl- edge the
supports from the Helmholtz International Center for FAIR within the
framework of the LOEWE program (Landes- offensive zur Entwicklung
Wissenschaftlich- {\"O}konomischer Exzellenz) launched by the State
of Hesse. L. He also ac- knowledges the support from the Alexander
von Humboldt Foundation.

\emph{Note Added} --- During the preparation of this manuscript, we
became aware of the recent paper by Vyasanakere and Shenoy
\cite{3DVS}, where similar results were reported.

\end{document}